\documentclass[sigconf,nonacm,review=false]{acmart}

\usepackage{amsmath,amsfonts}

\usepackage[linesnumbered,vlined,ruled]{algorithm2e}
\usepackage{subcaption}
\usepackage{multicol}
\usepackage{multirow}
\usepackage{pifont}

\usepackage{enumitem}
\usepackage{url}

\newcommand{\cmark}{\ding{51}}%
\newcommand{\xmark}{\ding{55}}%

\newcommand\vldbdoi{10.14778/3523210.3523218}
\newcommand\vldbpages{XXX-XXX}
\newcommand\vldbvolume{15}
\newcommand\vldbissue{7}
\newcommand\vldbyear{2022}
\newcommand\vldbauthors{\authors}
\newcommand\vldbtitle{\shorttitle} 
\newcommand\vldbavailabilityurl{https://github.com/RapidsAtHKUST/ContinuousSubgraphMatching}
\newcommand\vldbpagestyle{empty}

\begin{document}

\title{An In-Depth Study of Continuous Subgraph Matching}


\author{Xibo Sun}
\affiliation{%
  \institution{Hong Kong University of Science and Technology}
  }
\email{xsunax@cse.ust.hk}

\author{Shixuan Sun}
\authornote{Shixuan Sun is the corresponding author.}
\affiliation{%
  \institution{National University of Singapore}
  }
\email{sunsx@comp.nus.edu.sg}

\author{Qiong Luo}
\affiliation{%
  \institution{Hong Kong University of Science and Technology}
  }
\email{luo@cse.ust.hk}

\author{Bingsheng He}
\affiliation{%
  \institution{National University of Singapore}
  }
\email{hebs@comp.nus.edu.sg}

\begin{abstract}
Continuous subgraph matching (CSM) algorithms find the occurrences of a given pattern on a stream of data graphs online.
A number of incremental CSM algorithms have been proposed. However, a systematical study on these algorithms is missing to identify their advantages and disadvantages on a wide range of workloads. Therefore, we first propose to model CSM as incremental view maintenance (IVM) to capture the design space of existing algorithms. Then, we implement six representative CSM algorithms, including IncIsoMatch, SJ-Tree, Graphflow, IEDyn, TurboFlux, and SymBi, in a common framework based on IVM. We further conduct extensive experiments to evaluate
the overall performance of competing algorithms as well as study the effectiveness of individual techniques to pinpoint the key factors leading
to the performance differences. We obtain the following new insights into the performance: (1) existing algorithms start the search from an edge in the query graph that maps to an updated data edge, potentially leading to many invalid partial results; (2) all matching orders are based on simple heuristics, which appear ineffective at times; (3) index updates dominate the query time on some queries; and (4) the algorithm with constant delay enumeration bears significant index update cost. Consequently, no algorithm dominate the others in all cases. Therefore, we give a few recommendations based on our experiment results. In particular, the SymBi index is useful for sparse queries or long running queries. The matching orders of IEDyn and TurboFlux work well on tree queries, those of Graphflow on dense queries or when both query and data graphs are sparse, and otherwise, we recommend SymBi's matching orders.  
\end{abstract}

\maketitle

\pagestyle{\vldbpagestyle}
\begingroup\small\noindent\raggedright\textbf{PVLDB Reference Format:}\\
\vldbauthors. \vldbtitle. PVLDB, \vldbvolume(\vldbissue): \vldbpages, \vldbyear.\\
\href{https://doi.org/\vldbdoi}{doi:\vldbdoi}
\endgroup
\begingroup
\renewcommand\thefootnote{}\footnote{\noindent
This work is licensed under the Creative Commons BY-NC-ND 4.0 International License. Visit \url{https://creativecommons.org/licenses/by-nc-nd/4.0/} to view a copy of this license. For any use beyond those covered by this license, obtain permission by emailing \href{mailto:info@vldb.org}{info@vldb.org}. Copyright is held by the owner/author(s). Publication rights licensed to the VLDB Endowment. \\
\raggedright Proceedings of the VLDB Endowment, Vol. \vldbvolume, No. \vldbissue\ %
ISSN 2150-8097. \\
\href{https://doi.org/\vldbdoi}{doi:\vldbdoi} \\
}\addtocounter{footnote}{-1}\endgroup

\ifdefempty{\vldbavailabilityurl}{}{
\vspace{.3cm}
\begingroup\small\noindent\raggedright\textbf{PVLDB Artifact Availability:}\\
The source code, data, and/or other artifacts have been made available at \url{\vldbavailabilityurl}.
\endgroup
}

\setlength{\textfloatsep}{0pt}
\section{Introduction} \label{sec:introduction}

\begin{figure}[h]
	\setlength{\abovecaptionskip}{0pt}
	\setlength{\belowcaptionskip}{-6pt}
	\captionsetup[subfigure]{aboveskip=3pt,belowskip=2pt}
	\centering
	\begin{subfigure}[t]{0.22\textwidth}
		\centering
		\includegraphics[scale=0.34]{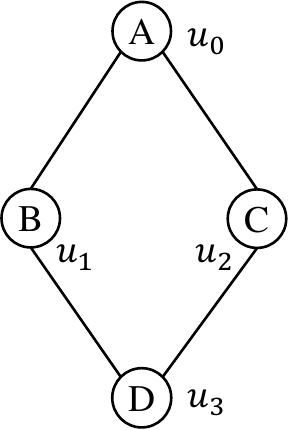}
		\caption{Query graph $Q$.}
		\label{fig:query_graph}
	\end{subfigure}
	\begin{subfigure}[t]{0.24\textwidth}
		\centering
		\includegraphics[scale=0.34]{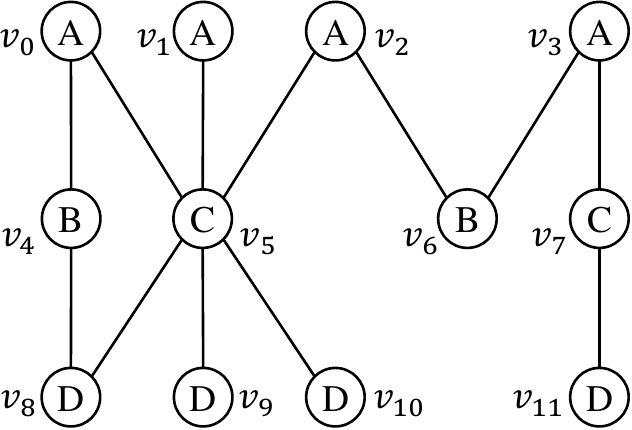}
		\caption{Data graph $G$.}
		\label{fig:data_graph}
	\end{subfigure}
	
	\begin{subfigure}[t]{0.22\textwidth}
		\centering
		\includegraphics[scale=0.34]{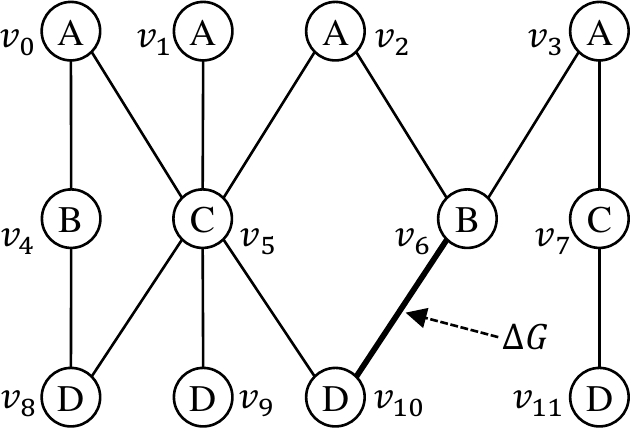}
		\caption{$G'$: Insert $e(v_6, v_{10})$ into $G$.}
		\label{fig:data_graph_insertion}
	\end{subfigure}
		\begin{subfigure}[t]{0.24\textwidth}
		\centering
		\includegraphics[scale=0.34]{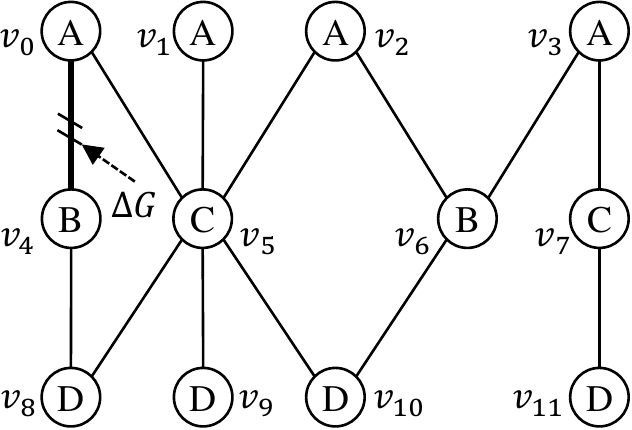}
		\caption{$G''$: Delete $e(v_0, v_4)$ from $G'$.}
		\label{fig:data_graph_deletion}
	\end{subfigure}
	\caption{Example graphs.}
	\label{fig:sample_graphs}
\end{figure}

\emph{Subgraph matching} (SM) is a fundamental operation in graph analysis,  finding all matches of a query graph $Q$ in a data graph $G$.
Extensive studies have been conducted on static graphs, utilizing pruning strategies, determining query plans, and proposing auxiliary data structures to
improve the performance~\cite{sun2020memory}. In contrast, as many real-world graphs change over time, \emph{continuous subgraph matching} (CSM)
reports matches in a graph stream. Specifically, for each graph update $\Delta G$ in the stream, CSM finds positive or negative matches
(called \emph{incremental matches}) on insertion or deletion of edges.
In this paper, we study the problem of exact CSM.
For example, $\{(u_0, v_0),(u_1, v_4), (u_2, v_5), (u_3, v_8)\}$ is a match given $Q$ and $G$ in Figure \ref{fig:sample_graphs}.
When the edge $e(v_6,v_{10})$ is inserted to $G$ in Figure \ref{fig:data_graph_insertion},
a positive match $\{(u_0, v_2), (u_1, v_6), (u_2, v_5), (u_3, v_{10})\}$ occurs in $G'$.
After that, $e(v_0,v_4)$ is deleted from $G'$ in Figure \ref{fig:data_graph_deletion}. A negative match
$\{(u_0, v_0),(u_1, v_4), (u_2, v_5), (u_3, v_8)\}$ disappears in $G''$.

The topic of CSM is essential in various scenarios.
For instance, social network providers detect the spread of rumors among users by matching rumor patterns in the streams of message
transmission graphs~\cite{wang2015detecting}. Similarly, financial administrators monitor cyclic patterns in transaction 
graphs for suspected money laundering offenses~\cite{qiu2018real}.
Another example is that IT departments continuously analyze system logs that record communications between computers to detect
system anomalies ~\cite{manzoor2016fast}.

\begin{figure}[h]\small
    \setlength{\abovecaptionskip}{3pt}
    \setlength{\belowcaptionskip}{-10pt}
    \includegraphics[width=0.475\textwidth]{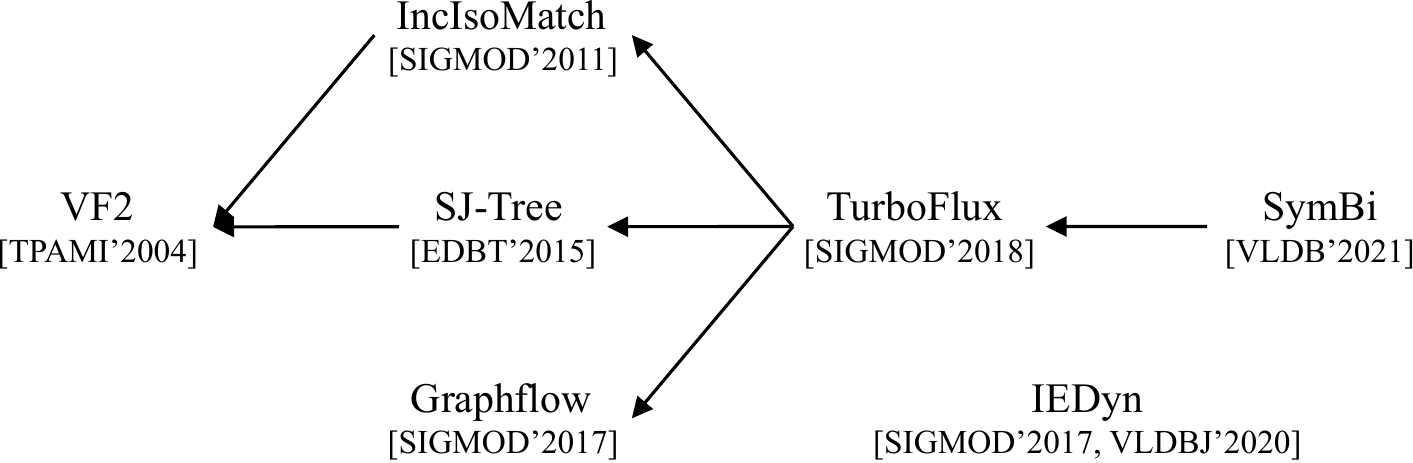}
    \centering
    \caption{Comparison of algorithms in previous experiments. An edge from \emph{A} to \emph{B} represents that the experiments of \emph{A} involved \emph{B}.
    VF2 denotes a re-computation method using VF2~\cite{cordella2004sub}, a classical SM algorithm.}
    \label{fig:algorithm_comparison}
\end{figure}

A number of exact CSM
algorithms have been proposed, including IncIsoMatch~\cite{fan2011incremental}, SJ-Tree~\cite{choudhury2015selectivity},
Graphflow~\cite{kankanamge2017graphflow}, IEDyn~\cite{idris2017dynamic, idris2020general}, TurboFlux~\cite{kim2018turboflux}, and SymBi~\cite{min2021symmetric}. To
find incremental matches efficiently, these algorithms all start the execution from the update edge in the data graph and recursively enumerate results
by mapping a \emph{query vertex} (i.e., a vertex in $Q$) to a \emph{data vertex} (i.e., a vertex in $G$) at a step. They further
proposed a variety of techniques to accelerate the enumeration and compared with previous algorithms in experiments.
Figure \ref{fig:algorithm_comparison} summarizes the timeline of these algorithms in previous experiments in the literature~\cite{fan2011incremental,choudhury2015selectivity,kankanamge2017graphflow,idris2017dynamic,idris2020general,kim2018turboflux,min2021symmetric}.

We make the following observations on existing work. First, no prior work has proposed a common framework to abstract the CSM problem and identify key performance factors. Second, existing work focused on evaluating the overall performance, but not individual strategies. Third, even though previous experiments have good coverage of algorithms (i.e., there is a path from the latest algorithm to other algorithms
in Figure \ref{fig:algorithm_comparison}), there is no comprehensive comparison between all these algorithms.  

\begin{figure}[h]\small
    \setlength{\abovecaptionskip}{3pt}
    \setlength{\belowcaptionskip}{-10pt}
    \includegraphics[scale=0.4]{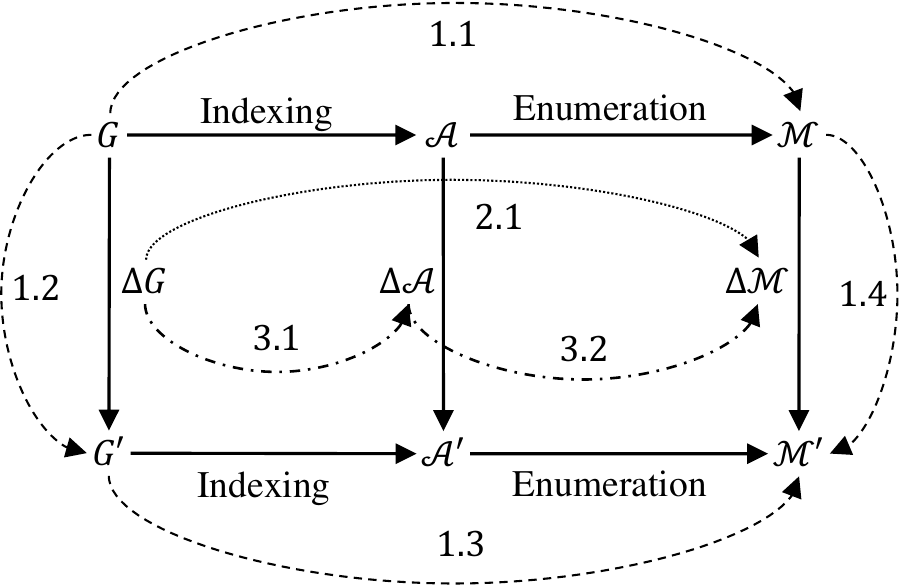}
    \centering
    \caption{Modeling continuous subgraph matching as incremental view maintenance. 1.1-1.4 show the execution flow of a recomputation-based method (IncIsoMatch);
    2.1 shows that of a direct-incremental algorithm (Graphflow); and 3.1-3.2 shows that of index-based incremental algorithms (SJ-Tree, IEDyn, TurboFlux and SymBi).}
    \label{fig:model}
\end{figure}

\textbf{Our Work.} We propose to conduct an in-depth study on sequential algorithms on exact CSM. We first model CSM as the \emph{incremental view maintenance} (IVM)
problem to capture the design space of existing algorithms. Figure \ref{fig:model} illustrates the general idea. Given a query graph $Q$, the
data graph $G$ is the base data. $\mathcal{M}$ is the set of matches in $G$, which is materialized. IVM aims to keep $\mathcal{M}$
consistent with the base data by computing and applying the incremental change $\Delta \mathcal{M}$ incurred by the base data update. A subtle
difference from IVM is that CSM computes $\Delta \mathcal{M}$, but does not store $\mathcal{M}$.

Based on this IVM model, we have three categories of existing CSM methods. The first kind is the \emph{recomputation-based} method, illustrated by 1.1-1.4 in Figure \ref{fig:model}. Given $\Delta G$ on $G$, this algorithm first computes the set $\mathcal{M}$
of matches in $G$, and then obtains $G'$ by applying $\Delta G$ to $G$. It next finds the set $\mathcal{M}'$ of matches in $G'$, and
finally gets $\Delta \mathcal{M}$ by computing the difference between $\mathcal{M}$ and $\mathcal{M}'$. It finds $\mathcal{M}$ and
$\mathcal{M}'$ with SM algorithms, which generally enumerate matches with the assistance of a lightweight index.

The other two kinds of methods are incremental. Specifically, the \emph{direct-incremental} method, illustrated by 2.1 in Figure \ref{fig:model}, computes $\Delta \mathcal{M}$
by directly searching matches containing edges in $\Delta G$ from the data graph. In contrast, the
\emph{index-based incremental} method, shown by 3.1-3.2 in Figure \ref{fig:model},
maintains a lightweight index $\mathcal{A}$ on which we can enumerate all matches of $Q$ in $G$. Given $\Delta G$, it will first update
$\mathcal{A}$ by computing and applying incremental changes $\Delta \mathcal{A}$ to $\mathcal{A}$, and then enumerate $\Delta \mathcal{M}$
based on the index. Thus, the lightweight index $\mathcal{A}$ is also a materialized view.

Based on our IVM model, we design a common framework for CSM and study six representative sequential algorithms on exact CSM, including IncIsoMatch, SJ-Tree, Graphflow, IEDyn, TurboFlux, and
SymBi, within the framework. These algorithms mainly differ in the method of building the index (if any)
and the method optimizing the matching order, so we introduce these algorithms regarding indexing and matching orders, and give an in-depth comparison and analysis.
Moreover, we compare the indexing methods with those in SM algorithms and discuss their commonalities and differences.

For a fair comparison, we re-implement the six algorithms in C++ within our framework and optimize them with our best efforts. This choice is because programming languages (Java or C++) as well as graph data structures (CSR, adjacency arrays, or others) affect the empirical performance. Moreover, only SJ-tree's source code is publicly available.  
With our implementation, We first evaluate the overall performance of competing algorithms in \emph{query time}, i.e., the elapsed time of answering a query given a graph stream, and the number of \emph{unsolved queries}, i.e., queries that cannot be completed within a time limit (one hour in our experiments). Moreover, we count the number of candidates in the index. Through experiments, we have the following results consistent with previous work.

\begin{enumerate}[leftmargin=*]
    \item The recomputation-based method is much slower than incremental methods~\cite{kim2018turboflux}.
    \item SJ-Tree runs out of memory in most cases due to the caching of partial results~\cite{kim2018turboflux}.
    \item SymBi is more stable than the other algorithms (i.e., it has fewer unsolved queries)~\cite{min2021symmetric}.
    \item The index reduces the number of data vertices involved in computation~\cite{kim2018turboflux,min2021symmetric},
    and the pruning power of the index in SymBi is stronger than that of TurboFlux~\cite{min2021symmetric}.
\end{enumerate}

However, we also find some new results: 1) the latest algorithms do not consistently outperform the old ones, 2) the algorithm with constant delay enumeration bears significant index update cost, and 3) the direct-incremental method runs faster than the index-based method in some cases. Specifically,

\begin{enumerate}[leftmargin=*]
    \item On tree queries, IEDyn and TurboFlux slightly outperform SymBi, and all of them run much faster than Graphflow.
    \item On sparse cyclic queries, both Graphflow and SymBi run faster than TurboFlux, and Graphflow performs better than SymBi
    on data graphs without dense substructures.
    \item On dense cyclic queries, Graphflow generally outperforms both TurboFlux and SymBi.
    \item If edges are updated at the sparse regions in the data graph, Graphflow runs faster than IEDyn, TurboFlux, and SymBi.
\end{enumerate}

Motivated by these new findings, we conduct more detailed experiments to pinpoint the factors leading to the performance differences among competing algorithms.
As the algorithms mainly differ in the indexes and matching orders, we evaluate their effectiveness individually.

Particularly, given a matching order, we compare the query performance using different indexes to evaluate
\emph{the effectiveness of the index}. Then, given the same index, we compare the query performance using different matching orders
to study \emph{the effectiveness of the matching order}. Based on the results, we conclude \emph{both are the key factors leading to the
performance differences}. Furthermore, in the experiments, we observe that some queries have a long running time; even worse, some queries cannot be completed within
the time limit by any algorithms. Therefore, we collect detailed metrics (e.g., the number of results and invalid partial results) and perform
case studies to answer the question: \emph{Where did time go in these queries?}
Based on our extensive experiments, we give a recommendation of algorithms on different workloads and discuss issues in existing algorithms.

\section{Background} \label{sec:background}

\subsection{Preliminaries} \label{sec:preliminaries}

This paper focuses on the undirected edge- and vertex-labeled graph $g = (V,E)$ where $V$ is a set of vertices and $E$ is a set of edges.
We also use $V(g)$ and $E(g)$ to denote the vertex set and edge set of graph $g$. 
Given $u \in V$, $N(u)$ denotes the neighbors of $u$, i.e., the vertices adjacent to $u$. $d(u)$ is the degree of $u$, i.e.,
$d(u) = |N(u)|$. $L_V$ (resp. $L_E$) is the function mapping from a vertex $u \in V$ (resp. an edge $e \in E$) to a label $l$ in a label
set $\Sigma_V$ (resp. $\Sigma_E$). We use $L$ to denote both mappings for brevity. $Q$ and $G$ denote the query graph and the data graph, respectively.
We use $e(u, u')$ to denote the edge between the vertices $u$ and $u'$.
We call vertices and edges of $Q$ (resp. $G$) query vertices and query edges (resp. data vertices and data edges). 

Definition \ref{def:hom} defines \emph{subgraph isomorphism} and \emph{subgraph homomorphism}. The difference between them
is that subgraph homomorphism allows duplicate vertices in a result, while subgraph isomorphism does not. We call a subgraph isomorphism
(or homomorphism) a \emph{match} for short. \emph{Subgraph matching} (SM) finds all matches $\mathcal{M}$ of $Q$ in $G$.

\begin{definition} \label{def:hom}
    Given graphs $g$ and $g'$, a subgraph isomorphism (resp. homomorphism) is an \textbf{injective function} (resp. \textbf{mapping}) $M$:
    $V(g) \rightarrow V(g')$ such that 1) $\forall u \in V(g), L(u) = L(M(u))$; 2) $\forall e(u, u') \in E(g), L(e(u, u')) = L(e(M(u), M(u')))$; and
    3) $\forall e(u, u') \in E(g), e(M(u), M(u')) \in E(g')$.
\end{definition}

The data graph $G$ is dynamic in this paper. $\Delta \mathcal{G}$ is a sequence of graph update operations $(\Delta G_1, \Delta G_2, ...)$ on $G$
where $\Delta G = \{ \Delta e\}$ is a set of edge insertions/deletions and $\Delta e = (+/-, e)$ is the insertion/deletion of an edge $e$.
Let $G'$ be the graph after applying the update $\Delta G$ on $G$. The \emph{incremental matches} $\Delta \mathcal{M}$
are the difference between $\mathcal{M}$ and $\mathcal{M}'$ where $\mathcal{M}$ and $\mathcal{M}'$ represent the matches of $Q$ in $G$
and $G'$, respectively. We call the results newly appearing the \emph{positive matches},
and that disappearing the \emph{negative matches}. We define the \emph{continuous subgraph matching} problem as follows.

\textbf{Problem Statement.} Given $Q$, $G$ and $\Delta \mathcal{G}$, continuous subgraph matching (CSM) finds incremental matches $\Delta \mathcal{M}$
for each $\Delta G \in \Delta \mathcal{G}$.

In the context of both subgraph isomorphism and homomorphism, SM and CSM are NP-hard~\cite{fan2011incremental}. 
A special case is that SM and CSM are
tractable for \emph{acyclic queries} (i.e., query graphs with no cycle) under the setting of subgraph homomorphism~\cite{yannakakis1981algorithms}.
Since subgraph isomorphism can be obtained by checking the injective constraint on subgraph homomorphism results, we use subgraph homomorphism as the default setting to illustrate the algorithms, following previous work. Nevertheless, we conduct experiments in both subgraph homomorphism and subgraph isomorphism settings.
We assume that all $\Delta e \in \Delta G$ have the same operation (either insertion or deletion) for ease of presentation. $\Delta G$ represents
the set of edges in the graph update, and $\Delta G.op$ denotes the operation. Given $\Delta G$ has mixed operations, we can handle $\Delta G$ as follows:
1) separate $\Delta G$ into $\Delta G_+$ and $\Delta G_-$ based on the operation; 2) delete edges $e$ from $\Delta G_+$ and $\Delta G_-$ if it exists
in both of them; and 3) first evaluate the query given $\Delta G_-$ and then evaluate it given $\Delta G_+$. Thus, the assumption does not break
the generality of our work.

\textbf{Graph Data Model.} We generally model the graph data frequently updated as \emph{dynamic graphs} or \emph{streaming graphs}~\cite{pacaci2020regular}.
Dynamic graphs model the graph data as a sequence of update operations on an initial data graph. In contrast, streaming graphs view the data
as a sequence of update operations and consider the graph constructed by the updates within a sliding window. The window can be defined by a time-constraint
or count-constraint. The old edges will be deleted with the sliding of the window. Furthermore, the models can be further divided into
\emph{single-update} ($|\Delta G| = 1$) based and \emph{batch-update} ($|\Delta G| > 1$) based according to the number of edges in
$\Delta G$~\cite{dhulipala2019low}. Existing CSM algorithms generally use the dynamic graph model with single-update~\cite{fan2011incremental,choudhury2015selectivity,kankanamge2017graphflow,kim2018turboflux,idris2017dynamic,idris2020general,min2021symmetric}. 

\textbf{Multi-way Join.} Given $Q$ and $G$, we can model the SM problem as a multi-way join $\mathcal{Q} = \Join_{e(u_x, u_y) \in E(Q)} R(u_x, u_y)$
where each vertex in $V(Q)$ corresponds to an attribute, each edge $e(u_x, u_y)\in E(Q)$ corresponds to a relation $R(u_x, u_y)$ and $R(u_x, u_y) = \{e(v_x, v_y) \in E(G) | L(u_x) = L(v_x) \bigwedge L(u_y) = L(v_y) \bigwedge L(e(u_x, u_y)) = L(e(v_x, v_y))\}$ \cite{aberger2017emptyheaded,mhedhbi2019optimizing,sun2020rapidmatch}. 
Worst-case optimal join (WCOJ)~\cite{ngo2018worst} is a class of join algorithms whose running time matches the largest query result size.
The latest algorithms~\cite{aberger2017emptyheaded,mhedhbi2019optimizing,sun2020rapidmatch} evaluate $\mathcal{Q}$ with WCOJ,
which extends partial results by vertex-at-a-time, instead of the traditional binary join, which extends partial results by edge-at-a-time.

\begin{algorithm}[b]
	\caption{Evaluating $\mathcal{Q}$ by vertex-at-a-time}
	\label{algo:wcoj}
	\footnotesize
	\SetKwFunction{Enumerate}{Enumerate}
	\SetKwProg{proc}{Procedure}{}{}
	\proc{\Enumerate{$\varphi, M, i$}}{
	 \lIf{$i = |\varphi| + 1$}{Output $M$, \KwRet}
	 \lElseIf{$i = 1$} {$u\leftarrow \varphi[i]$, $C_M(u) \leftarrow \bigcap_{u' \in N_- ^ \varphi (u)} \pi_{\{u\}} R(u, u')$}
	 \lElse{$u\leftarrow \varphi[i]$, $C_M(u) \leftarrow \bigcap_{u' \in N_+ ^ \varphi(u)} R(u':M(u'), u)$}
	 \ForEach{$v \in C_M(u)$}{
	        \If{$v$ is not visited}{
				Add $(u, v)$ to $M$\;
				\Enumerate{$\varphi, M, i + 1$}\;
				Remove $(u, v)$ from $M$\;
				}
		}
	}
\end{algorithm}

As shown in Algorithm \ref{algo:wcoj}, to enumerate all matches, WCOJ extends partial results by mapping a query vertex to a data vertex along an order of query vertices.
Algorithm \ref{algo:wcoj} presents the sketch of the procedure in the context of SM~\cite{sun2020rapidmatch}. $\varphi$ denotes the \emph{matching order},
which is a permutation of query vertices. $N_+ ^ {\varphi}(u)$ (resp. $N_- ^ {\varphi}$) denotes the neighbors of $u$ positioned before (resp. after)
$u$ in $\varphi$. $M$ is a mapping from query vertices to data vertices and $i$ records the recursion depth. Initially, $M=\{\}$ and $i=1$.
If all query vertices are mapped to data vertices, then output a result and return. Given $M$, we first compute the candidates $C_M(u)$ of the next
query vertex $u$ in $\varphi$ (Line 2). If $u$ is the first vertex, then $C_M(u)$ contains the common values of $u$ in relations that contain $u$ (Line 3).
Otherwise, $C_M(u)$ contains the common neighbors of data vertices mapped to query vertices $u'\in N_+ ^ {\varphi}(u)$ where $M(u')$ denotes
the data vertex mapped to $u'$ in $M$ and $R(u':M(u'), u)$ represents the neighbors of $M(u')$ in the relation $R(u', u)$ (Line 4). After that
we loop over $C_M(u)$ to map a query vertex to a data vertex and continue the search (Lines 5-9). If we want to find subgraph homomorphisms, we can 
remove the check at Line 6. To avoid Cartesian product in the enumeration, the matching order $\varphi$ is generally \emph{connected}, i.e.,
$N_+^{\varphi}(u) \neq \emptyset$ given $u \in V(Q)$ except the first vertex in $\varphi$.

\textbf{Incremental View Maintenance (IVM).} \emph{Materialized views} are queries whose results are stored~\cite{ramakrishnan2003database}.
We can use them to speed up query evaluation instead of computing from scratch. \emph{Incremental view maintenance} keeps
materialized views consistent with the base data by computing and applying the incremental changes incurred by
the base data update~\cite{griffin1995incremental}. The topic is well studied in relational databases for queries with different relational operators such as select-project-join~\cite{chirkova2011materialized}.
Given a multi-way join $\mathcal{Q}=\Join_{i \in [1, n]} R_i$, $\Delta R_i$ denotes the changes on the relation $R_i$, 
$R_i'$ is the relation before applying the changes, and $R_i$ is the relation
after applying the changes. The incremental results of $\mathcal{Q}$ (i.e., the difference between $\Join_{i \in [1, n]}R_i'$ and $\Join_{i \in [1, n]}R_i$)
can be computed by Equation \ref{eq:incremental_computation}~\cite{blakeley1986efficiently,gupta1993maintaining,griffin1995incremental}.
Join-based CSM methods~\cite{ammar2018distributed,kankanamge2017graphflow,mhedhbi2021optimizing} obtain $R$s and $\Delta R$s directly from the data graph $G$. In comparison, index-based algorithms \cite{idris2017dynamic, idris2020general, kim2018turboflux, min2021symmetric} apply pruning rules to filter the relations, store the results into indexes, and then obtain $R$s and $\Delta R$s from the indexes rather than from $G$.

\begin{equation} \label{eq:incremental_computation}
    \begin{aligned}
        \Delta \mathcal{Q}_1 &= \Delta R_1 \Join R_2' \Join ... \Join R_n',\\
        \Delta \mathcal{Q}_i &= R_1 \Join ... \Join R_{i - 1} \Join \Delta R_i \Join R_{i+1}' \Join ... \Join R_n', \\
        \Delta \mathcal{Q}_n &= R_1 \Join ... \Join R_{n - 1} \Join \Delta R_n,\\
        \Delta \mathcal{Q} &= \bigcup_i \Delta \mathcal{Q}_i.
    \end{aligned}
\end{equation}

\textbf{Constant Delay Enumeration.} Given $Q$, $G$, and $\Delta \mathcal{G}$ an algorithm finds $\mathcal{M}$ (or $\Delta \mathcal{M}$) in two phases: the preprocessing phase and the enumeration phase. In the latter, the algorithm reports each match in $\mathcal{M}$ (or $\Delta \mathcal{M}$) without repetition, followed by the end-of-enumeration message \texttt{EOE}. If the maximum time between the start of the enumeration phase and the output of the first match, between the output of two consecutive matches, and between the output of the last match and \texttt{EOE} only depends on the size of $Q$, then the enumeration is a \emph{constant delay} enumeration. Researchers proved that there exist an algorithm for \emph{acyclic queries} (i.e., query graphs with no cycle) under the setting of graph homomorphism which enumerates $\mathcal{M}$ with constant delay after $O(|V(Q)||E(G)|)$ time preprocessing \cite{bagan2007acylic, berkholz2020constant}.

\subsection{Related Work} \label{sec:related_work}

\textbf{Subgraph Matching.} Subgraph matching, which finds all matches of $Q$ of $G$, has been widely studied. Ullmann~\cite{ullmann1976algorithm} first proposed a graph exploration-based backtracking approach. 
Then, various approaches are presented~\cite{cordella2004sub, shang2008taming, bonnici2013subgraph, zhang2009gaddi, zhao2010graph, rivero2017efficient} to reduce the overall matching cost.
The most recent algorithms~\cite{han2013turboiso, bi2016efficient, bhattarai2019ceci, han2019efficient, kim2021versatile} summarized the data graph into an auxiliary data structure, which helps generate the query plan and enumerate subgraph matches. In contrast, another approach is to regard the graph as a database where attributes and relations correspond to vertices and edges and to conduct multi-way joins to find all matches~\cite{sun2020memory, aref2015design, aberger2017emptyheaded, tran2015fast, mhedhbi2019optimizing,sun2020rapidmatch}.
Nevertheless, both exploration-based and join-based algorithms follow the procedure in Algorithm \ref{algo:wcoj} to find all matches~\cite{sun2020rapidmatch}.

\textbf{Continuous Subgraph Matching.} In addition to CSM algorithms targeting arbitrary queries, a collection of studies conduct continuous subgraph matching on specific query types. \textsc{Casqd}~\cite{mondal2016casqd} was proposed to find cliques and stars in graph streams, and Qiu et al.~\cite{qiu2018real} presented GraphS that detects cycles. Subgraph isomorphism may be too restrictive for some applications, and therefore, Wang et al.~\cite{wang2009continuous} proposed to answer approximate subgraph containment queries that allow false-positive results.
Additionally, Fan et al.~\cite{fan2011incremental} and Song et al.~\cite{song2014event} worked on graph simulation~\cite{henzinger1995computing}, which relaxes the injective constraint. Other researchers found matches on the graph summarization instead of the data graph~\cite{zhao2011gsketch, tang2016graph}.
Additionally, Ammer et al.~\cite{ammar2018distributed} and Gao et al.~\cite{gao2016toward} detect patterns in the distributed environment.

\textbf{Multi-query Optimization for Subgraph Matching.} Several recent papers study the problem of matching a group of query graphs at one time. Specifically, Pugliese et al.~\cite{pugliese2014efficient} utilized a merged view of multiple query graphs to update results incrementally. Zervakis et al.~\cite{zervakisSTH20} designed a query graph clustering algorithm to handle a large number of continuous queries. Mhedhbi et al.~\cite{mhedhbi2021optimizing} proposed a general greedy optimizer to share computation among multiple instances of continuous queries. Currently, we focus on continuous subgraph matching for one query graph at a time.

\textbf{Subgraph Counting in Graph Streams.} In some applications, it is not necessary to explicitly enumerate all the subgraphs. Therefore, some work only counts the number of occurrences of the subgraph. Pavan et al. \cite{pavan2013counting}, Jha et al. \cite{jha2013space}, and Kara et al. \cite{kara2019counting} proposed to count triangles in graph streams, while Manjunath et al. \cite{manjunath2011approximate} focused on counting cycles of a given length. For counting arbitrary subgraphs, Kane et al. \cite{kane2012counting} proposed a novel method based on random vector, and Assadi et al. \cite{assadi2019simple} designed an efficient algorithm adopting edge sampling.

\section{A Generic Model} \label{sec:model}
To have a systematical study on existing continuous subgraph matching (CSM) algorithms and identify their advantages and disadvantages on a wide range of workloads, we attempt to develop a methodology to have common abstractions among different CSM algorithms. We find that existing CSM algorithms can be modelled as well studied incremental view maintenance (IVM) problems in relational databases~\cite{chirkova2011materialized}. Specifically, given $Q$ and $G$, we
can find all matches $\mathcal{M}$ with the multi-way join $\mathcal{Q}$, and
$\mathcal{Q}$ can be regarded as a materialized view over the
base data $G$. 
Thus, given $Q$ and $G$, and $\Delta G \in \Delta \mathcal{G}$, the CSM problem is equivalent to maintain the materialized view incrementally and the incremental matches $\Delta \mathcal{M}$ can be computed by Equation \ref{eq:incremental_computation} (i.e.,
$\Delta \mathcal{M} = \Delta \mathcal{Q}$).
A subtle difference is that IVM
maintains all results of $\mathcal{Q}$ incrementally by updating $\mathcal{M}$ with
$\Delta \mathcal{M}$, while CSM targets at $\Delta \mathcal{M}$ only. 
In the following, we present a common framework based on IVM for different CSM algorithms. In Section~\ref{sec:algorithms_study}, we use the common framework to compare the similarities and differences among the six algorithms under study.

As $\Delta G$ is relatively small compared with $G$,
using $E(G)$ in $\mathcal{Q}$ can result in many \emph{dangling tuples} (i.e., the tuples
do not appear in any join results). Therefore, we introduce an indexing phase before the enumeration.
Particularly, the indexing phase is to prune relations without breaking their completeness (Definition \ref{def:complete}).
We use $\mathcal{A}$ to denote the set of relations after pruning. However, the indexing phase also comes with overhead in its maintenance.
The relations in $\mathcal{A}$ are also materialized views of $G$. We must update it
to keep its completeness with respect to each data graph snapshot. To the end, the query evaluation
is divided into two phases: 1) update $\mathcal{A}$ given the graph update; and 2) enumerate incremental results using
$\mathcal{A}$.

\begin{definition} \label{def:complete}
    Given $Q$, $G$, and a query edge $e(u, u') \in E(Q)$, the relation corresponding to $e(u, u')$, $R(u, u')$, is a \emph{complete relation} if it contains all data edges $e(v, v') \in E(G)$ such that there exists a match of $Q$ in $G$ which maps $e(u, u')$ to $e(v, v')$.
\end{definition}

Algorithm \ref{algo:framework} presents a common framework of CSM based on IVM. Given $Q$ and $G$, we first generate an initial matching order $\varphi_0$ used in the indexing phase (Line 1). Then, we
build an index $\mathcal{A}$ in the offline processing stage (Line 2). In online processing,
we process the insertion by applying $\Delta G$ to $G$, updating the index, and finding incremental matches (Lines 4-7). In contrast,
we handle the deletion in the reverse order (Lines 8-11). Because positive matches appear in the updated graph and index, while negative
matches only exist in the data graph and index before the update. Intuitively, we find incremental results by enumerating the matches containing
edges in $\Delta G$ (Lines 12-19). We first relabel the query edges from $1$ to $n$, where $n$ is the number of query edges. 
For each relation $R_i$, we compute a subset $\Delta R_i$ containing only those edges updated (Line 13-14).
Then, we find results based on Equation \ref{eq:incremental_computation} (Lines 15-18).
Given $i \in [k + 1, n]$, $\Delta \mathcal{Q}_k$ uses $R_i - \Delta R_i$ to avoid reporting duplicate matches (Line 17).
We evaluate $\Delta \mathcal{Q}_k$ using Algorithm \ref{algo:wcoj}, or traditional binary joins. 
Finally, we output $\Delta \mathcal{M}$ as positive matches if $\Delta G.op$ is an insertion operation or negative matches otherwise (Line 19).
For direct-incremental methods, we ignore the execution of Lines 2, 6, and 10.
For recomputation-based methods, we use $\mathcal{Q}$ instead of $\Delta \mathcal{Q}$
to find matches in the data graph both before and after the update and get the incremental matches by computing the difference.
Proposition \ref{prop:correctness} shows the correctness of our framework. We put the proof of Proposition \ref{prop:correctness} and an example of the common framework in appendix.

\begin{algorithm}[t]\footnotesize
	\caption{A Common Framework of CSM using IVM}
	\label{algo:framework}
	\KwIn{a query graph $Q$, a data graph $G$, an update stream $\Delta \mathcal{G}$}
	\KwOut{incremental matches $\Delta \mathcal{M}$ for each $\Delta G \in \Delta \mathcal{G}$}
	\SetKwFunction{BuildInitialIndex}{BuildInitialIndex}
	\SetKwFunction{UpdateIndex}{UpdateIndex}
	\SetKwFunction{FindIncrementalMatch}{FindIncrementalMatch}
	\SetKwProg{proc}{Procedure}{}{}
	\tcc{Offline preprocessing}
	Generate an initial matching order $\varphi_0$\;
	$\mathcal{A} \gets$ \BuildInitialIndex{$Q, G, \varphi_0$}\;
	\tcc{Online processing}
	\ForEach{$\Delta G\in \Delta \mathcal{G}$}{
		\If{$\Delta G.op$ is $+$} {
			Apply $\Delta G$ to $G$\;
			\UpdateIndex{$Q, G, \mathcal{A}, \varphi_0, \Delta G$}\;
			\FindIncrementalMatch{$Q, G, \mathcal{A}, \Delta G$}\;      
		}
		\ElseIf{$\Delta G.op$ is $-$} {
			\FindIncrementalMatch{$Q, G, \mathcal{A}, \Delta G$}\;
			\UpdateIndex{$Q, G, \mathcal{A}, \varphi_0, \Delta G$}\;
			Apply $\Delta G$ to $G$\;
		}
	}
	\proc{\FindIncrementalMatch{$Q, G, \mathcal{A}, \Delta G$}}{
		\For{$k \gets 1$ to $n$ where $n = |E(Q)|$}{
			$\Delta R_k \leftarrow R_k \cap \Delta G$\;
		}
		$\Delta \mathcal{M} \leftarrow \{\}$\;
		\For{$k \gets 1$ to $n$ where $n = |E(Q)|$}{
			$\Delta \mathcal{Q}_k \leftarrow (\Join_{i \in [1, k - 1]}R_i) \Join \Delta R_k \Join (\Join_{i \in [k + 1, n]} (R_i - \Delta R_i))$\;
			$\Delta \mathcal{M}\leftarrow \Delta \mathcal{M} \cup \Delta \mathcal{Q}_k$\;
		}
		Output $\Delta \mathcal{M}$ as positive/negative matches if $\Delta G.op$ is $+$/$-$\;
	}
\end{algorithm}

\begin{proposition} \label{prop:correctness}
    The incremental matches output by our framework are correct. Namely, on each update, the $\Delta \mathcal{M}$ reported by Algorithm \ref{algo:framework} is exactly the difference of $\mathcal{M}$ before and after the update.
\end{proposition}

From the common framework, we can capture the key performance factors for CSM: 1) the efficiency and effectiveness of index,
which determines the maintenance cost of $\mathcal{A}$ and the relation cardinality in $\Delta \mathcal{Q}_k$; and 2)
the effectiveness of the join plan evaluating $\Delta \mathcal{Q}_k$. 
Existing algorithms mainly differ in the optimization of the two problems within the framework. We will present the details in the next section.

\section{Algorithms Under Study} \label{sec:algorithms_study}

\subsection{Overview}

This paper studies six sequential algorithms on exact CSM, including IncIsoMatch, SJ-Tree, Graphflow, IEDyn, TurboFlux, and SymBi. IncIsoMatch is recomputation-based, while
the other four are incremental methods. Our IVM-based framework can capture all CSM algorithms under study. 
First, it supports CSM algorithms with different enumeration methods, including vertex-at-a-time based and edge-at-a-time based.
Second, it supports algorithms with and without index. Particularly, if the CSM has no index, it directly obtains
relations from $G$. In contrast, it enumerates incremental matches based on the
index for indexing-based methods. Finally, our framework supports both algorithms that are of constant delay enumeration and those not.

Rather than storing relations in the index, native methods~\cite{idris2017dynamic,idris2020general,kim2018turboflux,min2021symmetric} generate a candidate vertex set $C(u)$ for each query vertex $u$. 
These algorithms can be seamlessly integrated into our framework. Particularly, given $e(u, u') \in E(Q)$,
if they maintain edges between candidates of $C(u)$ and $C(u')$, then the relation $R(u, u')$ is constructed by adding these
edges. Therefore, for ease of understanding, we introduce the indexing method of native algorithms in terms of the generation of candidate
vertex sets.

In the following, we first introduce the ordering, indexing, and enumeration techniques of each algorithm based
on our framework and then show their space and time complexity. The algorithms \texttt{BuildInitialIndex} and \texttt{UpdateIndex} for index-based methods, the implementation details within our framework, and the formal complexity analysis are presented in appendix.
Note that, within our framework, all algorithms under study can keep the same time and space complexity as the original papers.
Finally, we summarize the differences between these algorithms. 

In the original papers, the competing algorithms work on dynamic graphs with single-update. Without loss of generality,
we assume that $\Delta G = \{e(v_x, v_y)\}$, and the query edge $e(u_x, u_y)$ has the same label with $e(v_x, v_y)$, i.e.,
$L(u_x) = L(v_x), L(u_y)= L(v_y)$ and $L(e(u_x, u_y)) = L(e(v_x, v_y))$.  We focus on
the data structures and filtering rules in the indexing phase, and the join evaluation methods (especially the matching order)
in the enumeration phase (Line 17 in Algorithm \ref{algo:framework}) given $\Delta G$.

\subsection{IncIsoMatch~\cite{fan2011incremental}}

To the best of our knowledge, IncIsoMatch is the first CSM algorithm. It finds incremental matches by recomputation 
and does not generate the initial matching order $\varphi_0$ or index $\mathcal{A}$. 
On each update, IncIsoMatch extracts a subgraph $G'$ of $G$ based on $e(v_x, v_y)$
and enumerates results on $G'$ instead of $G$ to reduce the search space.

\textbf{Enumeration.} Let \emph{dia} denote the \emph{diameter} of $Q$, i.e., the length of the longest shortest path in $Q$. Given $\Delta G = \{e(v_x, v_y)\}$,
$V_{dia}$ contains the vertices in $G$ within $dia$ hops from both $v_x$ and $v_y$,
and $G_{dia}$ is the vertex-induced subgraph of $G$ on $V_{dia}$. IncIsoMatch computes $\Delta \mathcal{M}$ (Lines 15-18 in Algorithm \ref{algo:framework}) as follows: 
1) enumerate all matches $\mathcal{M}$ of $Q$ in $G_{dia}$ without $e(v_x, v_y)$;
2) enumerate all matches $\mathcal{M}'$ of $Q$ in $G_{dia}$ with $e(v_x, v_y)$; and 
3) computes the difference between $\mathcal{M}$ and $\mathcal{M}'$. 

\textbf{Complexity.} The time and space cost of extracting $G_{dia}$ by breath-first search is $O(|E(G)| + |V(G)|)$. IncIsoMatch is not worst-case optimal since the number of matches it finds may be greater than the number of incremental matches.

\subsection{Graphflow~\cite{kankanamge2017graphflow}}

Graphflow is a system that supports both SM and CSM queries with multi-way joins. 
It enumerates $\Delta \mathcal{M}$ without $\varphi_0$ or $\mathcal{A}$.

\textbf{Enumeration.} Given $\Delta G = \{e(v_x, v_y)\}$, Graphflow generates $\varphi$ as follows:
1) add $u_x, u_y$ to $\varphi$; and 2) repeatedly add a query vertex $u^*$, which is
not in $\varphi$ and with the maximum number of neighbors in $\varphi$, to $\varphi$ until all query vertices are selected.
If there are ties among vertices, Graphflow picks the vertex with a larger degree.
As the matching order for each query edge is fixed, Graphflow generates $\varphi$s offline
and directly retrieves it in online processing. Graphflow evaluates the join query by extending partial results along $\varphi$. 

\textbf{Complexity.} Graphflow finds $\Delta M$ with worst-case optimality.

\subsection{SJ-Tree~\cite{choudhury2015selectivity}}

SJ-Tree evaluates the join query with binary joins using the index.

\textbf{Ordering.} The initial matching order $\varphi_0$ contains a sequence of query edges. Given $Q$ and $G$, SJ-Tree defines the
selectivity of $e(u, u') \in E(Q)$ as the size of the relation $R(u, u')$ obtained from $G$. SJ-Tree generates $\varphi_0$ based on the selectivity:
1) add the edge with the minimum selectivity to $\varphi_0$; and 2) repeatedly add the edge, which
has the minimum selectivity among the edges with at least one endpoint in $\varphi_0$, to $\varphi_0$ until all query edges
are selected. 

\textbf{Indexing.} The index $\mathcal{A}$ is a left-deep tree, where the $i$th leaf node on the left maintains the relation associated with the $i$th query edge in $\varphi_0$, and an 
internal node records partial results of the query.
As SJ-Tree uses the hash join, the relation of a node $Q'$ is stored as a hash table.
The key is the common vertices between $Q'$ and $Q''$ where $Q''$ is the sibling node of $Q'$ in the left-deep tree. 

\textbf{Enumeration.} In online processing, SJ-Tree only supports the insertion operation. Given $\Delta G = \{e(v_x, v_y)\}$,
SJ-Tree inserts $e(v_x, v_y)$ into $R(u_x, u_y)$, and triggers the incremental computation along the
bottom-up order of the left-deep tree. The partial results of each join
operation are inserted into the relations of internal nodes. 
To fit the algorithm into our framework, SJ-Tree does not call the function \texttt{UpdateIndex} (Line 6) but directly enumerates incremental matches using the binary join (Line 17) and caches all partial results to the index. Similarly, offline processing is achieved by running SM on $G$ and caching the partial results.

\textbf{Complexity.} The space cost of the index, the time complexity of building and updating the index, and the time complexity of the incremental enumeration are $O(|E(G)| ^ {|E(Q)|})$ due to the edge-at-a-time join approach. Therefore, SJ-Tree is not worst-case optimal.

\begin{figure}[t]
	\setlength{\abovecaptionskip}{0pt}
	\setlength{\belowcaptionskip}{3pt}
	\captionsetup[subfigure]{aboveskip=3pt,belowskip=2pt}
	\centering
	\begin{subfigure}[t]{0.17\textwidth}
		\centering
    	\includegraphics[width=\textwidth]{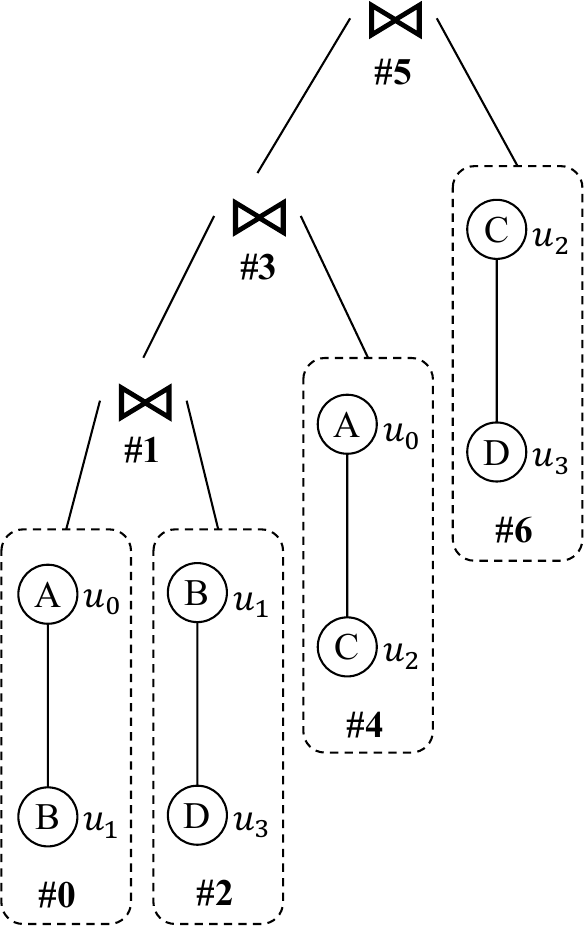}
		\caption{Left-deep tree.}
		\label{fig:sj-tree-example-1}
	\end{subfigure}
	\begin{subfigure}[t]{0.27\textwidth}
		\centering
    	\includegraphics[width=\textwidth]{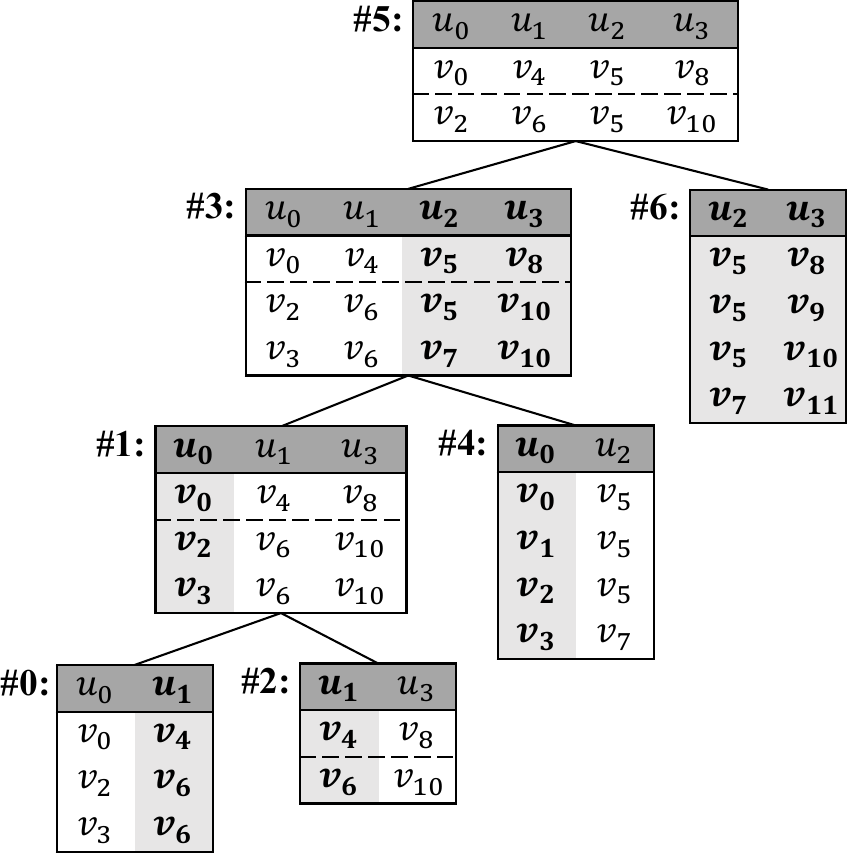}
		\caption{Index.}
		\label{fig:sj-tree-example-2}
	\end{subfigure}
	\caption{SJ-Tree for the update in Figure \ref{fig:data_graph_insertion}.}
	\label{fig:sj-tree-example}
\end{figure}

\begin{example}
    Figure \ref{fig:sj-tree-example-2} shows the index where each table corresponds to a node in the left-deep tree in Figure \ref{fig:sj-tree-example-1}
    and the join keys are grey shaded. The tuples below the dashed line in each table are newly added on the update operation. As $e(v_6, v_{10})$
    can be mapped to $e(u_1, u_3)$ based on their labels, it is first inserted into Table \#2. Then, the insertion triggers the join in the bottom-up
    order of the left-deep tree. The partial results generated by the join operation are stored in each table.
\end{example}

\subsection{TurboFlux~\cite{kim2018turboflux}}

The index of SJ-Tree takes exponential space. To solve the problem, TurboFlux proposes to store matches of paths in $Q$ without materialization. Furthermore, TurboFlux evaluates the query with the vertex-at-a-time method (Algorithm \ref{algo:wcoj}).

\textbf{Ordering.} Given $Q$ and $G$, TurboFlux first generates a spanning tree $Q_T$ of $Q$ rooted at $u_r$. $u_r$ is
selected as follows: 1) pick $e(u, u')$ appearing the least frequently in $G$; and 2) pick the endpoint of $e(u, u')$
appearing least frequently in $G$. Next, TurboFlux enlarges $Q_T$ by repeatedly adding the edge, which
has the fewest matches in $G$ and has one endpoint in $Q_T$, until all query vertices are selected.
$\varphi_0$ is a depth-first-order of all $v \in V(Q_T)$ starting from $u_r$. At Lines 2, 6, and 10 in Algorithm \ref{algo:framework}, TurboFlux pass $Q_T$ to the functions rather than $Q$.

\textbf{Indexing.} The index of TurboFlux, called DCG, has the same structure as $Q_T$, maintaining candidate vertex sets and data edges between candidates. 
Given $u \in V(Q_T)$, we denote $P_u$ as the path from $u_r$ to $u$ in $Q_T$ and $T_u$ as the subtree of $Q_T$ rooted at $u$.
DCG puts a data vertex into a candidate set based on Proposition \ref{prop:turboflux}.

\begin{proposition} \label{prop:turboflux}
    Given $v \in C(u)$, if the mapping $(u, v)$ appears in a match of $Q$, then
    $v$ satisfies that 1) it appears in a match of $P_u$; and 2) it appears in a match of $T_u$. 
\end{proposition}

TurboFlux also maintains $C_{im}(u)$, 
i.e., the \emph{implicit candidate} set of $u$ for each $u \in V(Q)$. $v \in C_{im}(u)$ if $v$ only satisfies the first condition.
In offline processing, DCG sets candidate sets in two phases. In the forward phase, DCG processes each $u$ along $\varphi_0$ by the following operations: given $v\in V(G)$, if $L(v) = L(u)$ and $v$ has a neighbor in $C_{im}(u')$ for each $u' \in N_+^{\varphi_0}(u)$, then insert $v$ to $C_{im}(u)$. Next, in the backward phase, DCG processes each $u$ along reversed order of $\varphi_0$ by the following operations: given $v\in V(G)$, if $L(v) = L(u)$, $v$ has a neighbor in $C(u')$ for each $u' \in N_-^{\varphi_0}(u)$ and $v \in C_{im}(u)$, then insert $v$ to $C(u)$. 

In the online phase, given $\Delta G = \{e(v_x, v_y)\}$, if $e(u_x, u_y)$ belongs to $E(Q_T)$, then we assume that $u_x$ precedes $u_y$ in $\varphi_0$ without loss of generality. DCG updates candidate sets using the same operations as those in offline processing starting from $u_y$. If $e(u_x, u_y)$ is a non-tree edge, TurboFlux does not update DCG. The index update for deletion is similar to that for insertion, and thus we omit the details.

\begin{figure}[t]
	\setlength{\abovecaptionskip}{0pt}
	\setlength{\belowcaptionskip}{3pt}
	\captionsetup[subfigure]{aboveskip=3pt,belowskip=2pt}
	\centering
	\begin{subfigure}[t]{0.21\textwidth}
		\centering
    	\includegraphics[width=\textwidth]{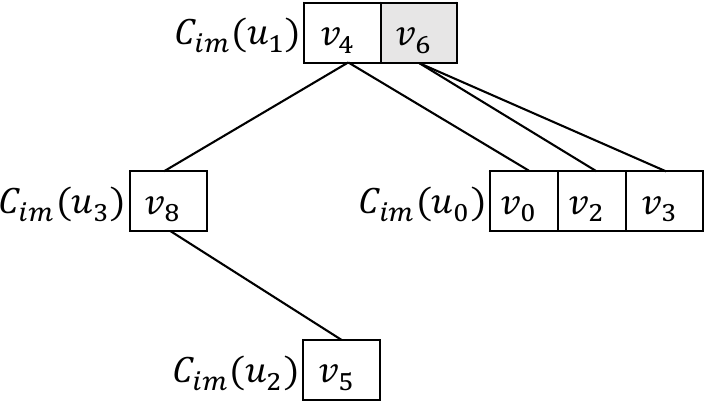}
		\caption{Forward phase.}
		\label{fig:turboflux-example-1}
	\end{subfigure}
	\hfill
	\begin{subfigure}[t]{0.21\textwidth}
		\centering
    	\includegraphics[width=\textwidth]{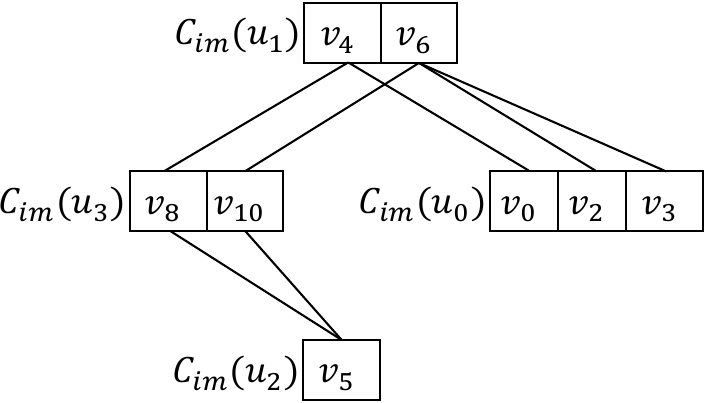}
		\caption{Backward phase.}
		\label{fig:turboflux-example-2}
	\end{subfigure}
	\caption{DCG for the update in Figure \ref{fig:data_graph_insertion}.}
	\label{fig:turboflux-example}
\end{figure}

\begin{example}
    Given $Q$ in Figure \ref{fig:query_graph}, suppose that $Q_T$ is rooted at $u_1$ with $e(u_0, u_2)$ as the non-tree edge.
    Figure \ref{fig:turboflux-example-1} shows the DCG in which each node records an implicit candidate set. A vertex $v$ is grey shaded if $v \in C_{im}(u)$ but $v \notin C(u)$. When $e(v_6,v_{10})$ is inserted into $G$,
    DCG first starts the forward phase. $v_{10}$ is inserted into $C_{im}(u_3)$ and the edge $e(v_{10}, v_5)$ is recorded.
    In the backward phase, $v_{10}$ is inserted to $C(u_3)$ since $v_{10}$ has a subtree matching $T_{u_3}$. After that, $v_6$ is inserted to $C(u_1)$ because $v_6$ has a subtree matching $T_{u_1}$.
\end{example}

\textbf{Enumeration.} The enumeration only considers the candidates in $C(u)$ for each $u \in V(Q)$ according to Proposition \ref{prop:turboflux}.
Given $Q$ and $G$, TurboFlux first generates $\varphi$ starting from $u_r$ as follows: 1) repeatedly delete the leaf node $u \in V(Q_T)$
with the minimum number of matches of $P_u$ in DCG; and 2) $\varphi$ is the reverse sequence of deletion.
After that, given $\Delta G = \{e(v_x, v_y)\}$, TurboFlux generates a matching order $\varphi'$ starting from $u_x$ and $v_y$ as follows: 1) add $u_x, u_y$ to $\varphi'$;
2) add the vertices in $P_{u_x}$ along the order from $u_x$ to $u_r$; and 3) add other vertices to $\varphi'$ along the order of the $\varphi$ that starts from $u_r$.
Because all matching orders are fixed in online processing, we can precompute them for each query edge.

\textbf{Complexity.} The space and time complexity of the DCG are both $O(|E(G)||V(Q)|)$ and the enumeration is wort-case optimal.

\subsection{SymBi~\cite{min2021symmetric}}

The index of TurboFlux does not use the non-tree edges to prune candidate vertex sets. To solve the problem, SymBi proposes to prune candidate vertex sets using all query edges.

\textbf{Ordering.} Given $Q$, SymBi first builds a DAG (directed acyclic graph) $Q_D$ of $Q$ by executing
a breadth-first-search from a root vertex $u_r$. Each edge is directed from the earlier visited vertex to the later visited one.
The height of $Q_D$ is the length of the longest path
from $u_r$ to sink vertices $u_s$. SymBi selects a query vertex as $u_r$ such that $Q_D$ has the highest height. $\varphi_0$ is the breadth-first order of $Q$ starting from $u_r$.

\textbf{Indexing.} SymBi builds an index called DCS, maintaining candidate vertex sets and data edges between candidates similar to TurboFlux. 
We denote $\mathcal{P}_u$ as the set of paths from $u_r$ to $u$ in $Q_D$ and $\mathcal{T}_u$
as the set of paths from $u$ to each $u_s$ in $Q_D$. 
SymBi puts a data vertex into a candidate set based on Proposition \ref{prop:symbi}.

\begin{proposition} \label{prop:symbi}
    Given $v \in C(u)$, if the mapping $(u, v)$ appears in a match of $Q$, then $v$ satisfies that 1) for each $P \in \mathcal{P}_u$,
    $(u, v)$ exists in a match of $P$; and 2) for each $P \in \mathcal{T}_u$, $(u, v)$ exists in a match of $P$.
\end{proposition}

The offline and online processing of SymBi is the same as that of TurboFlux, except that the first input parameter of function-calls at Lines 2, 6, and 10 in Algorithm \ref{algo:framework} are $Q$ rather than $Q_T$. 

\textbf{Enumeration.} The enumeration only considers the candidates $C(u)$ for each $u \in V(Q)$ according to Proposition \ref{prop:symbi}. 
We illustrate the matching order determination of SymBi based on Algorithm \ref{algo:wcoj}. Given an intermediate result $M$,
$N_M(u)$ is the set of neighbors of $u$ in $M$. $X_M$ is the set of query vertices not in $M$ but
$N_M(u) \neq \emptyset$. At Line 4, SymBi selects $u \in X_M$ with the minimum value of $|C_M(u)|$ as the next query vertex to extend $M$.
$C_M(u)$ is computed based on $N_M(u)$ instead of $N_+^{\varphi}(u)$.

\textbf{Complexity.} The space and time complexity of the DCS are both $O(|E(G)||E(Q)|)$ and the enumeration is worst-case optimal.

\begin{table*}[t]
\setlength{\abovecaptionskip}{2pt}
\setlength{\belowcaptionskip}{0pt}
\centering
\caption{Comparison of algorithms under study. \emph{Inc} denotes incremental computation. 
$+/-$ denotes single insertion/deletion.}
\small
\label{tab:comparison}
\resizebox{\textwidth}{!}{%
\begin{tabular}{c|l|cc|cc|cc|cc|c|c|c}
\hline
\multirow{3}{*}{\textbf{Algorithm}}
& \multirow{3}{*}{\begin{tabular}[c]{@{}c@{}}\textbf{Computation}\\ \textbf{Method}\end{tabular}} 
& \multicolumn{2}{c|}{\textbf{Index}} 
& \multicolumn{2}{c|}{\textbf{Enumeration}}
& \multicolumn{7}{c}{\textbf{Functionality}}\\ \cline{3-4} \cline{5-13}
&& \textbf{Space} & \textbf{Time} 
& \textbf{Worst-case} & \textbf{Constant}
& \multicolumn{2}{c|}{\textbf{Edge}} & \multicolumn{2}{c|}{\textbf{Vertex}}
& \textbf{Label} & \textbf{Batch} 
& \textbf{Early} \\
&& \textbf{Complexity} & \textbf{Complexity} 
& \textbf{Optimality}  & \textbf{Delay}
& \textbf{$+$} & \textbf{$-$} & \textbf{$+$} & \textbf{$-$}
& \textbf{Update} & \textbf{Update}
& \textbf{Termination}\\ \hline\hline

IncIsoMatch & Recomputation & N/A & N/A & \xmark & \xmark & \cmark & \cmark & \cmark & \cmark & \cmark & \xmark & \xmark \\ \hline
Graphflow & Direct Inc & N/A & N/A & \cmark & \xmark & \cmark & \cmark & \cmark & \cmark & \cmark & \cmark & \cmark \\ \hline
SJ-Tree& Index-based Inc& $O\left(|E(G)|^{|E(Q)|}\right)$ & $O\left(|E(G)|^{|E(Q)|}\right)$ & \xmark & \xmark & \cmark & \xmark & \cmark & \xmark & \xmark & \xmark & \xmark \\ \hline
IEDyn & Index-based Inc& $O(\left|E(G)| |V(Q)|\right)$ & $O\left(|E(G)| |V(Q)|\right)$ & \cmark & \cmark & \cmark & \cmark & \cmark & \cmark & \cmark & \cmark & \cmark \\ \hline
TurboFlux & Index-based Inc& $O\left(|E(G)| |V(Q)|\right)$ & $O\left(|E(G)| |V(Q)|\right)$ & \cmark & \xmark & \cmark & \cmark & \cmark & \cmark & \cmark & \xmark & \cmark \\ \hline
SymBi  & Index-based Inc& $O\left(|E(G)| |E(Q)|\right)$ & $O\left(|E(G)| |E(Q)|\right)$ & \cmark & \xmark & \cmark & \cmark & \cmark & \cmark & \cmark & \xmark & \cmark \\ \hline
\end{tabular}%
}
\end{table*}

\subsection{IEDyn~\cite{idris2017dynamic, idris2020general}} \label{sec:iedyn} 

IEDyn is a CSM algorithm for acyclic queries, achieving constant delay enumeration under the setting of graph homomorphism.

\textbf{Ordering.} Given $Q$ and $G$, IEDyn first selects an arbitrary $u_r \in V(Q)$ as the root of $Q$. 
Then, IEDyn generates $\varphi_0$ by executing a depth-first-search on $Q$ starting from $u_r$.

\textbf{Indexing.} The index of IEDyn maintains candidate vertex sets and data edges between candidates, similar to TurboFlux. 
IEDyn puts a data vertex into a candidate set based on Proposition \ref{prop:iedyn}.

\begin{proposition} \label{prop:iedyn}
    If the mapping $(u, v)$ appears in a match of $Q$ in $G$, then
    $v$ appears in a match of $T_u$.
\end{proposition}

IEDyn only performs the backward phase, and the operations are the same as those of TurboFlux. We call the index built in the offline phase the \emph{global index}. On each update, IEDyn can update the global index directly and find all matches $\mathcal{M}$ based on the global index. However, since CSM focuses on $\Delta M$, IEDyn maintains a \emph{local index}, storing an additional candidate set $C_\Delta(u)$ for each $u \in V(Q)$.
In online processing, given $\Delta G = \{e(v_x, v_y)\}$, we assume that $u_x$ precedes $u_y$ in $\varphi_0$. For each $u \notin P_{u_x}$, IEDyn assigns $C(u)$ to $C_\Delta(u)$ directly.
After that, IEDyn performs the same operations as those in offline processing starting from $u_x$ to set $C_\Delta(u)$ where $u \in P(u_x)$.

\textbf{Enumeration.} The matching order for enumeration $\varphi$ is the same as $\varphi_0$. The enumeration is performed on the local index. After the enumeration, the local index is merged into the global index.

\begin{figure}[t]
	\setlength{\abovecaptionskip}{0pt}
	\setlength{\belowcaptionskip}{3pt}
	\captionsetup[subfigure]{aboveskip=3pt,belowskip=2pt}
	\centering
	\begin{subfigure}[t]{0.1525\textwidth}
		\centering
    	\includegraphics[width=\textwidth]{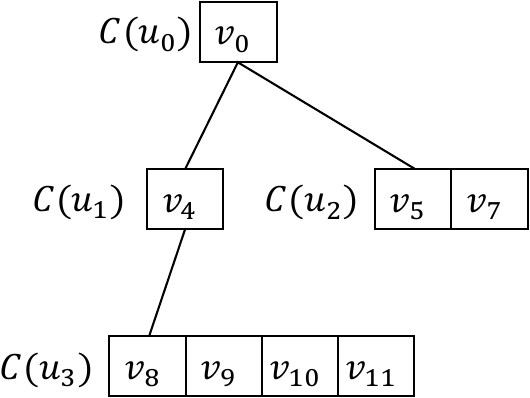}
		\caption{Before update.}
		\label{fig:iedyn-example-1}
	\end{subfigure}
	\begin{subfigure}[t]{0.1525\textwidth}
		\centering
    	\includegraphics[width=\textwidth]{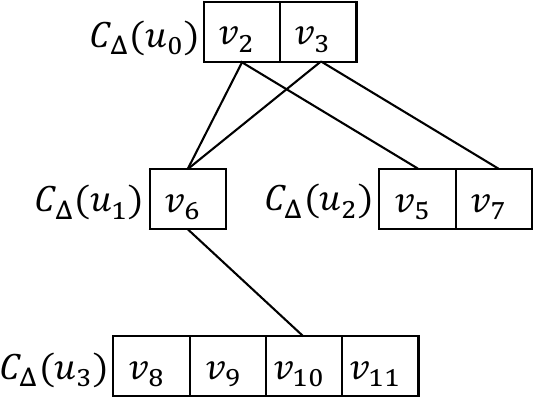}
		\caption{On update.}
		\label{fig:iedyn-example-2}
	\end{subfigure}
	\hfill
	\begin{subfigure}[t]{0.1625\textwidth}
		\centering
    	\includegraphics[width=\textwidth]{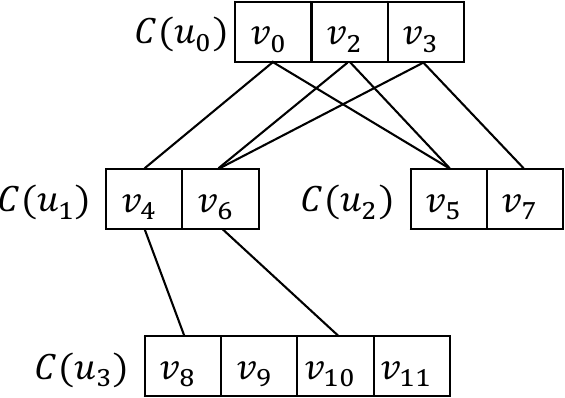}
		\caption{After update.}
		\label{fig:iedyn-example-3}
	\end{subfigure}
	\caption{Global and local index for update in Figure \ref{fig:data_graph_insertion}.}
	\label{fig:iedyn-example}
\end{figure}

\begin{example}
    Suppose we get $Q$ by removing the edge $e(u_2, u_3)$ in Figure \ref{fig:query_graph} and $G$ from Figure \ref{fig:data_graph}, Figure \ref{fig:iedyn-example-1} shows the global index before the update. When $e(v_6, v_{10})$ is inserted into $G$, it matches $e(u_1, u_3)$. IEDyn first assigns $C(u_2)$ to $C_\Delta(u_2)$ and $C(u_3)$ to $C_\Delta(u_3)$. Then, since $v_6$ has a subtree matching $T_{u_1}$, $v_6$ is inserted in $C_\Delta(u_1)$. After that, $v_2$ and $v_3$ are inserted in $C_\Delta(u_0)$ since both of them have neighbors in $C_\Delta(u_1)$ and $C_\Delta(u_2)$. The local index on update is shown in Figure \ref{fig:iedyn-example-2}. After the enumeration, the local index is merged into the global index, shown in Figure \ref{fig:iedyn-example-3}.
\end{example}

\textbf{Complexity.} The time and space complexity of the index of IEDyn are both $O(|E(G)||V(Q)|)$. The enumeration is worst-case optimal. Additionally, the algorithm can enumerate incremental results of acyclic queries under graph homomorphism with constant delay (Proposition \ref{prop:iedyn-constant-delay}). We put the proof in appendix.

\begin{proposition} \label{prop:iedyn-constant-delay}
    Given $Q$, $G$, and $\Delta G \in \Delta \mathcal{G}$, where $Q$ is a tree, IEDyn enumerates all incremental matches with constant delay after $O(|E(G)||V(Q)|)$ time of preprocessing.
\end{proposition}

\subsection{Comparison} \label{sec:comparison}

The space complexity of the index and the time complexity of updating the index on each update of competing algorithms are summarized in Table \ref{tab:comparison}. In the following, we compare the algorithms under study in their enumeration, and functionalities. 

\textbf{Enumeration.} Table \ref{tab:comparison} summarizes the worst-case optimality of each algorithm. IncIsoMatch is not worst-case optimal since the number of matches it finds may be greater than the number of incremental matches. SJ-Tree extends a partial match by an edge at a time, and thus not worst-case optimal either. Among all methods, only IEDyn supports constant delay enumeration under the setting of tree homomorphism, as stated in Proposition \ref{prop:iedyn-constant-delay}. The enumeration of IncIsoMatch is not of constant delay because the algorithm needs to find all matches in $G_{dia}$ to get the first incremental result. SJ-Tree does not support constant delay enumeration either. When an edge is inserted into a leaf node of the left-deep tree, the edge is joined with all partial results on its sibling node with the same key. However, there may be up to $V(G)$ partial results that do not lead to a match. TurboFlux and SymBi also fail to enumerate incremental matches with constant delay, because their indexes store both $C(u)$ and $C_{im}(u)$ for each $u \in V(Q)$. When the algorithms get the neighbors of a candidate (Line 3 or Line 4 in Algorithm \ref{algo:wcoj}), they may visit up to $V(G)$ vertices $v$ where $v \in C_{im}(u')$ but $v \notin C(u')$. Since our framework supports different indexing and ordering methods, it can accommodate both algorithms that are of constant delay enumeration and those not.

\textbf{Functionality.} Table \ref{tab:comparison} also illustrates the functionalities of the algorithms under study. All algorithms support edge insertion. All algorithms expect SJ-Tree support edge deletion. 
No algorithm provides specialized methods for vertex update. But a vertex insertion (resp. deletion) can be achieved by inserting (resp. deleting) the vertex and all the edges adjacent to the vertex. Therefore, all algorithms support vertex insertion, while SJ-Tree does not support vertex deletion because it cannot perform edge deletion. Similarly, no algorithm provides specialized methods for label updates of a vertex or an edge. But the label update of a vertex (resp. an edge) can be achieved by deleting the vertex (resp. the edge) and then inserting a vertex (resp. an edge) with the new label. Therefore, IncIsoMatch, Graphflow, IEDyn, TurboFlux, and SymBi support label updates because they can perform both the insertion and the deletion of a vertex or an edge. Furthermore, only Graphflow and IEDyn were proposed with batch update support initially. The index update operations of SJ-Tree, TurboFlux, and SymBi work for single edge insertion or deletion only. As IncIsoMatch extracts a subgraph of $G$ based on the distance to the updated edge, it processes a single update at a time. We also show the early termination support of each algorithm, namely the ability to return after finding a specific number of incremental results on each update. IEDyn, Graphflow, TurboFlux, and SymBi support early termination. IncIsoMatch cannot return before finding all incremental matches since it computes them based on the difference between results in $G_{dia}$ before and after the update. SJ-Tree cannot terminate early either, because its index records all the partial results of the join operation and must be kept consistent with each graph data snapshot.

\section{Experiment Setting} \label{sec:setting}

We study six algorithms in our experiments, including IncIsoMatch (IM), SJ-Tree (SJ), Graphflow (GF), IEDyn (DYN), TurboFlux (TF), and SymBi (SYM).
As most of those algorithms except SJ-Tree are not open-sourced, we have to use our homegrown algorithm based on the common IVM model. We implement all algorithms in C++ and optimize them with our best efforts. Our implementation of SJ-Tree is faster than the original version. To state the performance of those CSM algorithms, we also include the latest and open-sourced SM algorithm RapidMatch (RM) \cite{sun2020rapidmatch} in our experiments.
The graph is stored as adjacent arrays. Vertices in each neighbor array are sorted by their IDs. Therefore, an edge is inserted or deleted with the binary search of the arrays.
The code was compiled by g++ 8.3.1. We experiment on a Linux machine with two Intel Xeon Gold 5218 CPUs and 512GB RAM.

\textbf{Datasets.} Following previous studies~\cite{choudhury2015selectivity, kim2018turboflux, min2021symmetric}, we use Netflow and LSBench, two dynamic
graphs, in our experiments. Netflow is a real-world graph containing anonymized passive traffic traces~\cite{caida}. LSBench is a synthetic social network produced by the Linked Stream Benchmark data generator~\cite{le2012linked}. The insertion (resp. deletion) rate is defined as the ratio of the number of edge insertions (resp. deletions) to the total number of edges in the original dataset. We use data graphs with the insertion rate of 10\% by default. Specifically, we set the first 90\% edges of each dataset as the initial graph and the remaining 10\% as the insertion stream. Edges of LSBench and Netflow contain 44 and 7 labels, respectively. The label
distribution is skewed, e.g., 13.7\% edges in LSBench have the same label and as high as 70.9\% edges in Netflow have the same label.

In addition to the two commonly used datasets, we further generate dynamic graphs from static graphs. We evaluate static graphs including Amazon and LiveJournal by randomly sampling 10\% edges as the insertion stream as in related work~\cite{kankanamge2017graphflow, mhedhbi2021optimizing}. The original datasets are unlabeled, and we randomly assign one of 6 and 30 labels to each vertex
in Amazon and LiveJournal, respectively. 

\begin{table}[t]\small
\setlength{\abovecaptionskip}{2pt}
\setlength{\belowcaptionskip}{0pt}
\centering
\caption{Datasets.}
\resizebox{0.475\textwidth}{!}{%
\begin{tabular}{l|c|c|c|c|c|c|c} \hline
	\textbf{Datasets}       & $|V|$ & $|E|$ & $|\Sigma_V|$  & $|\Sigma_E|$  & $d_{avg}$   & $d_{max}$ & $c_{max}$\\\hline\hline
	Amazon (\emph{az})      & 0.4M  & 2.4M  & 6             & 1             & 12.2        & 0.2M      & 10  \\\hline
	LiveJournal (\emph{lj}) & 4.9M  & 42.9M & 30            & 1             & 18.1        & 4.3M      & 350 \\\hline
	Netflow (\emph{nf})    & 3.1M  & 2.9M  & 1             & 7             & 2.0         & 0.2M      & 8   \\\hline
	LSBench (\emph{ls})     & 5.2M  & 20.3M & 1             & 44            & 8.2         & 2.3M      & 27  \\\hline
\end{tabular}
\label{table:datasets}
}
\vspace{3pt}
\end{table}

Table \ref{table:datasets} summarizes the graph size, number of labels, the average degree $d_{avg}$,
the maximum degree $d_{max}$, and the maximum core number $c_{max}$ of each dataset. Netflow and Amazon are sparse and have no dense
sub-structures, while Livejournal and LSBench are relatively dense. In summary, our datasets cover a wide range of settings, e.g., the distribution of labels
and the density of graphs. In addition, we perform sensitivity studies on those data sets, such as altering the insertion and deletion rate in our experiments.

\textbf{Query graphs.} Following previous work~\cite{kim2018turboflux,min2021symmetric}, we generate query graphs by randomly extracting subgraphs
from the data graph. We divide query graphs into three types based on the density: tree, sparse ($d_{avg} \leqslant 3$) and dense ($d_{avg} > 3$).
Both sparse and dense are cyclic queries.
For each type, we generate query graphs with $V(Q)$ varied from 4 to 12 in an increment of two. We generate 100 query graphs of each size
and each query type as a query set. Due to the space limit, we report the results on query sets containing query graphs with 6 vertices by default.

\textbf{Metrics.} We measure the \emph{query time}, which is the elapsed time of the online processing given a graph update stream. We exclude the time for data graph update because it is the same for all algorithms in our framework. For the index-based method, the query time consists of the \emph{indexing time},
which is the time on updating the index, and the \emph{enumeration time}, which is the time on enumerating results. To complete the experiments in a reasonable time,
we set the time limit for processing a query to one hour. We say that a query is \emph{unsolved} if the execution exceeds the time limit. Given
an unsolved query, if the algorithm finds fewer than $10 ^ 9$ incremental matches within the time limit, then we call it a \emph{hard unsolved} query
because the algorithm achieves considerable performance if it can find many results despite that it runs out of time.
Additionally, we count the number of candidates for each query vertex to compare the pruning power of indexes.

By default, we report the average value of a query set. To compare the performance of two methods $A$ and $B$ on an individual query, we examine the 
\emph{individual speedup} of $A$ over $B$, which is $\frac{1}{|\mathbf{Q}|}\sum_{Q \in \mathbf{Q}} \frac{t_B(Q)}{t_A(Q)}$ where
$\mathbf{Q}$ is a query set and $t(Q)$ is the query time on $Q$. To compare the relative performance of multiple methods on an individual query in
terms of a metric $X$, we compute the \emph{relative performance} of each method $A$, which is
$\frac{1}{|\mathbf{Q}|}\sum_{Q \in \mathbf{Q}} \frac{X_A(Q)}{X^*(Q)}$ where $X_A(Q)$ is the value of $A$ and $X ^ *(Q)$ is the maximum value
among competing algorithms.
\section{Experimental Results} \label{sec:results}

We first evaluate the overall performance of competing algorithms, and then examine the effectiveness of indexes and matching orders.

\subsection{Overall Comparison} \label{sec:overall_comparison}

\begin{figure*}[t]
	\setlength{\abovecaptionskip}{0pt}
	\setlength{\belowcaptionskip}{-6pt}
	\centering
	\includegraphics[width=\textwidth]{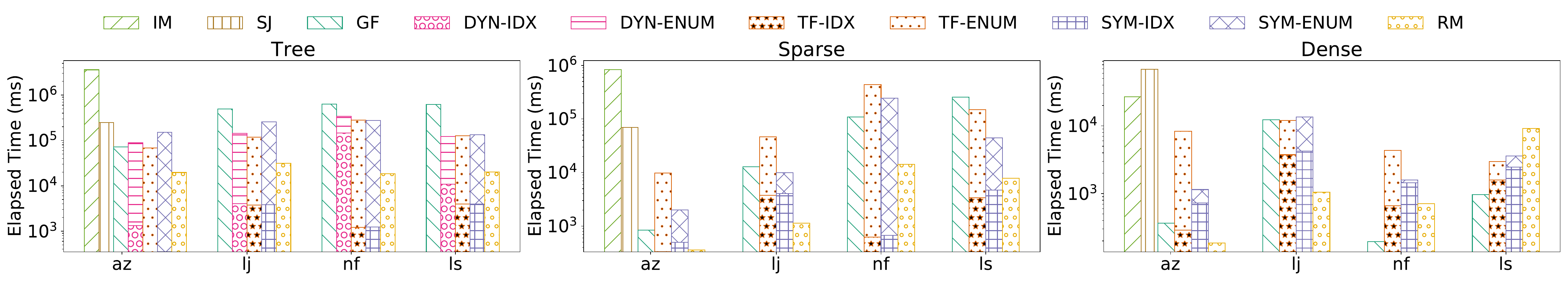}
	\caption{Comparison of competing algorithms on query time under subgraph isomorphism.}
	\label{fig:exp-execution-breakdown}
\end{figure*}

Figure \ref{fig:exp-execution-breakdown} presents the query time breakdown of competing algorithms under subgraph isomorphism. 
For an index-based method, the suffixes "-IDX" and "-ENUM" represent the indexing and enumeration time 
in the incremental matching phase, respectively. We only report
the results of IM and SJ in \emph{az} because, in other datasets, IM frequently runs out of time, while SJ runs out of memory.
The figure shows that IM is much slower than incremental algorithms due to the recomputation. 
In the following, we focus on GF, DYN, TF, SYM, and RM, which have valid experiment results in all datasets.
We also show the query time under subgraph homomorphism in appendix. There is little difference in performance between the two settings. Consequently, in the remaining experiments, we only report the results under subgraph isomorphism, a common practice in real-world applications \cite{bindschaedler2021tesseract, wang2015detecting, qiu2018real, manzoor2016fast}. Nevertheless, our findings applies to both subgraph isomorphism and homomorphism.

On tree queries, TF and DYN are slightly better than SYM, all of which outperform GF. In contrast, GF runs faster than others on dense queries
because the indexing time can dominate the cost. SYM and GF generally run much faster than TF on sparse queries.
As the time complexity of indexing of TF is lower than that of SYM, TF takes less indexing time than SYM. 
DYN spends more time on indexing than TF and SYM since it builds the local index and merges it back to the global index on each update.
Nevertheless, no CSM algorithm can dominate all others in each case. Even though DYN is of constant delay enumeration, it does not outperform all other competitors.

RM runs faster than CSM algorithms in most cases because (1) All CSM algorithms force the execution starting from the edge updated, whereas RM chooses a matching order optimized based on the entire data graph; and (2) CSM algorithms may join the same relation on multiple updates, while RM performs the join operation only once. However, RM may underperform CSM algorithms in short-running queries because the time RM spends on finding matches in the initial graph may offset the query time on subsequent updates.

Table \ref{tab:unsolved_query_count} lists the number of unsolved queries.
GF and DYN have more unsolved queries than others on tree queries, while TF has more than GF and SYM
on cyclic queries. SYM is more robust than others, with fewer unsolved queries,
especially the hard unsolved ones. To further investigate the index cost, we ran the queries without enumerating any match. We found that 18 queries in DYN did not finish in time whereas the other schemes finished all queries.

\begin{table}[t]\small
	\setlength{\abovecaptionskip}{2pt}
	\setlength{\belowcaptionskip}{0pt}
\caption{The number of unsolved queries (U.) and hard unsolved queries (HU.). The unsolved queries include hard unsolved ones. Dataset \emph{az} and Algorithm RM are omitted as there is no unsolved query.}
\resizebox{0.475\textwidth}{!}{%
\begin{tabular}{c|c|cccccc|cc}
\hline
\textbf{Query} & \multirow{2}{*}{\textbf{Method}} & \multicolumn{2}{c}{\textit{\textbf{lj}}} & \multicolumn{2}{c}{\textit{\textbf{nf}}} & \multicolumn{2}{c|}{\textit{\textbf{ls}}} & \multicolumn{2}{c}{\textbf{Total}} \\ \cline{3-10} 
\textbf{Structure} & & \textbf{U.} & \textbf{HU.} & \textbf{U.} & \textbf{HU.} & \textbf{U.} & \textbf{HU.} & \textbf{U.} & \textbf{HU.}     \\ \hline\hline
\multirow{6}{*}{Tree} 
& DYN & 4 & 0 & 57 & 18 & 11 & 0 & 72 & 18 \\
& TF & 3 & 0 & 38 & 3 & 11 & 0 & 52 & 3 \\
& GF & 6 & 0 & 55 & 19 & 11 & 4 & 72 & 23 \\
& O-DYN & 3 & 0 & 45 & 4 & 11 & 0 & 59 & 4 \\
& O-TF & 3 & 0 & 38 & 2 & 11 & 0 & 52 & 2 \\
& O-GF & 3 & 0 & 45 & 7 & 12 & 0 & 60 & 7 \\
& SYM & 4 & 0 & 41 & 1 & 12 & 0 & 57 & 1 \\ \hline
\multirow{5}{*}{Sparse} 
& TF & 0 & 0 & 20 & 15 & 35 & 32 & 55 & 47 \\
& GF & 0 & 0 & 2 & 2 & 33 & 30 & 35 & 32 \\
& O-TF & 0 & 0 & 20 & 14 & 29 & 26 & 49 & 40 \\
& O-GF & 0 & 0 & 2 & 1 & 17 & 14 & 19 & 15 \\
& SYM & 0 & 0 & 11 & 6 & 4 & 1 & 15 & 7 \\  \hline
\multirow{5}{*}{Dense} 
& TF & 0 & 0 & 1 & 1 & 2 & 2 & 3 & 3 \\
& GF & 0 & 0 & 0 & 0 & 0 & 0 & 0 & 0 \\
& O-TF & 0 & 0 & 0 & 0 & 0 & 0 & 0 & 0 \\
& O-GF & 0 & 0 & 0 & 0 & 0 & 0 & 0 & 0 \\
& SYM & 0 & 0 & 0 & 0 & 0 & 0 & 0 & 0 \\ \hline\hline
\multirow{5}{*}{Total} 
& TF & 3 & 0 & 59 & 19 & 48 & 34 & 110 & 53 \\
& GF & 6 & 0 & 57 & 21 & 44 & 34 & 107 & 55 \\
& O-TF & 3 & 0 & 58 & 16 & 40 & 26 & 101 & 42 \\
& O-GF & 3 & 0 & 47 & 8 & 29 & 14 & 79 & 22 \\
& SYM & 4 & 0 & 52 & 7 & 16 & 1 & 72 & 8 \\ \hline
\end{tabular}
}
\label{tab:unsolved_query_count}
\end{table}

\textbf{Summary.} According to the experiment results, we have the following findings on the overall performance. 

\begin{enumerate}[leftmargin=*]
    \item The recomputation-based method is much slower than incremental methods. SJ-Tree runs out of memory in most cases.
    \item Although SYM is more stable than other algorithms, no algorithm can dominate others in each case.
    \item On tree queries, TF and DYN are slightly better than SYM, all of which run much faster than GF.
    \item On sparse queries, TF performs worse than SYM and GF, and GF runs faster than SYM in sparse data graphs; otherwise, SYM has the best performance.
    \item On dense queries, GF runs faster than SYM and TF.
    \item Index update in DYN bears more significant overhead than the other schemes, probably due to the maintenance for constant delay enumeration.
    \item RM outperforms CSM methods in most cases.
\end{enumerate}

\subsection{Effectiveness of Individual Techniques}

\subsubsection{Effectiveness of Indexes} We examine the pruning power and the impact of the index on
the overall performance.

\begin{figure}[b]
	\setlength{\abovecaptionskip}{0pt}
	\setlength{\belowcaptionskip}{0pt}
	\centering
	\includegraphics[width=0.475\textwidth]{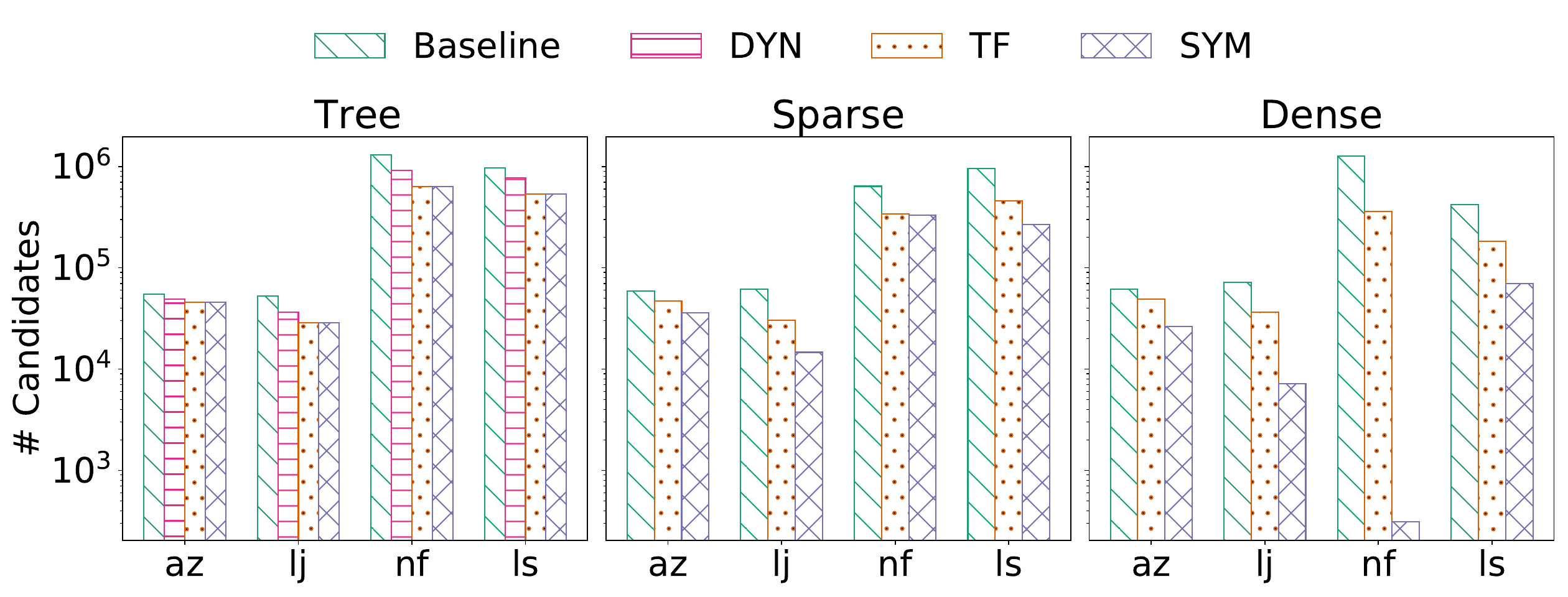}
	\caption{Number of candidates generated by different indexing methods.}
	\label{fig:num-candidates}
\end{figure}

\textbf{Pruning Power.} Figure \ref{fig:num-candidates} shows the number of candidates generated by different indexing methods.
\emph{Baseline} denotes the number of candidates obtained based on the vertex and edge labels. We can see that DYN, TF, and SYM can significantly
reduce the number of candidates. On tree queries, the pruning power of TF is competitive with that of SYM, 
and they have a smaller number of candidates than DYN and Baseline. In contrast, SYM outperforms
TF on cyclic queries because SYM utilizes the non-tree edges to filter invalid candidates. For the same reason, the gap between SYM and
TF enlarges with the increase of the query graph density.

\begin{figure}[h]
	\setlength{\abovecaptionskip}{0pt}
	\setlength{\belowcaptionskip}{3pt}
	\centering
	\includegraphics[width=0.475\textwidth]{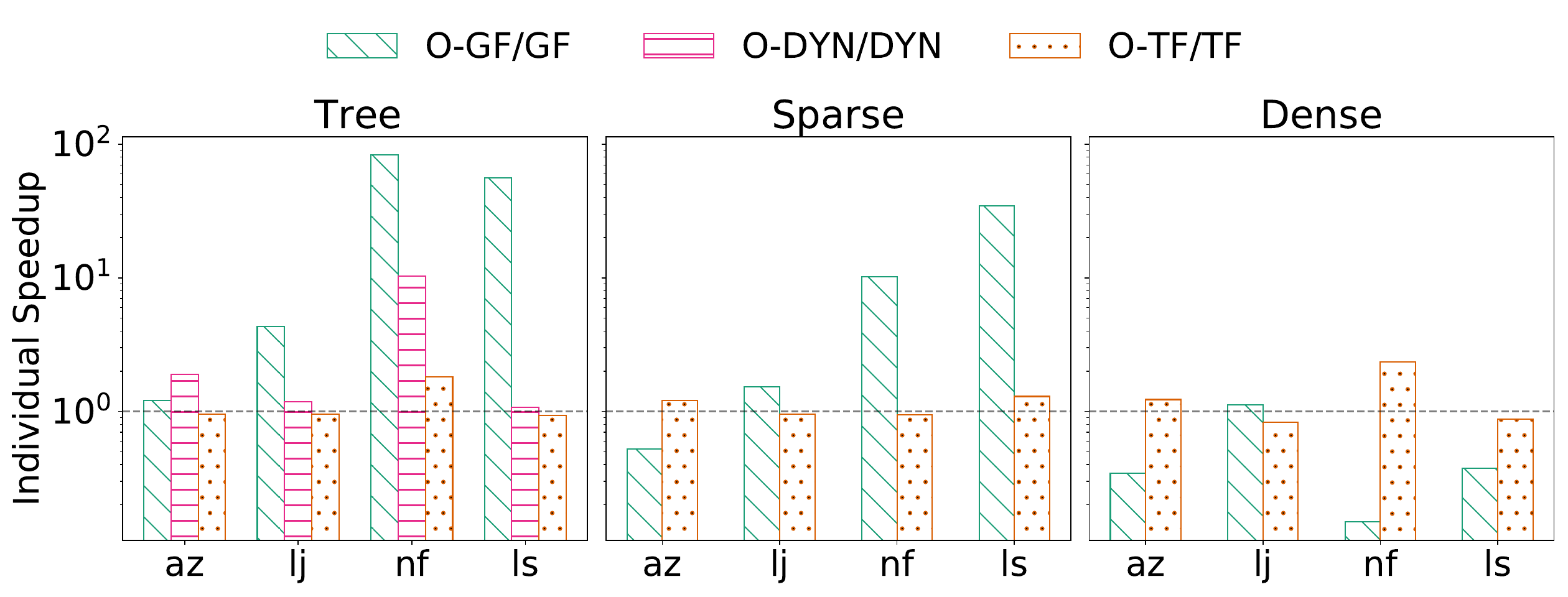}
	\caption{Individual speedup using the index of SYM.}
	\label{fig:exp-index-throughput}
\end{figure}

\textbf{Impact on Overall Performance.} As the index of SYM has the best pruning power, we optimize GF, DYN, and TF by integrating
their matching orders with the index of SYM. We call the optimized methods O-GF, O-DYN, and O-TF for short.
Then, we compare the overall performance of O-GF, O-DYN, and O-TF, with that of GF, DYN, and TF, to examine the impact of the index. Figure \ref{fig:exp-index-throughput}
shows the individual speedup. With the index, O-GF achieves significant speedup over GF on tree and sparse queries.
However, the index sometimes degrades the performance, e.g., the queries in \emph{az} and on dense queries, because the indexing time
is non-trivial compared with the query time (see Figure \ref{fig:exp-execution-breakdown}). Additionally, the speedup on long-running workloads (e.g.,
queries in \emph{nf} and \emph{ls}) is higher than that on short-running (e.g., queries in \emph{az} and \emph{lj}).
As DYN and TF initially have indexes, the speedup of O-DYN over DYN and O-TF over TF is lower than that of O-GF over GF.

Moreover, we report the number of unsolved queries on O-GF, O-DYN, and O-TF in Table \ref{tab:unsolved_query_count}. We can see that the number of unsolved queries is
reduced. In particular, the index significantly reduces the number of hard unsolved queries of GF on tree queries, because the index can rule out most invalid
candidates on acyclic queries, as discussed in Section \ref{sec:comparison}. 
Despite that the index of SYM does not support constant delay enumeration, it still reduces the number of unsolved queries of DYN since no local index is needed.

\textbf{Summary.} We have the following answer for the question: \emph{What is the effectiveness of the index?}

\begin{enumerate}[leftmargin=*]
    \item The index of SYM achieves the best pruning power.
    \item The index can reduce the number of candidates and improve the performance, especially on tree and sparse queries.
    \item The index provides more benefits on long-running queries, while the update of the index can dominate the query time on short-running queries.
    \item Although the index of SYM does not support constant delay enumeration, it speeds up the incremental matching of DYN.
\end{enumerate}

\subsubsection{Effectiveness of Matching Orders}

For a fair comparison, we evaluate the performance of O-GF, O-DYN, O-TF, and SYM, which have the same index.

\begin{figure}[b]
	\setlength{\abovecaptionskip}{0pt}
	\setlength{\belowcaptionskip}{0pt}
	\centering
	\includegraphics[width=0.475\textwidth]{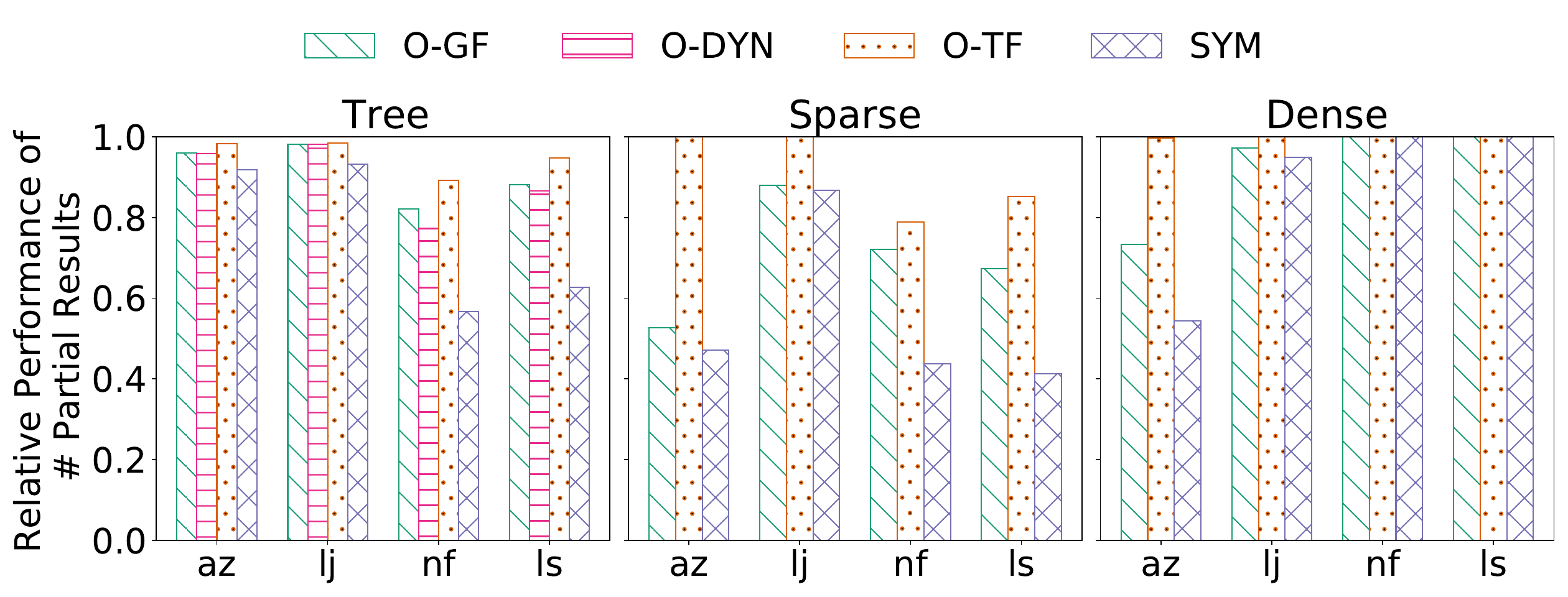}
	\caption{Relative performance of \#partial results using different matching orders given the index of SYM.}
	\label{fig:exp-order-intermediate_results}
\end{figure}

\textbf{Impact on \#partial results.} Figure \ref{fig:exp-order-intermediate_results} presents the relative performance
of the number of partial results. A smaller value indicates that the method generates fewer partial results.
SYM has the fewest partial results among all methods due to the dynamic matching order selection. 
O-TF performs much worse than O-GF and SYM on sparse and dense queries because its ordering method
ignores the effect of non-tree edges, as discussed in Section \ref{sec:comparison}. As there are few incremental matches
on dense queries against \emph{nf} and \emph{ls}, the relative performance of the three methods is one in these cases.

\begin{figure}[t]
	\setlength{\abovecaptionskip}{0pt}
	\setlength{\belowcaptionskip}{3pt}
	\centering
	\includegraphics[width=0.475\textwidth]{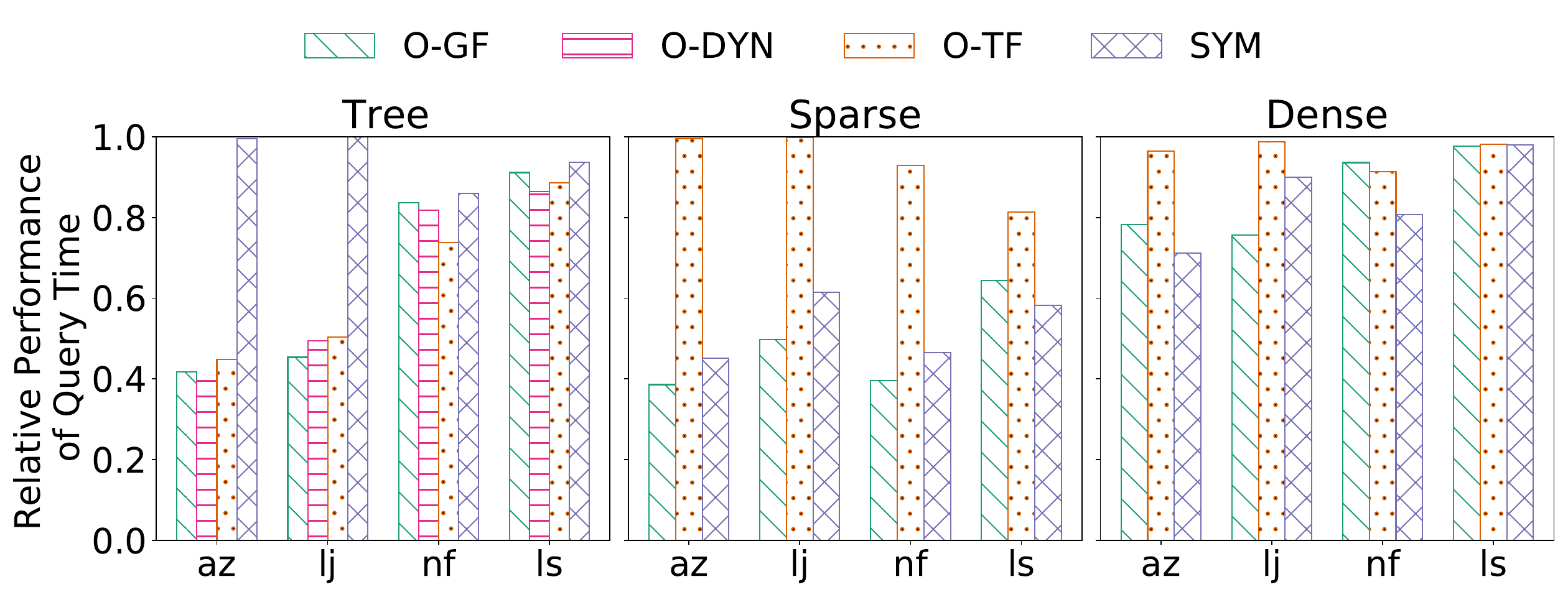}
	\caption{Relative performance of query time using different matching orders given the index of SYM.}
	\label{fig:exp-order-throughput}
\end{figure}

\textbf{Impact on Overall Performance.} Figure \ref{fig:exp-order-throughput} presents the relative performance of the query time.
A smaller value indicates that the method has a shorter query time. We can see that O-GF, O-DYN, and O-TF outperform SYM on tree queries
despite that SYM results in the fewest partial results. Because the overhead of optimizing matching orders at runtime offsets the benefit.
In contrast, SYM and O-GF are competitive on cyclic queries. O-TF performs worse than other algorithms on sparse and dense queries
due to many partial results. As shown in Table \ref{tab:unsolved_query_count}, SYM has fewer hard unsolved queries than O-GF, O-DYN,
and O-TF, which demonstrates that the ordering method of SYM is more robust. On sparse queries in the sparse dataset (e.g., \emph{nf}), O-GF
has much fewer hard unsolved queries than other methods. In contrast, all the four methods have only a few hard unsolved queries
on tree queries because the index can reduce many invalid candidates on acyclic queries, as discussed in Section \ref{sec:comparison}.
Nevertheless, all methods have hard unsolved queries because they all optimize matching orders based on simple heuristics,
e.g., the number of candidates and the degree of query vertices.

\textbf{Detailed Metrics.} In order to get more insights, we collect the detailed metrics of GF, TF, and SYM
in \emph{ls}. The results in other datasets are similar. Due to the space limit, we focus on
sparse queries on which competing algorithms have several unsolved queries. We relabel query \emph{ID}s
in the ascending order of the query time. For each query, we report the number of results (denoted by \emph{\#RES}), and
the number of \emph{invalid partial results} (denoted by \emph{\#INV}), i.e., the partial
results from which we cannot generate final results. According to Algorithm \ref{algo:wcoj}, a
partial result is invalid 
because 1) the local candidate
vertex set is empty; or 2) the data vertex has been mapped. We count the number of
invalid partial results caused by the two conditions, which are denoted by \emph{EMP}
and \emph{VIS}.

\begin{figure}[b]
	\setlength{\abovecaptionskip}{0pt}
	\setlength{\belowcaptionskip}{0pt}
	\centering
	\includegraphics[width=0.475\textwidth]{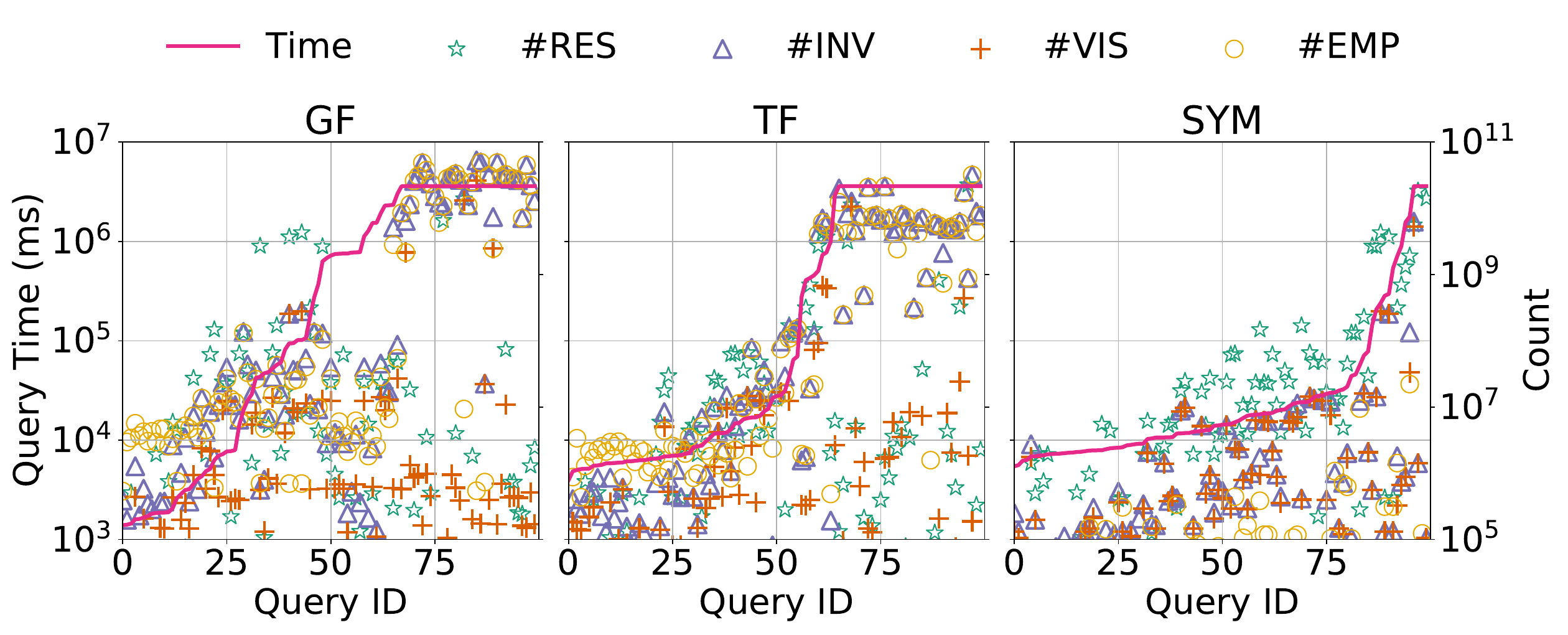}
	\caption{Detailed metrics of sparse queries in \emph{ls}.}
	\label{fig:exp-detail}
\end{figure}

As shown in Figure \ref{fig:exp-detail}, the query time is closely related to the number of results and invalid partial results. 
Benefiting from the index, SYM generates fewer invalid partial results than GF and TF, 
and it runs out of time because of the large result size. 
In comparison, unsolved queries of GF are generally due to the huge number of invalid partial results, as there is no index in GF. 
Despite that TF utilizes the index, the non-tree edges of the query graph are not considered by its filtering rules. 
Therefore, many unsolved queries are caused by numerous invalid partial results. 
Furthermore, we can see that most invalid partial results of SYM are caused by \emph{VIS}. This is because (1) in $\emph{ls}$, all data vertices and query vertices have the same label, and a data vertex can appear
in the candidate vertex sets of multiple query vertices; and (2) the index of SYM is constructed based on graph homomorphism, without concerning \emph{VIS} at all. 
GF and TF also have invalid partial results caused by \emph{VIS}, but \emph{\#EMP} dominates \emph{\#INV} due to the low pruning power of the algorithm.

\begin{figure}[t]
	\setlength{\abovecaptionskip}{0pt}
	\setlength{\belowcaptionskip}{3pt}
	\centering
	\includegraphics[width=0.36\textwidth]{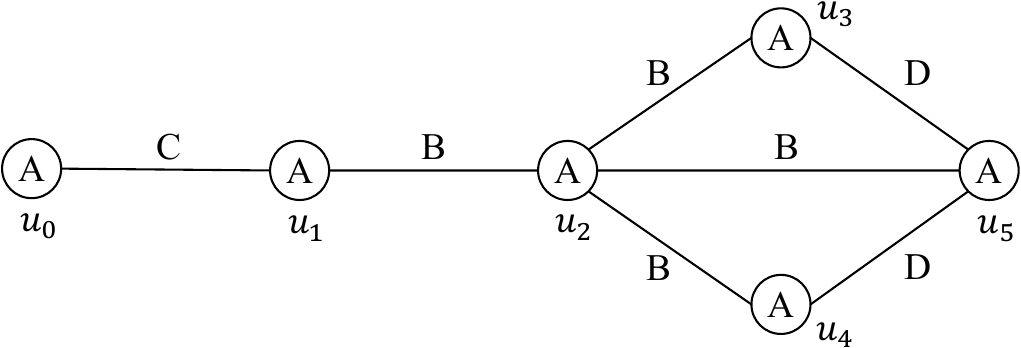}
	\caption{A test query graph in \emph{ls}.}
	\label{fig:case-study}
\end{figure}

\textbf{Case Study.} As discussed in the above experiments, algorithms run out of time on some queries because of the huge number of results. Additionally,
we observe that some queries cannot be completed by any CSM algorithm within the time limit, and only a few results are found. In contrast,
RM can enumerate all matches of $Q$ on the snapshot of $G$ after applying the entire stream $\Delta \mathcal{G}$
quickly. This motivates us to conduct a study on these cases. Figure \ref{fig:case-study} shows such a query in \emph{ls}. O-GF, O-TF, and SYM
cannot find even one result within one hour. Our investigation finds that the poor performance is caused by the ineffective matching order. Specifically,
given the insertion of a data edge with label \emph{C}, all competing algorithms will start the search by mapping it to $e(u_0, u_1)$.
For example, O-GF evaluates the query with the matching order $\varphi = (u_1, u_0, u_2, u_5, u_3, u_4)$.
The vertex-induced subgraph of $Q$ on the first four vertices is a tree, with many matches. When trying to further extend them,
the local candidate vertex set of $u_3$ (or $u_4$) is usually empty. Consequently, there is a large number of invalid partial results due to
\emph{EMP}. SM algorithms generally start the enumeration from the dense part of $Q$, which can terminate such invalid search paths at
an early stage~\cite{bi2016efficient,sun2020rapidmatch}. Unfortunately, the existing CSM framework forces the search to start from the query edge
mapped to the updated data edge $e$ to ensure the matches reported contain $e$.

\textbf{Summary.} Based on the results, we have the following answer to the question: \emph{What is the effectiveness
of the matching orders?}

\begin{enumerate}[leftmargin=*]
    \item On tree queries, the ordering methods of DYN and TF are generally better than those of other algorithms.
    \item On sparse queries, the ordering method of GF performs well on sparse data graphs; otherwise, the ordering method of SYM achieves the best performance.
    \item On dense queries, the ordering method of GF generally outperforms those of TF and SYM.
    \item Although the ordering method of SYM is more robust than those of other algorithms,
    all methods generate ineffective matching orders at times because they are based on simple heuristics.
    \item Forcing the search from the query edge mapped to the data edge potentially leads to many invalid partial results.
\end{enumerate}

Moreover, according to the findings on indexes and matching orders, we can obtain the answer to the question: \emph{What is the key factors leading to
the performance difference?}

\begin{enumerate}[leftmargin=*]
    \item DYN, TF, and SYM run faster than GF on tree queries because the index significantly reduces the invalid partial result size.
    \item TF has a poor performance on cyclic queries because its matching order does not consider the impact of non-tree edges.
    \item SYM has fewer hard unsolved queries than others because its matching order is robust.
    \item GF has a better performance than SYM on dense queries because the indexing cost can dominate the query time.
\end{enumerate}

Additionally, we have the following answer to the question: \emph{Where did time go in these queries?}

\begin{enumerate}[leftmargin=*]
    \item The query time is closely related to the number of results and invalid partial results.
    \item The competing algorithms can find a few results using a long time on some queries because of the large number of invalid partial results,
    which are incurred by \emph{VIS} and \emph{EMP}.
    \item The query vertices with the same label
    can result in many \emph{VIS}s, while the ineffective index can lead to many \emph{EMP}s.
    \item The tree queries have a long running time and many unsolved queries because they have many results in the data graph.
    \item The sparse queries have more hard unsolved queries than other types because they have a mixed
    sub-structure (e.g., a path with a diamond in Figure \ref{fig:case-study}), and the simple heuristics in existing ordering methods easily fail to process them.
    \item The dense queries generally have a shorter running time than other types because they have fewer results, and the invalid search
    paths can be terminated at an early stage due to constraints of query edges. 
\end{enumerate}

\textbf{Omitted Experiment Results.} We evaluate the standard deviation, response time, memory usage, offline indexing time, detailed metrics, and scalability of competing algorithms.
We also report the performance of the algorithms with various query graph properties~\cite{bonifati2017analytical}, data graph properties, and updated edge properties.
The trends of the experimental results are similar to those under our default setting. As such, we put the results and findings in appendix due to the space limit.

\section{Conclusion} \label{sec:conclusion}

In this paper, we conduct an in-depth study on the continuous subgraph matching (CSM) problem.
We first propose to model the problem as incremental view maintenance to capture the design
space of existing algorithms. Then, we design a common framework based on the model for CSM to depict, analyze and implement six CSM algorithms. 
Finally, we conduct extensive experiments to evaluate competing algorithms and give an in-depth analysis.

\textbf{Recommendation.} According to our experiments, we make the following recommendation of choices of existing indexing and ordering methods.
For the index, use the index of SymBi if the query graph is sparse; or the query takes a long running time.
Otherwise, use the direct-incremental methodology. For the matching order, if the query graph is a tree,
then use the matching order of IEDyn or TurboFlux; if the query graph is dense or both query and data graphs are sparse,
then use the matching order of Graphflow; otherwise, use the matching order of SymBi.

\textbf{Issues.} Our experiments on studying the effectiveness of individual techniques find severe issues
in existing CSM algorithms. First, to keep the reported matches containing the updated data edge $e$, the existing framework starts the search
from the query edge mapped to $e$, leading to many invalid partial results. Second, all matching orders
are based on simple heuristics, which can generate ineffective matching orders. Third, although the index generally improves the performance
of the query, the overhead of updating the index can offset its benefit on some short-running queries. Due to
these issues, existing algorithms encounter performance issues even on small queries.

\begin{acks}
This work was partially supported by grant 16209821 from the Hong Kong
Research Grants Council.
\end{acks}

\bibliographystyle{ACM-Reference-Format}

\clearpage
\bibliography{reference}

\appendix

\section{Supplement Examples}

\subsection{Example of the Common Framework}

Example \ref{exa:framework-example} illustrates the execution of a direct-incremental method within our common framework (Algorithm \ref{algo:framework}).

\begin{example} \label{exa:framework-example}
    Given $Q$ and $G$ in Figure \ref{fig:query_graph} and \ref{fig:data_graph}, the algorithm generates four relations $R_i'$ where $i \in \{0,1,2,3\}$ before the update, each of which is associated to a query edge:
    
    \begin{enumerate}
        \item $R_0'= R'(u_0, u_1) = \{(v_0, v_4), (v_2, v_6), (v_3, v_6)\}$,
        \item $R_1'= R'(u_0, u_2) = \{(v_0, v_5), (v_1, v_5), (v_2, v_5), (v_3, v_7)\}$,
        \item $R_2'= R'(u_1, u_3) = \{(v_4, v_8)\}$, and
        \item $R_3'= R'(u_2, u_3) = \{(v_5, v_8), (v_5, v_9), (v_5, v_{10}), (v_7, v_{11})\}$.
    \end{enumerate}
    
    The set of matches of $Q$ in $G$ is $\{(u_0, v_0), (u_1, v_4), (u_2, v_5), (u_3, v_8)\}$, so all those four relations are complete relations. When edge $e(v_6, v_{10})$ is inserted into $G$, the algorithm computes $R_k$ for each $k$ after the update (Line 5), and $\Delta R_k \subset R_k$ for each $k$ containing only the edge updated (Line 14):
    
    \begin{enumerate}
        \item $\Delta R_0 = \Delta R_1 = \Delta R_3 = \phi$, $\Delta R_2 = \{(v_6, v_{10})\}$,
        \item $R_0 = R_0', R_1 = R_1', R_3 = R_3'$, $R_2 = \{(v_4, v_8), (v_6, v_{10})\}$.
    \end{enumerate}
    
    Specifically, $e(v_6, v_{10})$ is added to $R_2'$ since $L(u_1) = L(v_6)$, $L(u_3) = L(v_{10})$, and $L(e(u_1, u_3)) = L(e(v_6, v_{10})$. Then, the algorithm performs the join (Line 17) based on IVM. $\Delta \mathcal{Q}_2 = \{(u_0, v_2), (u_1, v_6), (u_2, \\v_5), (u_3, v_{10})\}$ and $\Delta \mathcal{Q}_0 = \Delta \mathcal{Q}_1 = \Delta \mathcal{Q}_3 = \phi$. Therefore, a positive match $\{(u_0, v_2), (u_1, v_6), (u_2, v_5), (u_3, v_{10})\}$ is reported (Line 19).
\end{example}

\section{Algorithms and Analysis}

\subsection{IncIsoMatch}

\textbf{Implementation.} IncIsoMatch uses VF2~\cite{cordella2004sub}, which is a classical SM algorithm, to find $\mathcal{M}$ and $\mathcal{M}'$.
Since several SM algorithms have been proposed after VF2, we utilize a recent algorithm,
RapidMatch~\cite{sun2020rapidmatch}, to enumerate matches.

\textbf{Analysis.} RapidMatch is worst-case optimal. Therefore, on each update, the time complexity of the enumeration is $O(|\mathcal{M}| + |\mathcal{M}'|)$, where $\mathcal{M}$ and $\mathcal{M}'$ are the sets of matches before and after the update in $G_{dia}$ respectively.

\subsection{Graphflow}

\textbf{Analysis.} Our framework utilizes WCOJ to perform the enumeration in Graphflow. Thus, the time complexity of Graphflow within our framework matches the original implementation.

\subsection{SJ-Tree}

\textbf{Analysis.} Our framework extends a partial match in SJ-Tree by an edge at a time. As a result, our framework can achieve the same time complexity of enumeration as the original method.

\subsection{TurboFlux}

\begin{algorithm}[b]\footnotesize
	\caption{BuildInitialIndex for TurboFlux}
	\label{algo:turboflux-initial}
	\KwIn{a query graph $Q$, a data graph $G$, an initial matching order $\varphi_0$}
	\KwOut{an index $\mathcal{A}$}
	\SetKwFunction{InsertForward}{InsertForward}{}{}
	\SetKwFunction{InsertBackward}{InsertBackward}{}{}
	\SetKwProg{proc}{Procedure}{}{}
	$FrontiersF \gets [], FrontiersB \gets []$\;
	\tcc{Initialize the index $\mathcal{A}$}
	$C_{im}(u) \gets \phi, C(u) \gets \phi$ for each $u \in V(Q)$\;
	\ForEach{$v \in V(G)$ where $L(v) = L(u_r)$}{
		$FrontiersF$.push($(u_r, v)$)\;
	}
	\InsertForward{$Q, G, \mathcal{A}, \varphi_0, FrontiersF, FrontiersB$}\;
	\InsertBackward{$Q, G, \mathcal{A}, \varphi_0, FrontiersB$}\;
	\Return $\mathcal{A}$\;
	\proc{\InsertForward{$Q, G, \mathcal{A}, \varphi_0, FrontiersF, FrontiersB$}}{
		\ForEach{$(u, v) =$ $FrontiersF$.pop()}{
		    $C_{im}(u) = C_{im}(u) \cup \{v\}$\;
			\If{$\forall u' \in N_-^{\varphi_0}(u), C(u') \cap N(v) \neq \phi$}{
				$FrontiersB$.push($(u, v)$);
			}
			\ForEach{$u' \in N_-^{\varphi_0}(u)$}{
				\ForEach{$v' \in N(v)$,  $L(v') = L(u')$, and $v' \notin C_{im}(u')$}{
					\If{$\forall u'' \in N_+^{\varphi_0}(u'), C_{im}(u'') \cap N(v') \neq \phi$}{
						$FrontiersF$.push($(u', v')$)\;
					}
				}
			}
		}
	}
	\proc{\InsertBackward{$Q, G, \mathcal{A}, \varphi_0, FrontiersB$}}{
		\ForEach{$(u, v) =$ $FrontiersB$.pop()}{
			$C(u) = C(u) \cup \{v\}$\;
			\ForEach{$u' \in N_+^{\varphi_0}(u)$}{
				\ForEach{$v' \in N(v)$,  $L(v') = L(u')$, and $v' \notin C(u')$}{
					\If{$\forall u'' \in N_-^{\varphi_0}(u'), C(u'') \cap N(v') \neq \phi$}{
					    \If{$v' \in C_{im}(u')$}{
						    $FrontiersB$.push($(u', v')$)\;
						}
					}
				}
			}
		}
	}
\end{algorithm}

\textbf{Algorithms.} The \texttt{BuildInitialIndex} function of TurboFlux is shown in Algorithm \ref{algo:turboflux-initial}. DCG maintains a queue $FrontiersF$ to record each pair $(u,v)$ where $u \in V(Q)$, $v \in V(G)$, and $v$ should be inserted into $C_{im}(u)$. Similarly, DCG also maintains a queue $FrontiersB$ to record each pair $(u,v)$ where $u \in V(Q)$, $v \in V(G)$, and $v$ should be inserted into $C(u)$. 
In offline processing, DCG sets candidate sets in $\mathcal{A}$ in two phases. In the forward phase, DCG first pushes each pair $(u_r, v)$ to the queue $FrontierF$ if $L(v) = L(u_r)$ (Lines 3-4). For each pair $(u, v)$ in $FrontierF$, the function \texttt{InsertForward} first inserts $v$ into $C_{im}(u)$ and then checks whe\-ther $v$ has a neighbor in $C(u')$ for each $u' \in N_-^{\varphi_0}(u)$ (Line 11). If so, the function pushes $(u, v)$ to $FrontierB$ (Line 12). After that, \texttt{InsertForward} loops over each $u' \in N_-^{\varphi_0}(u)$ and checks whe\-ther each neighbor $v'$ of $v$, with the same label as $u'$, has a neighbor in $C(u'')$ for each $u'' \in N_+^{\varphi_0}(u')$ (Lines 13-15). If so, the function pushes the pair $(u', v')$ to $FrontiersF$ (Lines 16).

In the backward phase, \texttt{InsertBackward} processes each pair $(u, v) \in FrontierB$. The function inserts $v$ into $C(u)$ (Line 19). After that, the function loops over each $u' \in N_+^{\varphi_0}(u)$ and checks whether each neighbor $v'$ of $v$, with the same label as $u'$, has a neighbor in $C(u'')$ for each $u'' \in N_-^{\varphi_0}(u')$, such that $v' \in C_{im}(u')$ (Lines 20-23). If so, the function pushes the pair $(u', v')$ to $FrontiersB$ (Lines 24).

\begin{algorithm}[b]\footnotesize
	\caption{UpdateIndex for TurboFlux}
	\label{algo:turboflux-update}
	\KwIn{a query graph $Q$, a data graph $G$, an index $\mathcal{A}$, an initial matching order $\varphi_0$, and a graph update $\Delta G$}
	\SetKwFunction{InsertForward}{InsertForward}{}{}
	\SetKwFunction{InsertBackward}{InsertBackward}{}{}
	\SetKwFunction{DeleteForward}{DeleteForward}{}{}
	\SetKwFunction{DeleteBackward}{DeleteBackward}{}{}
	\SetKwProg{proc}{Procedure}{}{}
	$FrontiersF \gets []$, $FrontiersB \gets []$\;
	\uIf{$\Delta G.op$ is $+$}{
    	\If{$v_x\in C_{im}(u_x)$, $v_y\notin C_{im}(u_y)$, and $\forall u \in N_+^{\varphi_0}(u_y), C_{im}(u) \cap N(v_y) \neq \phi$}{
    		$FrontiersF$.push($(u_y, v_y)$)\;
    	}
    	\If{$v_y\in C(u_y)$, $v_x\notin C(u_x)$, and $\forall u \in N_-^{\varphi_0}(u_x), C(u) \cap N(v_x) \neq \phi$}{
    		$FrontiersB$.push($(u_x, v_x)$)\;
    	}
    	\InsertForward{$Q, G, \mathcal{A}, \varphi_0, FrontiersF, FrontierB$}\;
    	\InsertBackward{$Q, G, \mathcal{A}, \varphi_0, FrontierB$}\;
	}
    \Else{
    	\If{$v_x\in C_{im}(u_x)$, $v_y\in C_{im}(u_y)$, and $C_{im}(u_x) \cap N(v_y) = \{v_x\}$}{
    		$FrontiersF$.push($(u_y, v_y)$)\;
    	}
        \If{$v_y\in C(u_y)$, $v_x\in C(u_x)$, and $C(u_y) \cap N(v_x) = \{v_y\}$}{
    		$FrontiersB$.push($(u_x, v_x)$)\;
    	}
    	\DeleteForward{$Q, G, \mathcal{A}, \varphi_0, FrontiersF, FrontierB$}\;
    	\DeleteBackward{$Q, G, \mathcal{A}, \varphi_0, FrontierB$}\;
    }
    \proc{\DeleteForward{$Q, G, \mathcal{A}, \varphi_0, FrontiersF, FrontiersB$}}{
		\ForEach{$(u, v) =$ $Frontiers$.pop()}{
		    remove $v$ from $C_{im}(u)$\;
			\If{$v \in C(u)$}{
				$FrontiersB$.push($(u, v)$);
			}
			\ForEach{$u' \in N_-^{\varphi_0}(u)$}{
				\ForEach{$v' \in N(v) \cap C_{im}(u')$}{
					\If{$\exists u'' \in N_+^{\varphi_0}(u'), C_{im}(u'') \cap N(v') = \phi$}{
						$FrontiersF$.push($(u', v')$)\;
					}
				}
			}
		}
	}
	\proc{\DeleteBackward{$Q, G, \mathcal{A}, \varphi_0, FrontiersB$}}{
		\ForEach{$(u, v) =$ $FrontiersB$.pop()}{
    		remove $v$ from $C(u)$\;
			\ForEach{$u' \in N_+^{\varphi_0}(u)$}{
				\ForEach{$v' \in N(v) \cap C(u')$}{
					\If{$\exists u'' \in N_-^{\varphi_0}(u'), C(u'') \cap N(v') = \phi$}{
						$FrontiersB$.push($(u', v')$);
					}
				}
			}
		}
	}
\end{algorithm}

In online processing, given $\Delta G = \{e(v_x, v_y)\}$, if $e(u_x, u_y)$ belongs to $E(Q_T)$, then we assume that $u_x$ precedes $u_y$ in $\varphi_0$ without
loss of generality. DCG updates candidate sets through Algorithm \ref{algo:turboflux-update}. If $\Delta G.op$ is an insertion operation, DCG updates candidate sets using the same operations as those in offline processing but starting the forward phase from $u_y$ (Lines 2-8). 
The index update for deletion is similar to that for insertion, and therefore we omit the details (Lines 9-15).
If $e(u_x, u_y)$ does not belongs to $E(Q_T)$, DCG does not modify the candidate sets.

\textbf{Implementation.} Similar to the original implementation, our framework maintains two indexes, $MatchForward$ and $MatchBack\-ward$. For each $v' \in V(G)$ and $u' \in V(Q)$ where $L(v') = L(u')$, $MatchForward$ (resp. $MatchBackward$) stores the cardinality of $N(v') \cap C(u'')$ (resp. $N(v') \cap C_{im}(u'')$) for each $u'' \in N_-^{\varphi_0}(u')$ (resp. $N_+^{\varphi_0}(u')$), and the number of those cardinalities that are not zero. In both Algorithm \ref{algo:turboflux-initial} and Algorithm \ref{algo:turboflux-update}, $MatchForward$ and $MatchBackward$ are updated with DCG. With the two indexes, Lines 15 and 22 of Algorithm \ref{algo:turboflux-initial} and Lines 23 and 30 of Algorithm \ref{algo:turboflux-update} can be executed in $O(1)$ time.

\textbf{Analysis.} In the worst case, $C_{im}(u) = C(u)= V(G)$ for each $u \in V(Q)$. 
In the forward phase of offline processing, given $u \in V(Q)$, DCG will check whether each neighbor $v'$ of $v \in C(u)$ should be inserted into $C(u')$ where $u' \in N_+^{\varphi_0}(u)$. For all $v \in C(u)$, the total size of $N(v)$ is $2V(G)$. Additionally, since $Q$ is a tree and $\varphi_0$ is the depth-first order of $V(Q)$, for all $u \in V(G)$, the total size of $N_+^{\varphi_0}(u)$ is $V(Q)$. The backward phase have the same time complexity. Additionally, in the beginning of Algorithm \ref{algo:turboflux-initial}, Lines 3-4 add each $v \in V(G)$ to $C(u_r)$ if $L(v) = L(u_r)$, taking $O(|V(G)||V(Q)|)$ time. In summary, the total time of offline processing is $O(|E(G)||V(Q)|)$. 
The index needs $O(|V(G)||V(Q)|)$ space to store candidates 
and $O(|E(G)||V(Q)|)$ space to store edges. Taking into account $MatchForward$ and $MatchBackward$, both taking $O(|V(G)|(|E(Q)| + |V(Q)|))$ space in the worst case, the space complexity of the index is $O(|E(G)||V(Q)|)$. 
In online processing, the time complexity of the index update is the same as that of the index construction, 
but the actual cost is small because only a small portion of vertices are inserted into the candidate sets due to an edge insertion or deletion.

The original version of TurboFlux~\cite{kim2018turboflux} utilizes the enumeration method similar to TurboIso~\cite{han2013turboiso}, which is proved not worst-case optimal \cite{ammar2018distributed}. However, since our framework is proposed to systematically compare the indexing and ordering strategy and eliminate other differences, we optimize the enumeration of TurboFlux, making it worse-case optimal.

\subsection{SymBi}

\textbf{Algorithms and Implementation.} The offline and online processing of SymBi is the same as TurboFlux, except that the first input parameters of function-calls on Lines 2, 6, and 10 in Algorithm \ref{algo:framework} are $Q$ rather than $Q_T$.

\textbf{Analysis.} Compared to TurboFlux, DCS maintains all edges $e \in E(Q)$. Therefore, for all $u \in V(G)$, the total size of $N_+^{\varphi_0}(u)$ and $N_-^{\varphi_0}(u)$ are both $E(Q)$ rather than $V(Q)$. Consequently, the time and space complexity of DCS are both $O(|E(G)||E(Q)|)$. The time complexity for updating index is also $O(|E(G)||E(Q)|)$, but the actual cost is small. The enumeration of SymBi can achieve worst-case optimality in our framework.

\subsection{IEDyn} \label{sec:more-iedyn}

\textbf{Algorithms.} The \texttt{BuildInitialIndex} function of IEDyn is shown in Algorithm \ref{algo:iedyn-initial}. The index of IEDyn maintains a queue $Frontiers$ to record each pair $(u,v)$ where $u \in V(Q)$, $v \in V(G)$, and $v$ should be inserted into $C(u)$. In offline processing, IEDyn only performs the backward phase for indexing. Specifically, IEDyn adds each $v \in V(G)$ to $C(u)$ for each leaf $u \in V(Q)$ if $L(v) = L(u)$, and then pushes the pair $(u, v)$ to the queue $Frontiers$. \texttt{UpdateBack\-ward} is reused from Algorithm \ref{algo:turboflux-initial}, but Line 23 should be removed since there is no implicit candidate set in IEDyn.

In online processing, given $\Delta G = \{e(v_x, v_y)\}$, where $e(v_x, v_y)$ matches $e(u_x, u_y)$, we assume that $u_x$ precedes $u_y$ in $\varphi_0$ without
loss of generality. IEDyn generates an additional candidate set for each $u \in V(Q)$, namely $C_\Delta(u)$, storing in the local index.
The build of $C_\Delta(u)$ for each $u \in V(Q)$ uses the same operations as those of the offline processing but starts from $u_x$. As shown in Algorithm \ref{algo:iedyn-update}, if $\Delta G.op$ is an insertion operation, the algorithm assigns $C(u)$ to $C_\Delta(u)$ directly for each $u \notin P_{u_x}$ (Line 3), and starts the update from $u_x$. Compared with the index construction (Algorithm \ref{algo:iedyn-initial}) which \texttt{UpdateBackward} executes based on the global index, \texttt{UpdateBackward} at Line 7 in Algorithm \ref{algo:iedyn-update} executes based on the local index.

IEDyn was originally designed for finding all matches $M$ of $Q$ in the entire data graph on each update. However, the authors presented an implementation method to make the algorithm report incremental matches only on each edge insertion \cite{idris2020general}, which is illustrated in Algorithm \ref{algo:iedyn-update}. This method cannot enumerate $\Delta M$ on edge deletion, so we do not introduce the behavior of IEDyn when $\Delta G.op$ is $-$.

\begin{algorithm}[t]\footnotesize
	\caption{BuildInitialIndex for IEDyn}
	\label{algo:iedyn-initial}
	\KwIn{a query graph $Q$, a data graph $G$, an initial matching order $\varphi_0$}
	\KwOut{an index $\mathcal{A}$}
	\SetKwFunction{UpdateBackward}{UpdateBackward}{}{}
	\SetKwProg{proc}{Procedure}{}{}
	\tcc{Initialize the global index}
	$C(u) \gets \phi$ for each $u \in V(Q)$\;
	$Frontiers \gets []$\;
	\ForEach{$u \in V(Q)$ where $N_-^{\varphi_0}(u) = \phi$}{
		\ForEach{$v \in V(G)$ where $L(v) = L(u)$}{
			$Frontiers$.push($(u, v)$)\;
		}
	}
	\UpdateBackward{$Q, G, \mathcal{A}, \varphi_0, Frontiers$}\;
	\Return $\mathcal{A}$\;
\end{algorithm}

\begin{algorithm}[t]\footnotesize
	\caption{UpdateIndex for IEDyn}
	\label{algo:iedyn-update}
	\KwIn{a query graph $Q$, a data graph $G$, an index $\mathcal{A}$, an initial matching order $\varphi_0$, and a graph update $\Delta G$}
	\If{$\Delta G.op$ is $+$}{
    	\tcc{Initialize the local index}
    	$C_\Delta(u) \gets \phi$ for each $u \in P_{u_x}$\;
    	$C_\Delta(u) \gets C(u)$ for each $u \notin P_{u_x}$\;
    	$Frontiers \gets []$\;
    	\If{$v_y\in C(u_y)$ and $\forall u \in N_-^{\varphi_0}(u_x), C_\Delta(u) \cap N(v_x) \neq \phi$}{
    		$Frontiers$.push($(u_x, v_x)$)\;
    	}
    	\UpdateBackward{$Q, G, \mathcal{A}, \varphi_0, Frontiers$}\;
	}
\end{algorithm}

\textbf{Implementation.} IEDyn only needs to maintain $MatchForward$ for the incremental matching.

\textbf{Analysis.} Similar to TurboFlux, the total time for offline processing is $O(|E(G)||V(Q)|)$. 
The index and $MatchForward$ take $O(|E(G)||V(Q)|)$ space in total. 
In online processing, the index update complexity is the same as the construction. Still, the actual cost is small because only a small portion of vertices are inserted into the candidate sets due to an edge insertion or deletion. The enumeration of IEDyn can achieve worst-case optimality and constant delay in our framework.

\section{More on Comparison}

\textbf{Indexing.} Latest SM algorithms prune candidate vertex sets based on Proposition \ref{prop:sm_complete}~\cite{sun2020memory}.

\begin{proposition} \label{prop:sm_complete}
    Given $v \in C(u)$, if the mapping $(u, v)$ appears in a match of $Q$, then $v$ must satisfy that given $u' \in N(u)$,
    $v$ has at least a neighbor in $C(u')$.
\end{proposition}

Regarding the pruning power of CSM and the latest SM algorithms, we give the following proposition.

\begin{proposition} \label{prop:pruning-power}
    The pruning power of the Propositions has the following relationship: Proposition \ref{prop:sm_complete} > Proposition \ref{prop:symbi} > Proposition \ref{prop:turboflux} > Proposition \ref{prop:iedyn}.
\end{proposition}

\begin{proof}[Proof of Proposition \ref{prop:pruning-power}]

Suppose that each candidate $v \in C(u)$ satisfies the condition in Proposition \ref{prop:sm_complete}. We first prove that $v$ satisfies conditions in
Proposition \ref{prop:symbi}. Given $P = (u_r=u_0,..u_i,...,u_k = u)$ where $P \in \mathcal{P}_u$, we can recursively construct a match of $P$ in the order from $u$ to $u_r$
as follows: given $(u_{i-1}, u_i) \in P$, if $u_i$ is mapped to $v$, then select a neighbor $v'$ of $v$ in $C(u_{i-1})$. Based on Proposition \ref{prop:sm_complete},
$v'$ must exist. With the same method, we can obtain a match of $P \in \mathcal{T}_u$. Thus, each candidate in $C(u)$ satisfies the conditions
in Proposition \ref{prop:symbi}. Next, we compare Proposition \ref{prop:turboflux} and Proposition \ref{prop:symbi} Suppose that the spanning tree $Q_T$ of TurboFlux is generated from $Q_D$ of SymBi with the same root $u_r$.
Then, the path $P_u$ from $u_r$ to $u$ in $Q_T$ belongs to $\mathcal{P}_u$ in $Q_D$, and the root-to-leaf paths of $T_u$ in $Q_T$ belong to $\mathcal{T}_u$
in $Q_D$. Therefore, Proposition \ref{prop:turboflux} is a special case of Proposition \ref{prop:symbi}, and if each candidate $v \in C(u)$ satisfies Proposition \ref{prop:sm_complete}, then $v$ must meet the conditions in
Proposition \ref{prop:turboflux}. Finally, since one of the two conditions in Proposition \ref{prop:turboflux} is the same as the condition of Proposition \ref{prop:iedyn}, each candidate that satisfies Proposition \ref{prop:turboflux}, satisfies Proposition \ref{prop:iedyn}.
\end{proof}

When $Q$ is acyclic (i.e., has no cycle), TurboFlux and SymBi have similar pruning power because there is no non-tree edge in $Q$, and both of them
prune candidate vertex sets in two phases: forward phase and backward phase.
Inspired by Yannakakis algorithm~\cite{yannakakis1981algorithms}, the classical dangling tuple elimination algorithm for acyclic join
query in the relational database, we find that TurboFlux and SymBi can ensure candidate vertex sets to satisfy Proposition \ref{prop:modified_procedure}
by simply changing the pruning order. The \emph{modified pruning procedure} is listed in the following.

\begin{enumerate}[leftmargin=*]
    \item Set $C(u)$ to $\{v \in V(G) | L(v) = L(u)\}$ given $u \in V(Q)$;
    \item Bottom-up phase: Prune $C(u)$ along the bottom-up order of $Q$ by removing $v \in C(u)$ if there exists a child $u'$ of $u$ such that $v$ has no neighbor in $C(u')$;
    \item Top-down phase: Refine $C(u)$ along the top-down order of $Q$ by removing $v \in C(u)$ if $v$ has no neighbor in the candidate vertex set
    of the parent of $u$.
\end{enumerate}

\begin{proposition} \label{prop:modified_procedure}
    Given $Q$ and $G$, suppose $Q$ is a tree and candidate vertex sets are generated by the modified pruning procedure.
    Then, given $u \in V(Q)$ and $v \in C(u)$, $(u, v)$ appears in a match of $Q$ in $G$.
\end{proposition}

\begin{proof}[Proof of Proposition \ref{prop:modified_procedure}]

Suppose the modified procedure generates that candidate vertex sets. To prove the proposition, we first show that
$v$ must have a neighbor in $C(u')$ given $u' \in N(u)$ by contradiction. Particularly, given $v\in C(u)$ and $u' \in N(u)$,
assume that $v$ has no neighbor in $C(u')$. If $u'$ is the parent of $u$, then $v$ must have a neighbor in $C(u')$ because of the top-down pruning. Otherwise, $u'$ is a child of $u$. In that case, $v$ must have a neighbor
in $C(u')$ after the bottom-up pruning. Based on the filtering method, the neighbor of $v$ in $C(u')$ will not be removed in the top-down
pruning. As such, $v$ still has a neighbor in $C(u')$ after the top-down pruning. In the end, $v$ has a neighbor in $C(u')$ in both cases,
contradicting the assumption. Then, the statement is proved.

Based on the statement, we prove the proposition with the construction method. Specifically, given $v \in C(u)$, we first add $(u, v)$
to the partial result $M$. Let $u^*$ be a query vertex not in $M$ but having a neighbor $u'$ in $M$. We extend $M$ to generate a match of $Q$
as follows: 1) given $u'$, select a neighbor $v'$ of $M(u ^ *)$ in $C(u')$ and add $(u', v')$ to $M$; and 2) repeat Step 1 until all query vertices
added to $M$. Thus, the proposition is proved.
\end{proof}

Despite that both TurboFlux and SymBi ignore the impact of the filtering order, their pruning methods can remove many invalid candidates of acyclic
queries. Therefore, these queries can benefit a lot from indexes. However, the proposition cannot hold for cyclic queries, and we empirically study
the impact of indexes on query performance. Note that Proposition \ref{prop:modified_procedure} is proved in the context of subgraph homomorphism,
while it cannot hold for subgraph isomorphism.

\emph{Commons and Differences with SM.} CSM algorithms prune candidate vertex sets $C(u)$ using the similar method with SM algorithms: remove $v$ from $C(u)$
if $v$ has no neighbor in $C(u')$ where $u'$ is a selected neighbor of $u$. 
For example, CFLMatch~\cite{bi2016efficient}
builds a tree-structure index with a top-down generation procedure and a bottom-up filtering procedure. DP-iso~\cite{han2019efficient}
maintains candidate data edges for each query edge, and prunes candidate vertex sets along a sequence of query vertices. 
The difference is
that SM algorithms keep one candidate set for each query vertex. In contrast, CSM algorithms maintain multiple candidate sets for each query vertex to perform the incremental modification efficiently.

\begin{figure*}[t]
	\setlength{\abovecaptionskip}{3pt}
	\setlength{\belowcaptionskip}{0pt}
	\centering
	\includegraphics[width=\textwidth]{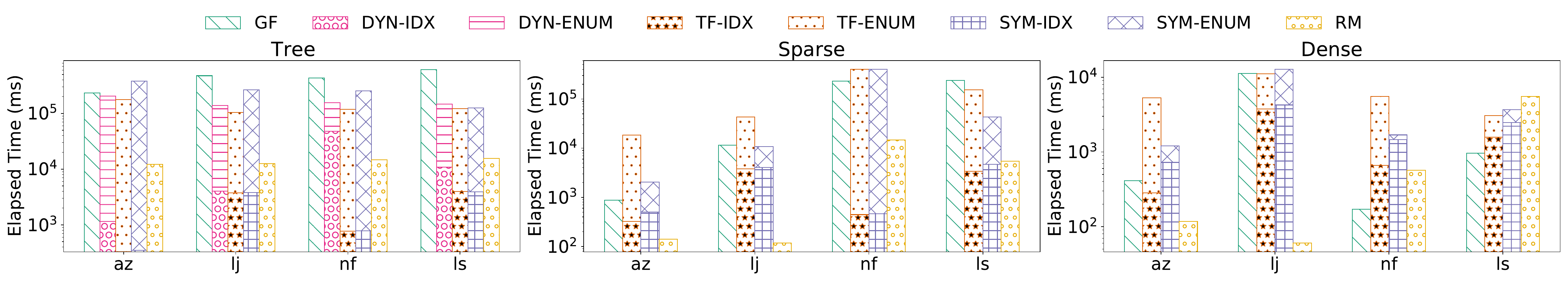}
	\caption{Comparison of competing algorithms on query time under graph homomorphism.}
	\label{fig:exp-homo-breakdown}
\end{figure*}

\textbf{Support of Sliding-window Graph Models.} In the sliding-window graph model, old edges will be deleted periodically. Therefore, if an algorithm supports deletion operation, it can work on sliding-window graphs. As a result, even though there is no specialized algorithm for the sliding-window graph model, IncIsoMatch, Graphflow, IEDyn, TurboFlux, and SymBi can work on the model with minor modification: recording the insertion timestamp of each inserted vertex or edge, and deleting a vertex or an edge if its timestamp is out of the range of the sliding window.

\section{Proofs of Propositions} \label{sec:proofs}

\begin{proof}[Proof of Proposition \ref{prop:correctness}.]
\textbf{Insertion}: For the $k$th relation, we use $R_k'$ and $R_k$ to denote the relation before and after the insertion, use $\Delta R_k$ to denote a subset of $R_k$ that only contains those edges inserted, and use $\mathcal{M}'$ and $\mathcal{M}$ to denote the set of matches before and after the insertion, namely $\mathcal{M}' = \Join_{k \in [1, N]}R_k'$ and $\mathcal{M} = \Join_{k \in [1, N]}R_k$. We also denote $\mathcal{M}''$ as $\Join_{k \in [1, N]}(R_k - \Delta R_k)$. 
Suppose the updates of the graph and index are correct. On the one hand, all $R_k'$s are complete relations before the insertion, and thus $\mathcal{M}'$ contains all matches before the insertion. Since the matches in $\mathcal{M}''$ contains no inserted edge, any match appearing in $\mathcal{M}''$ must appear in $\mathcal{M}'$. Therefore, $\mathcal{M}'' \subset \mathcal{M}'$.
On the other hand, all $R_k$s are complete relations after the insertion, and thus $\mathcal{M}$ contains all matches after the insertion. Compared to $\mathcal{M}$, $\mathcal{M}''$ only misses those matches containing at least one inserted edge, which do not belong to $\mathcal{M}'$. Therefore, any match in $\mathcal{M}'$ must appear in $\mathcal{M}''$, namely $\mathcal{M}' \subset \mathcal{M}''$.
In summary, $\mathcal{M}'=\mathcal{M}''$. Then, the set of all positive matches can be computed by $\Delta \mathcal{M} = \mathcal{M} - \mathcal{M}'' = \Join_{k \in [1, N]}R_k - \Join_{k \in [1, N]}(R_k - \Delta R_k)$. Lines 16-18 can computes the correct $\Delta \mathcal{M}$ according to Equation \ref{eq:incremental_computation}.

\textbf{Deletion}: For the $k$th relation, we use $R_k$ and $R_k'$ to denote the relation before and after the deletion, use $\Delta R_k$ to denote a subset of $R_i$ that only contains those edges to be deleted, and use $\mathcal{M}$ and $\mathcal{M}'$ to denote the set of matches before and after the deletion, namely $\mathcal{M} = \Join_{k \in [1, N]}R_k$ and $\mathcal{M}' = \Join_{k \in [1, N]}R_k'$. We also denote $\mathcal{M}''$ as $\Join_{k \in [1, N]}(R_k - \Delta R_k)$. 
Suppose the updates of the graph and index are correct. On the one hand, all $R_k$'s are complete relations after the deletion, and thus $\mathcal{M}'$ contains all matches after the deletion. Since the matches in $\mathcal{M}''$ contains no edge to be deleted, any match appearing in $\mathcal{M}''$ must appear in $\mathcal{M}'$. Therefore, $\mathcal{M}'' \subset \mathcal{M}'$.
On the other hand, all $R_k$s are complete relations before the deletion, and thus $\mathcal{M}$ contains all matches before the deletion. Compared to $\mathcal{M}$, $\mathcal{M}''$ only misses those matches containing at least one edge to be deleted, which do not belong to $\mathcal{M}'$. Therefore, any match in $\mathcal{M}'$ must appear in $\mathcal{M}''$, namely $\mathcal{M}' \subset \mathcal{M}''$.
In summary, $\mathcal{M}'=\mathcal{M}''$. Similarly, Lines 16-18 can computes the correct $\Delta \mathcal{M}$ according to Equation \ref{eq:incremental_computation}.
\end{proof}

\begin{proof}[Proof of Proposition \ref{prop:iedyn-constant-delay}.]
The time complexity of indexing in IEDyn is $O(|E(G)||V(Q)|)$ as described in Appendix \ref{sec:more-iedyn}. Suppose the index is correctly built and updated, we illustrate the enumeration of IEDyn based on Algorithm \ref{algo:wcoj}. IEDyn does not explicitly build $C_M(u)$ of each $u \in V(Q)$ because they can be obtained from the local index directly. Specifically, for $u_r$, $C_\Delta(u_r)$ contains all $v \in V(G)$ which have a neighbor in $C(u')$ for each $u' \in N_-^{\varphi_0}(u)$, matching the definition of $C_M(u_r)$ (Line 3). For each $u \in V(Q)$ where $u \neq u_r$, $|N_+^{\varphi_0}(u)| = 1$, because each query vertex except for $u_r$ has exactly one parent. The index stores both candidates and edges between candidates, so $C_M(u)$ where $u \neq u_r$ can be obtained directly from the index as well (Line 4).
In each step of the extension, the algorithm extends a partial match by a vertex $v \in C_M(u)$ until all query vertices are mapped (Lines 5-9). A vertex $v$ is inserted to $C_\Delta(u)$ only if $(u,v)$ appears in a match of $Q$ in $G$ (Proposition \ref{prop:iedyn}). Therefore, except for $u_r$, no $C_M(u)$ is empty at Line 5. If there exist incremental results, according to Algorithm \ref{algo:wcoj}, the maximum time between the start of the enumeration phase and the output of the first match is at most $V(Q)$; The maximum time between the output of two consecutive matches is at most $2V(Q)$; and the maximum time between the output of the last match and the termination of the algorithm is at most $V(Q)$. If there is no incremental match, $C_M(u_r)$ is empty and IEDyn can terminate the enumeration in $O(1)$ time. In summary, IEDyn enumerates all incremental matches with constant delay after $O(|E(G)||V(Q)|)$ time of preprocessing on each update.
\end{proof}

\section{Supplement Experiments}

\subsection{Overall Comparison}

\begin{figure*}[t]
	\setlength{\abovecaptionskip}{3pt}
	\setlength{\belowcaptionskip}{0pt}
	\centering
	\includegraphics[width=\textwidth]{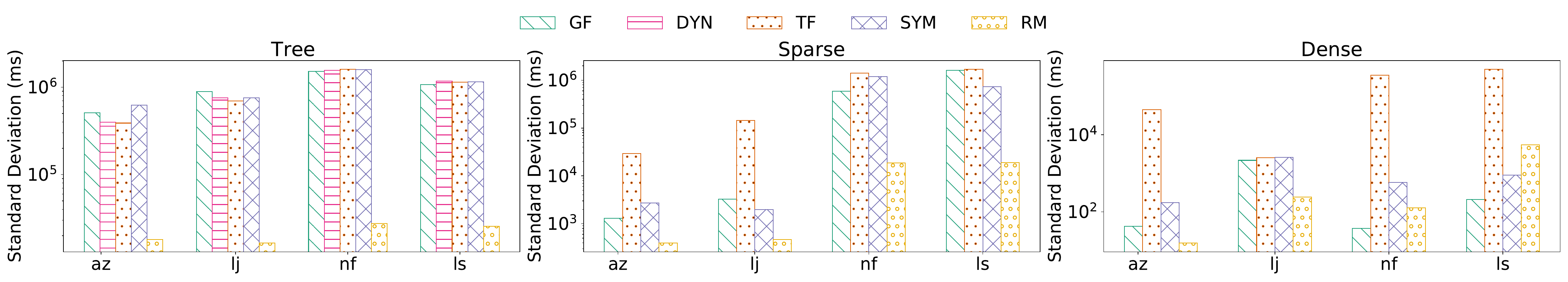}
	\caption{Comparison of competing algorithms on the standard deviation of query time.}
	\label{fig:exp-execution-std}
\end{figure*}

\begin{figure*}[t]
	\setlength{\abovecaptionskip}{3pt}
	\setlength{\belowcaptionskip}{0pt}
	\centering
	\includegraphics[width=\textwidth]{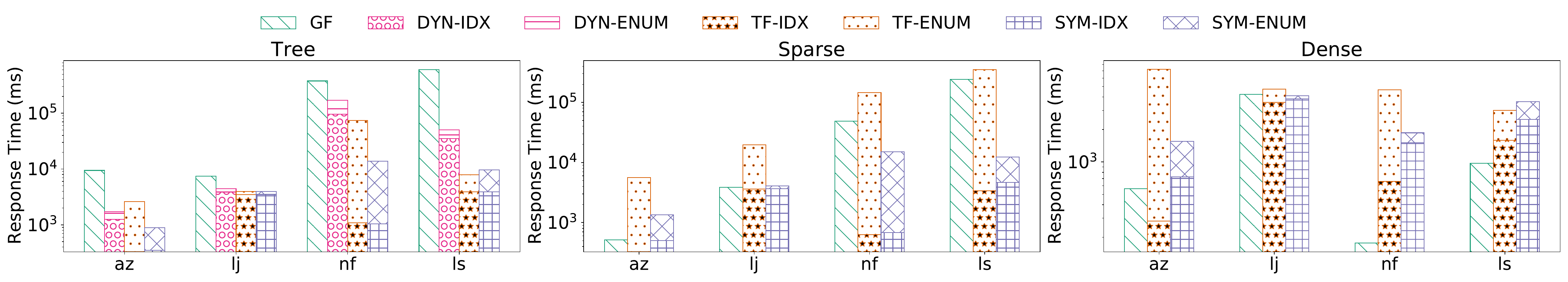}
	\caption{Comparison of competing algorithms on response time.}
	\label{fig:exp-response-breakdown}
\end{figure*}

In this subsection, we evaluate the query time under subgraph homomorphism, the standard deviation of query time, the response time, the offline indexing time, and the memory usage of algorithms under study.

\textbf{Query Time under Subgraph Homomorphism.} Figure \ref{fig:exp-homo-breakdown} shows the query time breakdown of competing algorithms under the setting of subgraph homomorphism. The indexing time is the same as that under subgraph isomorphism described in Section \ref{sec:overall_comparison}, because the algorithms perform the same operations to the indexes to keep them consistent with the data graph. However, compared to subgraph isomorphism, the enumeration time is slightly higher because the set of matches is larger. On tree queries, DYN and TF perform slightly better than SYM, and all three methods run faster than GF. However, GF is better than other competitors on dense queries. SYM is the top performer on sparse queries. In conclusion, the relative performance of competing algorithms is the same as that in Figure \ref{fig:exp-execution-breakdown}.

\textbf{Standard Deviation Comparison.} In addition to reporting the query time of the algorithms under study, we show the standard deviation of the query time in various datasets in Figure \ref{fig:exp-execution-std}. In most cases, the latest SM algorithm RM has a lower standard deviation than the other CSM algorithms. On tree queries, all CSM algorithms have a similar standard deviation. On sparse queries, GF and SYM generally have a higher standard deviation than TF. In comparison, on dense queries, GF has the lowest standard deviation among all CSM algorithms.

\textbf{Response Time Evaluation.} There may be a large number of incremental matches, but many online scenarios require to return after finding a specific number of incremental results~\cite{bindschaedler2021tesseract}. Therefore, in this experiment, we examine the \emph{response time} of competing algorithms, which is the total time of finding the first incremental match on each update. Figure \ref{fig:exp-response-breakdown} shows the response time breakdown on various types of queries and different datasets.

Compared with the overall query time shown in Figure 7, the first-result time is shorter, and the time difference between getting the first results and getting all results depends on both the data graphs and the queries. Nevertheless, the relative performance of getting the first result is the same as getting all results among all algorithms.

\begin{figure}[h]
	\setlength{\abovecaptionskip}{0pt}
	\setlength{\belowcaptionskip}{-15pt}
	\centering
	\includegraphics[width=0.475\textwidth]{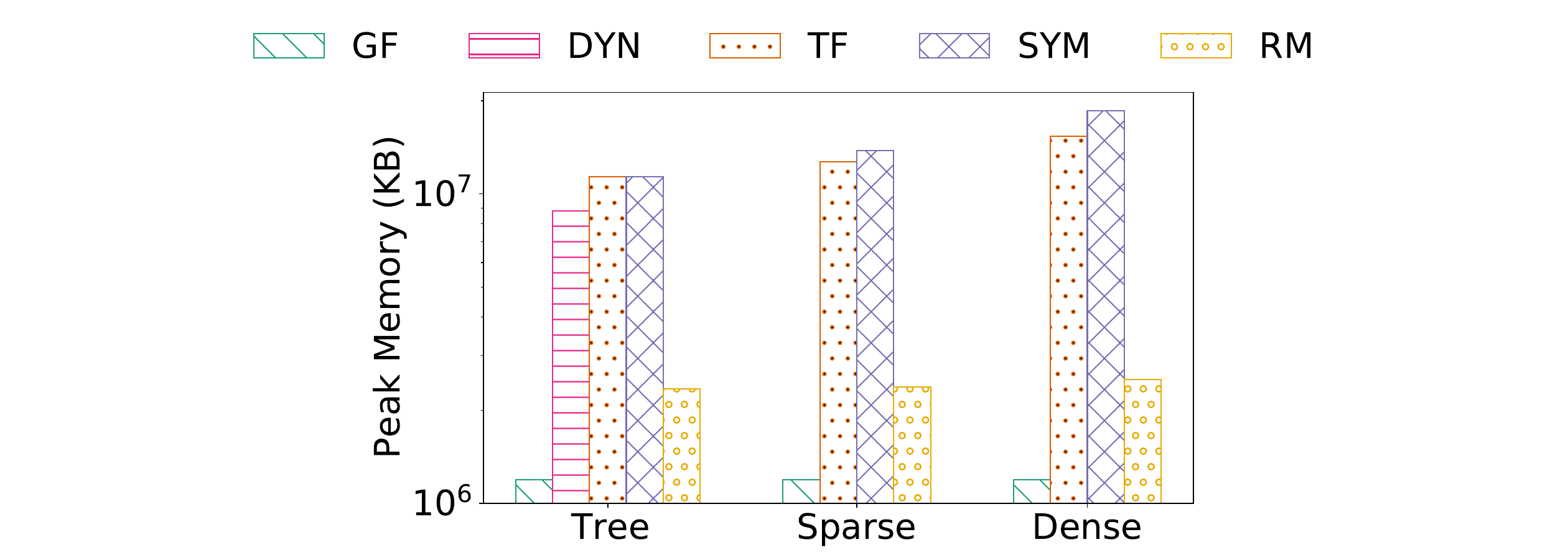}
	\caption{Comparison on memory usage.}
	\label{fig:1-memory-solved}
\end{figure}

\textbf{Memory Usage Comparison.} In this experiment, we report the memory usage of competing algorithms in \emph{ls}. Before exiting each query, we read the peak virtual memory (VmPeak) of the process from the \texttt{proc} filesystem. As shown in Figure \ref{fig:1-memory-solved}, GF uses the least memory among all algorithms because it obtains relations from the data graph, and no index is stored. Other CSM algorithms consume more memory than RM since CSM methods keep multiple candidate sets for each query vertex for the incremental matching. On tree queries, DYN spends slightly less memory than TF and SYM because our framework does not store the index \texttt{MatchBackward} for DYN. On sparse and dense queries, SYM takes more memory space than TF because SYM needs to store all data edges that match each non-tree edge of the query graph in the index.

\subsection{Effectiveness of Individual Techniques}

\begin{figure}[h]
	\setlength{\abovecaptionskip}{0pt}
	\setlength{\belowcaptionskip}{-5pt}
	\centering
	\includegraphics[width=0.475\textwidth]{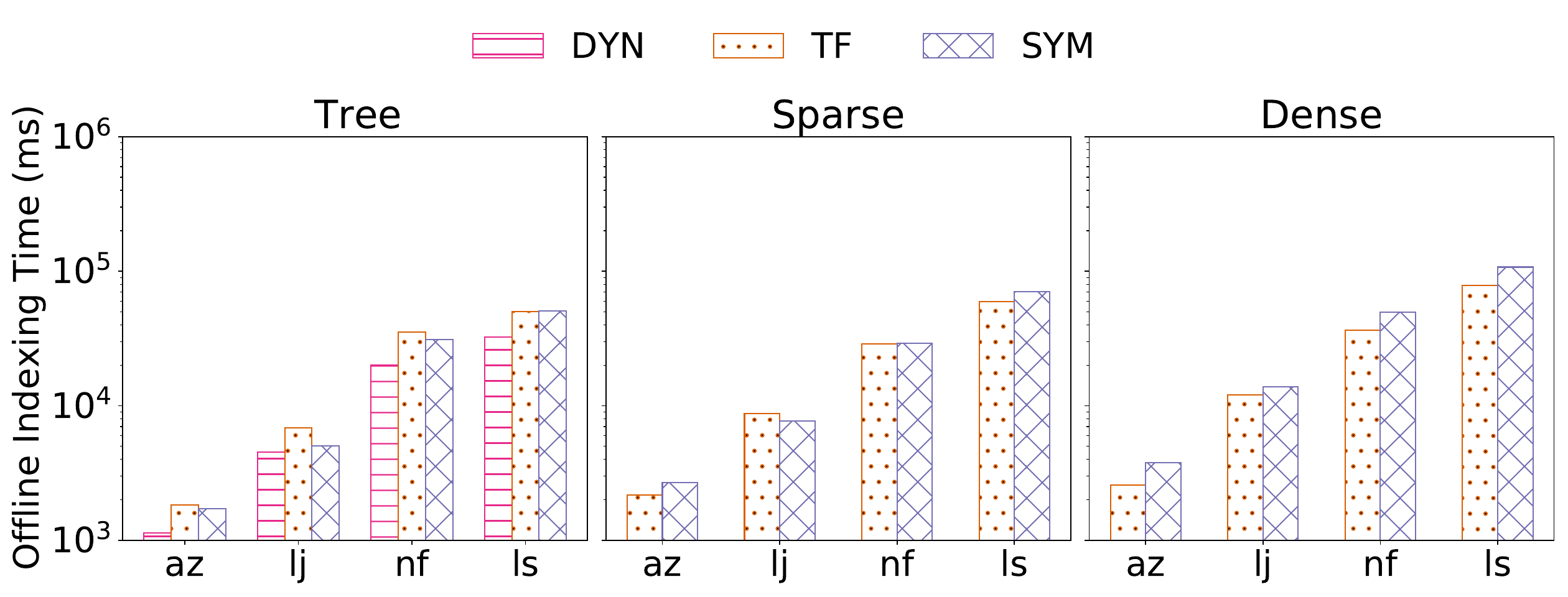}
	\caption{Comparison on offline indexing time.}
	\label{fig:2-offline}
\end{figure}

\textbf{Offline Indexing Time Evaluation.} Figure \ref{fig:2-offline} illustrates the offline indexing time of the index-based algorithms, including DYN, TF, and SYM. On tree queries, DYN spends less time than TF and SYM since DYN only performs the backward phase in the index construction, while TF and SYM conduct both the forward phase and the backward phase. TF takes slightly more time than SYM because the matching orders $\varphi$s of TF for the incremental matching are precomputed based on the enumeration of the index. In comparison, SYM does not need such computation because its matching orders are determined dynamically at the enumeration. However, on sparse queries, the performance gap between TF and SYM decreases, and on dense queries, SYM takes more time than TF. Because, compared to TF, SYM needs to store all data edges that match each non-tree edge of the query graph in the index.

\begin{figure}[h]
	\setlength{\abovecaptionskip}{0pt}
	\setlength{\belowcaptionskip}{3pt}
	\centering
	\includegraphics[width=0.475\textwidth]{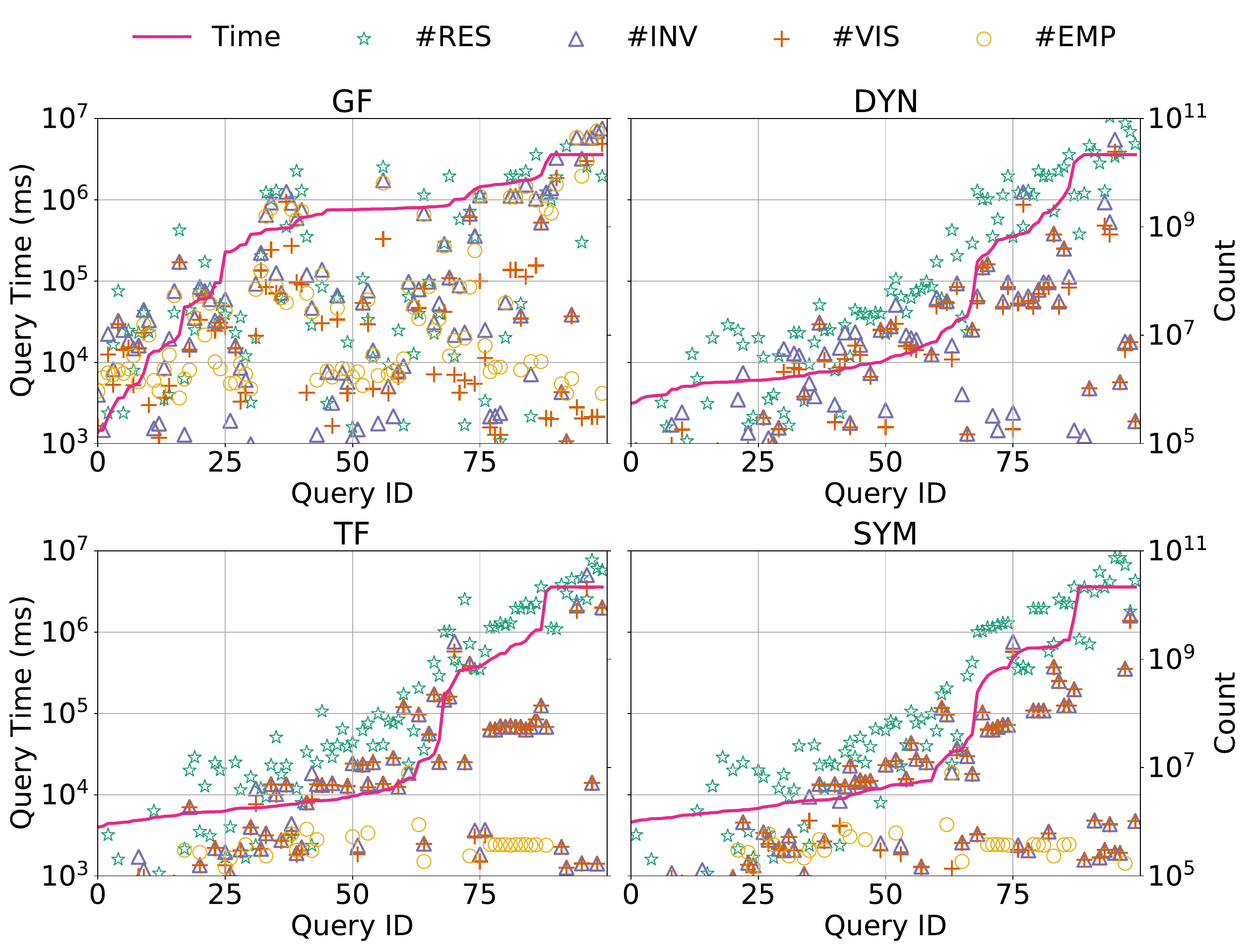}
	\caption{Detailed metrics of tree queries in \emph{ls}.}
	\label{fig:exp-detail-tree}
\end{figure}

\textbf{Detailed Metrics.} We also present the detailed metrics of GF, DYN, TF, and SYM on tree queries in \emph{ls} in Figure \ref{fig:exp-detail-tree}. GF and SYM show similar performance trends to those on sparse queries shown in Figure \ref{fig:exp-detail}. Compared with sparse queries, where the query time of TF is related to the number of invalid partial results, the query time is related to the number of incremental matches on tree queries. Because there is no non-tree edge and the indexes of TF and SYM have similar pruning power. In DYN, there is no \emph{EMP}. Because no local candidate vertex set is empty (Proof of Proposition \ref{prop:iedyn-constant-delay}).

\begin{figure}[h]
	\setlength{\abovecaptionskip}{0pt}
	\setlength{\belowcaptionskip}{-10pt}
	\centering
	\includegraphics[width=0.475\textwidth]{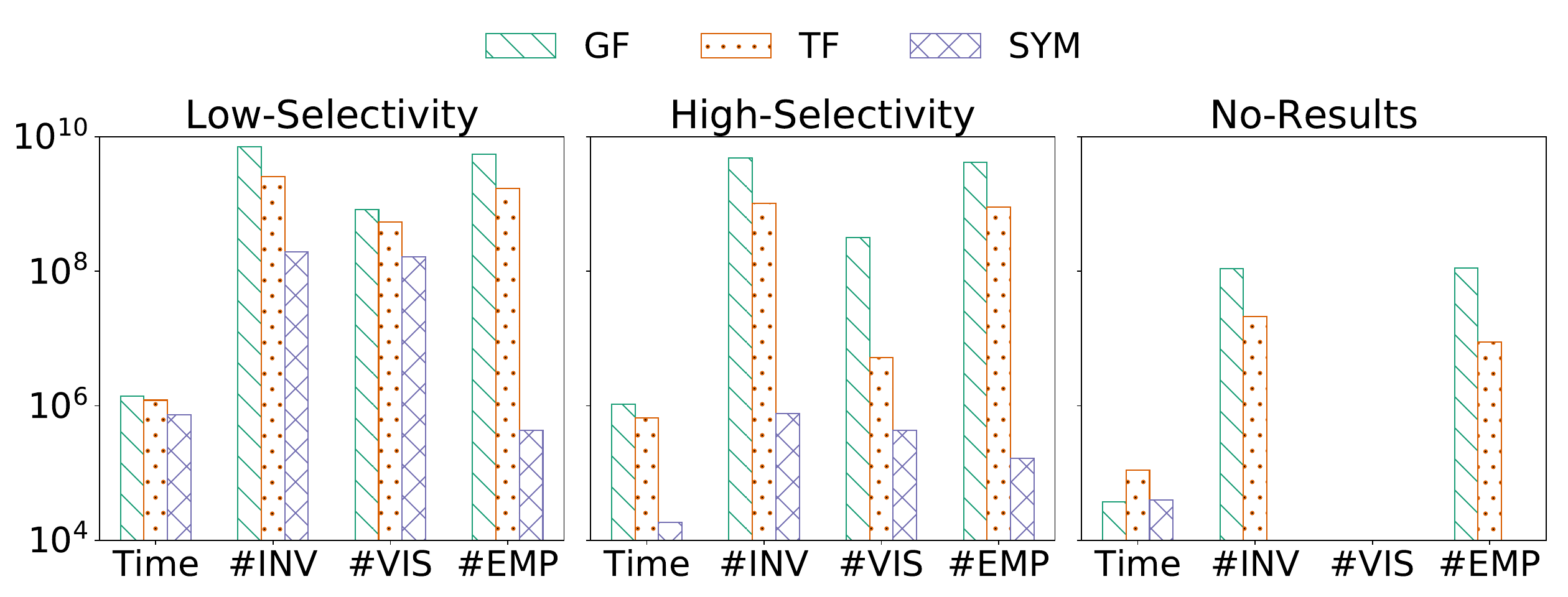}
	\caption{Detailed metrics in \emph{ls}.}
	\label{fig:exp-detail-bar}
\end{figure}

After that, we gather all the 300 queries in \emph{ls} (100 tree queries, 100 sparse queries, and 100 dense queries) together, and create a group containing all the queries with no incremental results, denoted as \emph{No-Result}. For the remaining queries, we get the median of the number of incremental matches and put all queries with the number of incremental matches greater (resp. less) than the median into the \emph{Low-Selectivity} (resp. \emph{High-Selectivity}) group. The detailed metrics of GF, TF, and SYM on the three groups of queries are shown in Figure \ref{fig:exp-detail-bar}, where \emph{Time} denotes the query time of the incremental matches, measured in millisecond. As the number of incremental results becomes smaller, the query time and the number of invalid partial results decrease. The decrease rate of SYM is greater than that of other algorithms, benefiting from its index. However, the relative performance between algorithms is the same across the three groups, with SYM being the best on each selectivity group of queries. Furthermore, for each group, \emph{\#EMP} dominates \emph{\#INV} in GF and TF while \emph{\#VIS} dominates \emph{\#INV} in SYM, which is consistent with the results shown in Figure \ref{fig:exp-detail}.

\subsection{Scalability Evaluation}

In this subsection, we evaluate the scalability of existing algorithms. 

\subsubsection{Varying Query Graph Property} First of all, we analyze the performance given various query graphs.

\begin{figure}[h]
	\setlength{\abovecaptionskip}{0pt}
	\setlength{\belowcaptionskip}{3pt}
	\centering
	\includegraphics[width=0.475\textwidth]{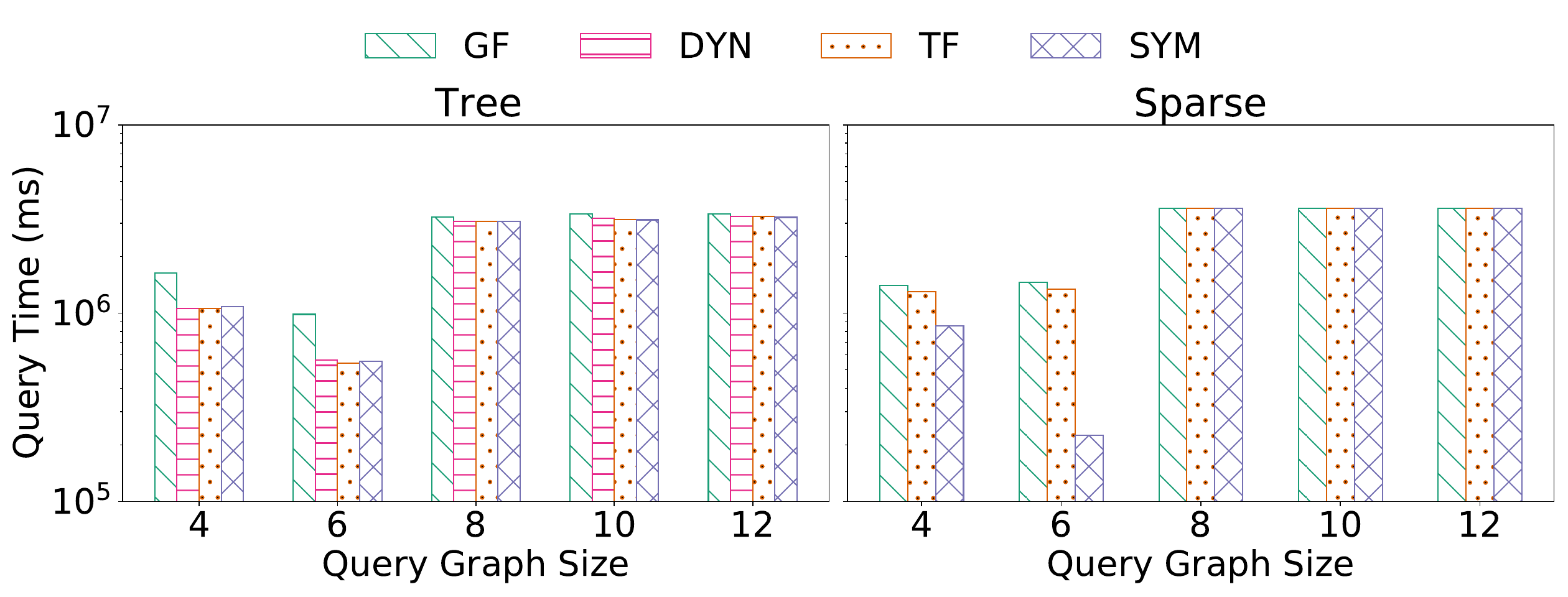}
	\caption{Varying query graph size.}
	\label{fig:exp-varying-query-graph-size}
\end{figure}

\textbf{Varying Query Graph Size.} In this experiment, we perform tree and sparse queries in \emph{ls} with the query graph size varying from 4 to 12 in 2 increments to evaluate the performance trends. Results are shown in Figure \ref{fig:exp-varying-query-graph-size}. On tree queries, DYN, TF, and SYM run faster than GF, while on sparse queries, SYM outperforms other algorithms. Generally, the query time grows as the query graphs become larger, especially for tree queries. However, for each query type, the top performer does not change due to the growth of query graphs.

\begin{figure}[h]
	\setlength{\abovecaptionskip}{0pt}
	\setlength{\belowcaptionskip}{-10pt}
	\centering
	\includegraphics[width=0.475\textwidth]{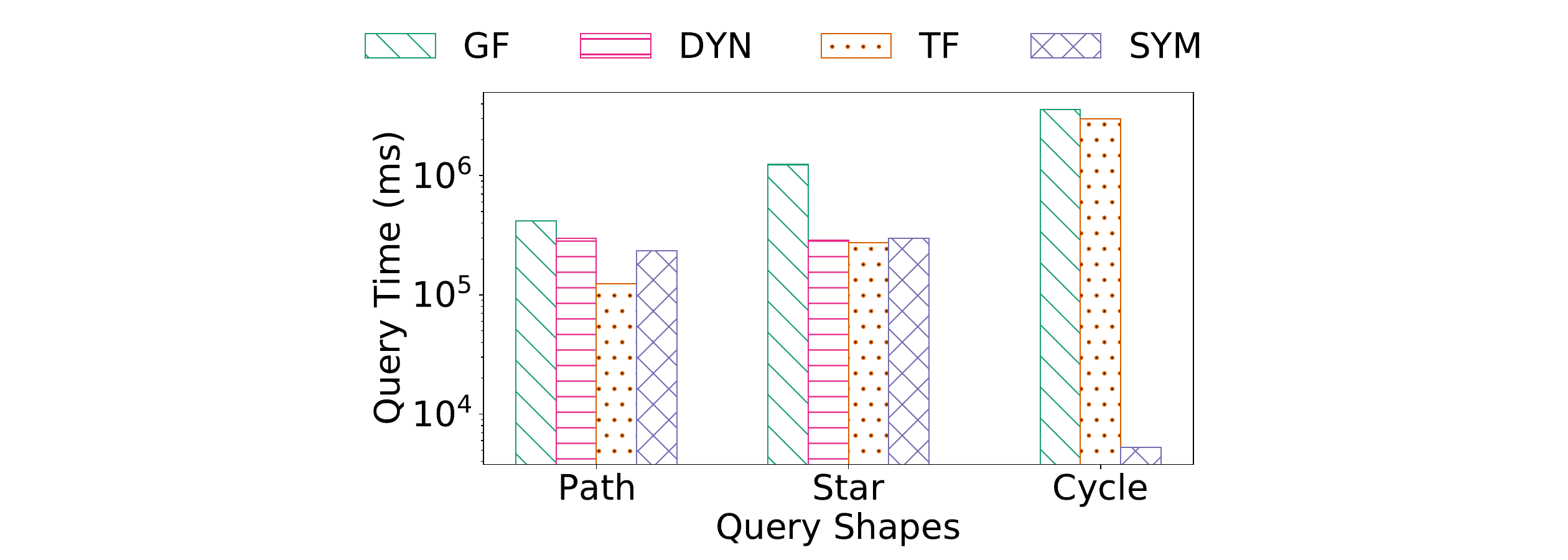}
	\caption{Varying query graph shape.}
	\label{fig:exp-varying-query-graph-shape}
\end{figure}

\textbf{Varying Query Graph Shape.} Previous work has identified the common query shapes in RDF data, including path, star, cycle, and flower shapes~\cite{bonifati2017analytical}. Specifically, a path is a tree where each vertex has at most one child; A star is a tree with two levels; In a cycle, all the vertices have the degree of two; And a graph of the flower shape consists of a vertex with three types of attachments: a path, a tree, and a \emph{petal}, i.e., a graph consisting of a source vertex $s$, target vertex $t$, and a set of at least two node-disjoint paths from $s$ to $t$. Most query graphs in our sparse queries are of the flower shape. Therefore, in this experiment, we evaluate competing algorithms on path, star, and cycle queries. Specifically, we generate 100 query graphs of each shape by random walking in \emph{ls}. The query time is shown in Figure \ref{fig:exp-varying-query-graph-shape}. On path and star queries, the relative performance is similar to that on tree queries shown in Figure \ref{fig:exp-execution-breakdown} since a path or a star is a special case of a tree. On cycle queries, SYM shows significant speedup over GF and TF because its index stores data edges matching the non-tree edges of the query graph, thus reducing the number of invalid partial results.

\subsubsection{Varying Data Graph Property}

Next, we evaluate the competing approaches given data graphs with different properties.

\begin{figure}[h]
	\setlength{\abovecaptionskip}{0pt}
	\setlength{\belowcaptionskip}{3pt}
	\centering
	\includegraphics[width=0.475\textwidth]{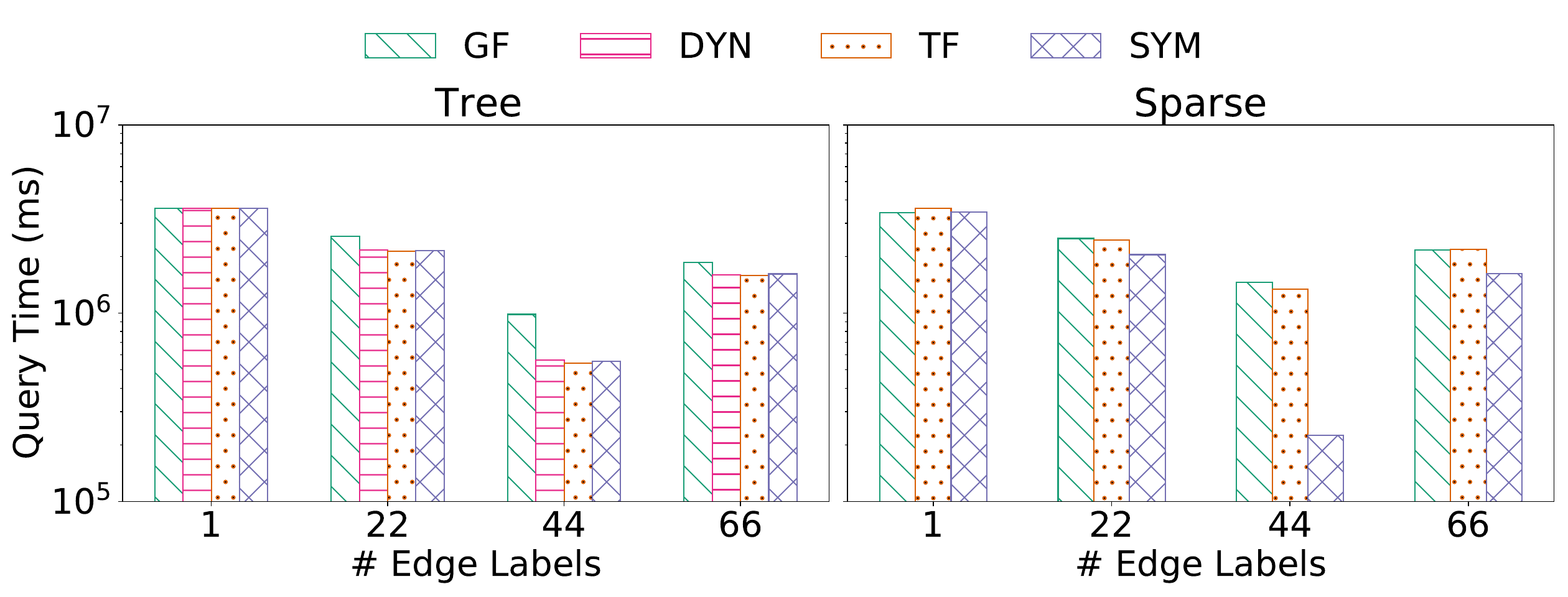}
	\caption{Varying number of labels.}
	\label{fig:exp-varying-number-of-labels}
\end{figure}

\textbf{Varying Number of Labels.} In this experiment, we examine the impact of the number of labels in the data graph. \emph{ls} has 44 edge labels originally. To conduct our experiment, we create other versions of \emph{ls} with 1, 22, and 66 edge labels, respectively. First, to generate a data graph with one edge label, we re-assign all edges by a single label; Next, to generate a data graph with 22 edge labels, we randomly pair all the 44 labels into 22 groups and then use a new label to denote both labels in each group; Finally, to generate a data graph with 66 labels, we first randomly select 22 labels in the original data graph. Then, the edges with each selected label are randomly re-assigned by one of the two new labels. Figure \ref{fig:exp-varying-number-of-labels} presents the performance trends given various number of edge labels. As the number of labels grows, both tree and sparse queries take less time as the local candidate sets represented by Line 5 in Algorithm \ref{algo:wcoj} shrink. Nevertheless, the relative performance between competing algorithms remains stable.

\begin{figure}[h]
	\setlength{\abovecaptionskip}{0pt}
	\setlength{\belowcaptionskip}{-5pt}
	\centering
	\includegraphics[width=0.475\textwidth]{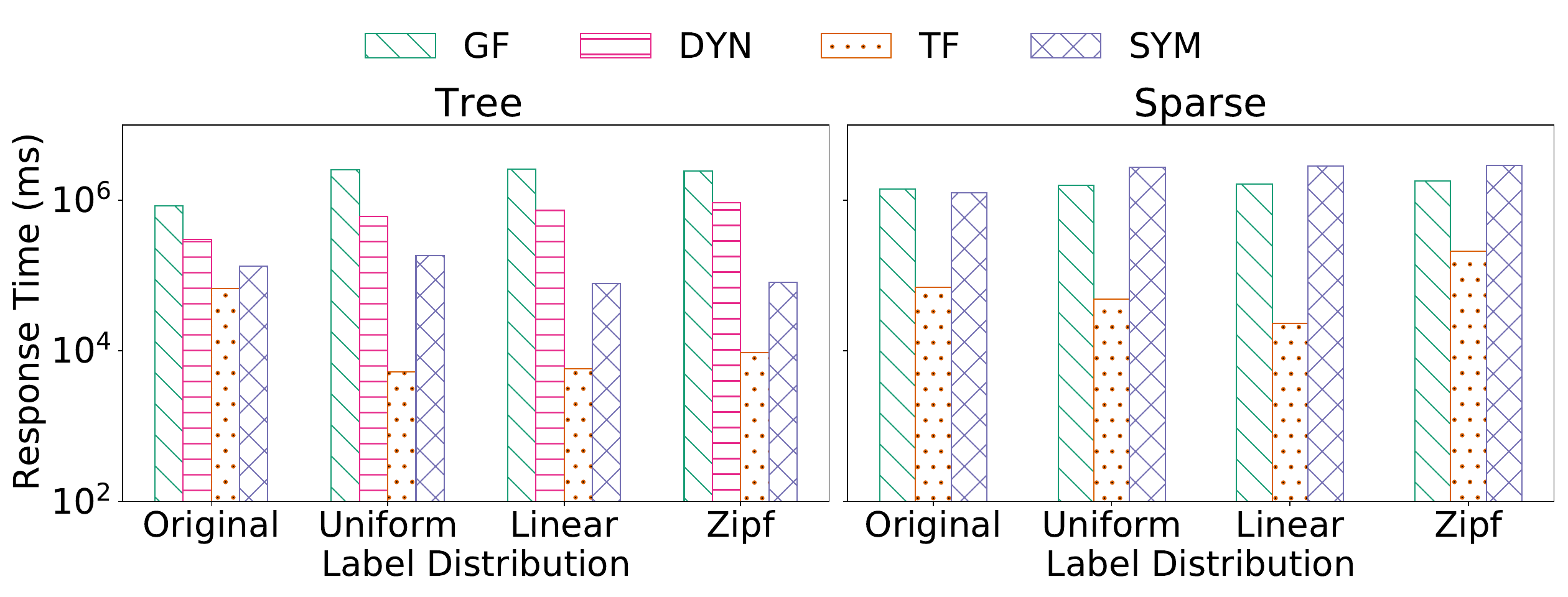}
	\caption{Varying label distribution.}
	\label{fig:exp-varying-label-distribution}
\end{figure}

\textbf{Varying Label Distribution.} Besides the number of labels in the data graph, the label distribution may also affect the query performance. Therefore, in this experiment, we generate other versions of \emph{ls} with the same number of edge labels but different label distribution from the original. Specifically, we consider the uniform, linear, and Zipfian distributions, common in real-world and synthetic graphs. \emph{ls} has 44 edge labels originally. Therefore, to generate a data graph with a certain label distribution, we randomly generate 44 numbers $p_1, p_2,\cdots p_{44}$ that follow such distribution and sum up to 1. Then each edge is re-assigned by a label, where $p_i$ is the probability of label $i$. Figure \ref{fig:exp-varying-label-distribution} shows the response time of competing algorithms on tree and sparse queries. As we can see, the response time fluctuates under different label distributions. However, the relative performance among all algorithms is similar. 

\begin{figure}[h]
	\setlength{\abovecaptionskip}{0pt}
	\setlength{\belowcaptionskip}{-5pt}
	\centering
	\includegraphics[width=0.475\textwidth]{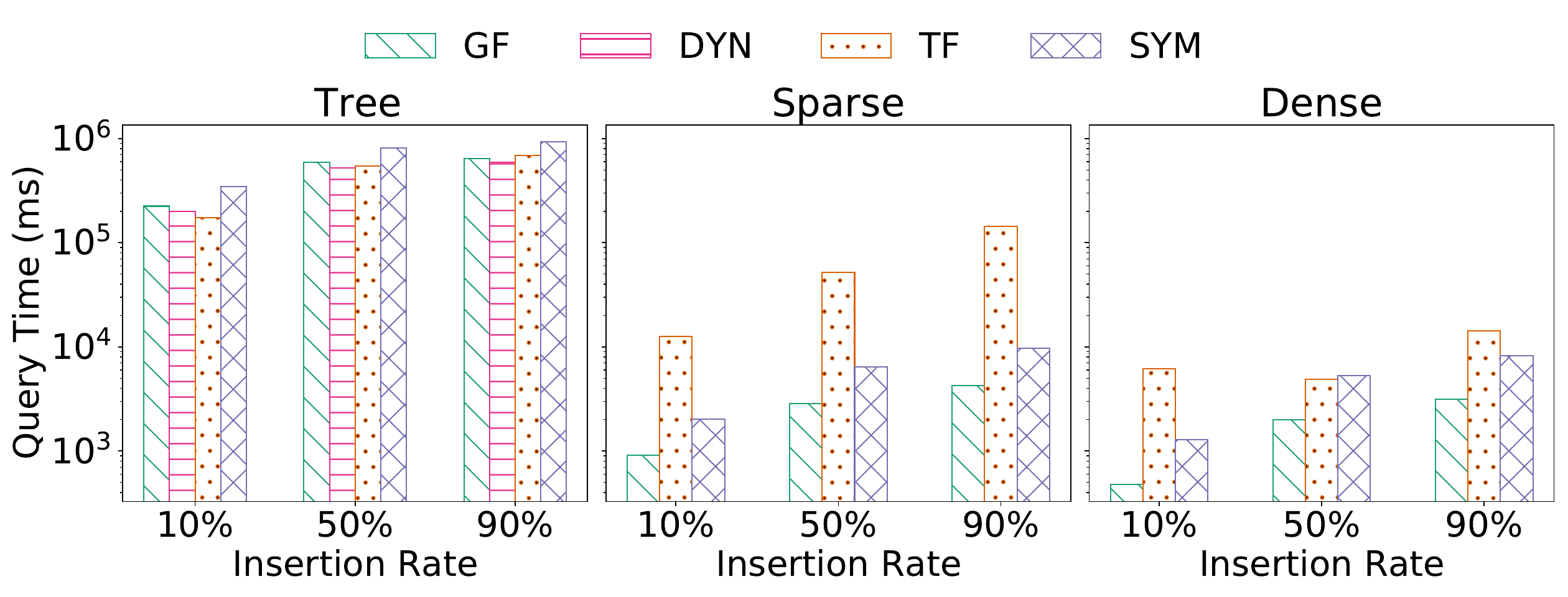}
	\caption{Varying insertion rate.}
	\label{fig:exp-varying-ins-rate}
\end{figure}

\textbf{Varying Insertion Rate.} In this experiment, we check the performance of the algorithms under various insertion rates. The results on \emph{az} with the insertion rate varying from 10\% to 90\% are shown in Figure \ref{fig:exp-varying-ins-rate}. Since the total number of edges in \emph{az} is fixed, the number of graph update operations increases as the insertion rate grows. Therefore all algorithms take more time on incremental matching. Nevertheless, the relative performance of competing algorithms under different insertion rates is similar.

\begin{figure}[t]
	\setlength{\abovecaptionskip}{0pt}
	\setlength{\belowcaptionskip}{3pt}
	\centering
	\includegraphics[width=0.475\textwidth]{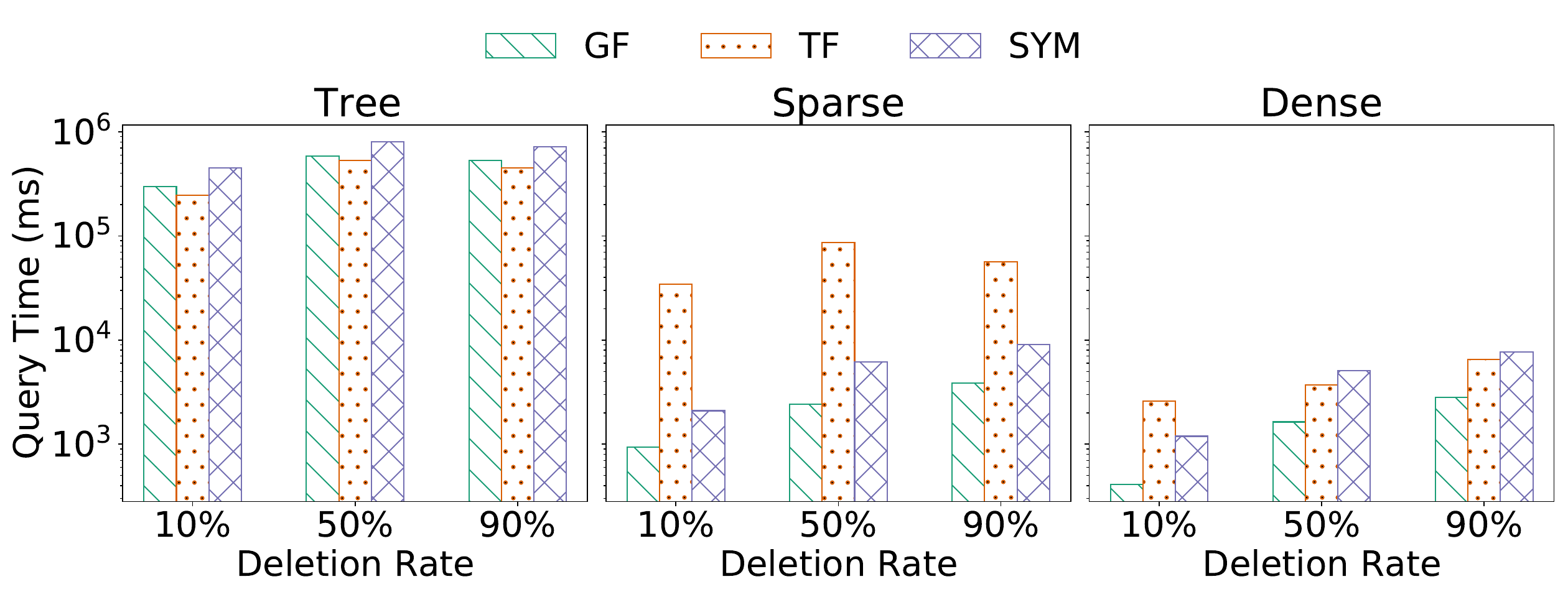}
	\caption{Varying deletion rate.}
	\label{fig:exp-varying-del-rate}
\end{figure}

\textbf{Varying Deletion Rate.} Deletion operation is essential in real-world scenarios, although all previous experiments emphasize edge insertions. To make our experiments extensive, we compare the algorithms on deletion-only dynamic graphs of \emph{az} with the deletion rate varying from 10\% to 90\%. Since DYN cannot report $\Delta M$ on edge deletion, we only experiment on GF, TF, and SYM. Results are presented in Figure \ref{fig:exp-varying-del-rate}. Similarly, as the number of graph update operations increases, all algorithms take more time on the incremental matching. Furthermore, the algorithms show similar performance trends as in the insertion because the insertion and deletion operations are symmetric in the three competing algorithms, shown in Section \ref{sec:algorithms_study}. 

\subsubsection{Varying Updated Edge Property.}

Then, we analyze the impact of the data graph with various types of edges updated.

\begin{figure}[h]
	\setlength{\abovecaptionskip}{0pt}
	\setlength{\belowcaptionskip}{-5pt}
	\centering
	\includegraphics[width=0.475\textwidth]{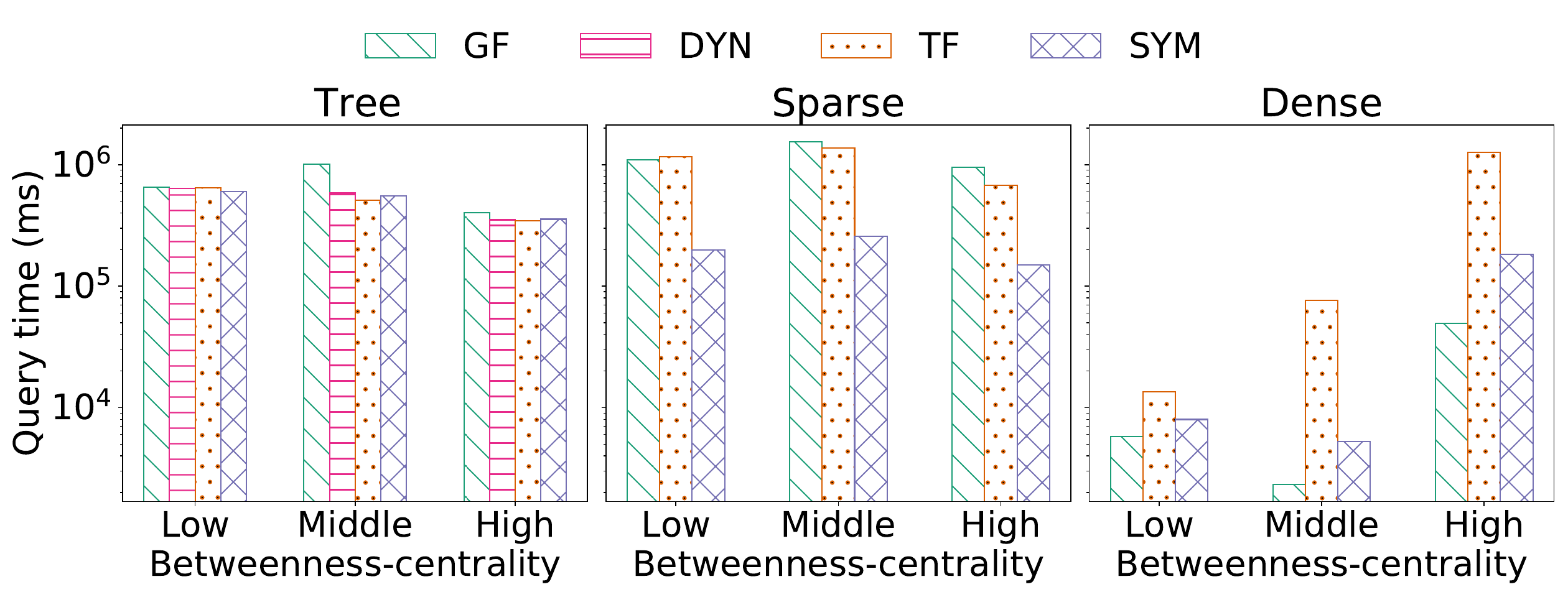}
	\caption{Varying betweenness-centrality.}
	\label{fig:exp-varying-centr}
\end{figure}

\textbf{Varying Betweenness-centrality.} In this experiment, we first compute the betweenness-centrality of each edge in \emph{ls}. Then, we sample edges with high, middle, and low betweenness-centrality as the inserted edges and put all the remaining edges of \emph{ls} to the initial graphs, respectively. Therefore, we get three dynamic graph datasets, of which all the inserted edges are with high, middle, and low betweenness-centrality, denoted as \emph{High}, \emph{Middle}, and \emph{Low}. Figure \ref{fig:exp-varying-centr} shows the query time of competing algorithms. On tree queries, DYN, TF, and SYM perform better than GF. However, SYM is the top performer on sparse queries, and GF is the fastest on dense queries among all algorithms. Although the query time fluctuates in different datasets, the relative performance of competing algorithms remains stable. And the performance trends are consistent with that in the original \emph{ls} dataset, shown in Figure \ref{fig:exp-execution-breakdown}.

\begin{figure}[h]
	\setlength{\abovecaptionskip}{0pt}
	\setlength{\belowcaptionskip}{-5pt}
	\centering
	\includegraphics[width=0.475\textwidth]{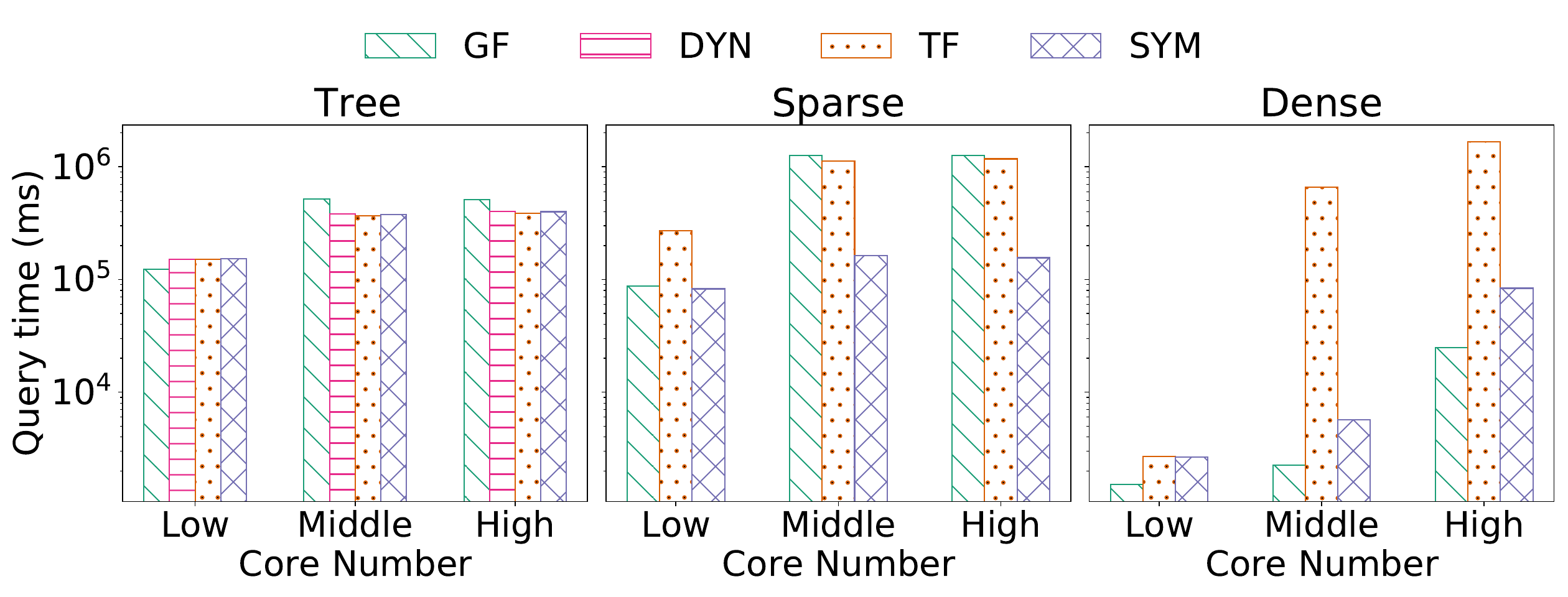}
	\caption{Varying density.}
	\label{fig:exp-varying-core}
\end{figure}

\textbf{Varying Density.} We also study the performance when edges are inserted from regions with various densities. We first extract \linebreak\newpage \noindent $k$-cores with different $k$ values in the graph \emph{ls}. Then, we sample edges in the $k$-cores with high, middle, and low $k$s as the inserted edges and put all the remaining edges of \emph{ls} to the initial graphs, respectively. Therefore, we get three dynamic graph datasets, of which all inserted edges are from $k$-cores with high, middle, and low $k$s, denoted as \emph{High}, \emph{Middle}, and \emph{Low}. Figure \ref{fig:exp-varying-core} shows the query time of competing algorithms. In \emph{Middle} and \emph{High}, the relative performance of competing algorithm is the same as the original \emph{ls} dataset. However, in \emph{Low}, GF outperforms other competitors on all types of queries because the graph structure around an edge sampled from $k$-cores with low $k$ is sparse, and the number of data vertices pruned by the indexes of TF and SYM is limited.

\begin{figure}[h]
	\setlength{\abovecaptionskip}{0pt}
	\setlength{\belowcaptionskip}{0pt}
	\centering
	\includegraphics[width=0.475\textwidth]{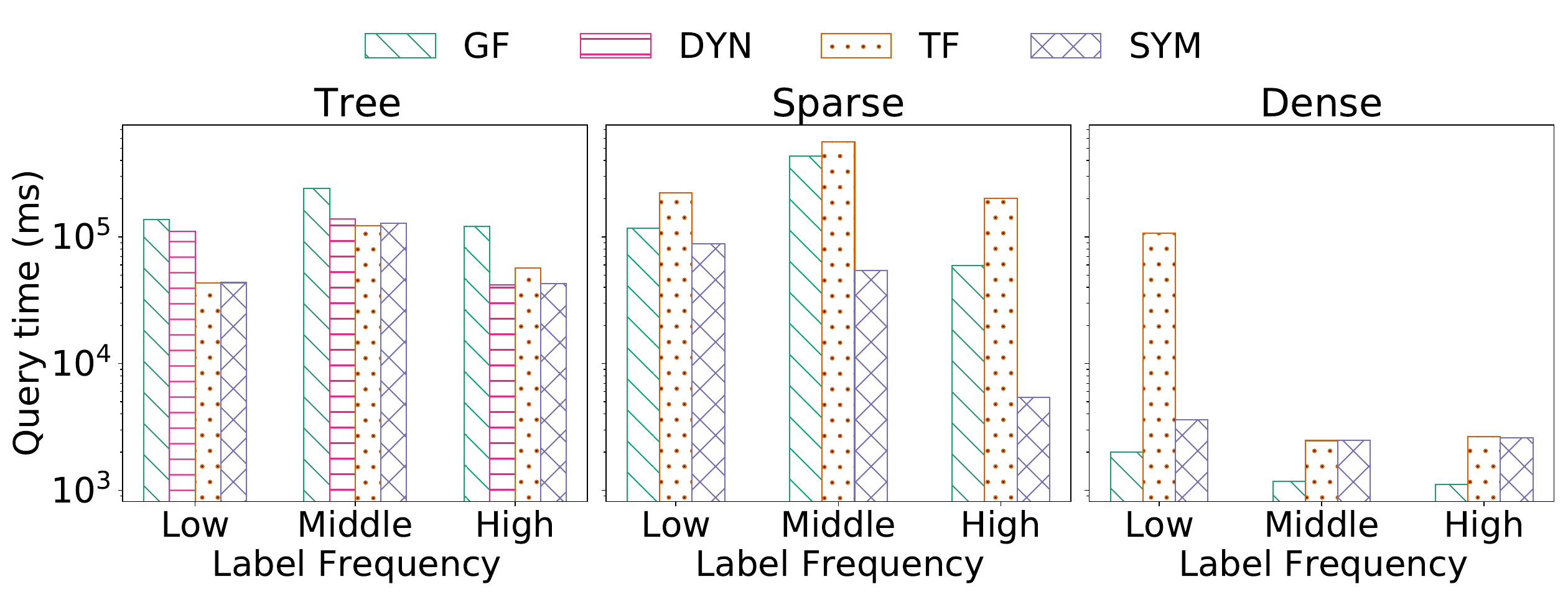}
	\caption{Varying label frequency.}
	\label{fig:exp-varying-label-freq}
\end{figure}

\textbf{Varying Label Frequency.} Additionally, we study the impact of the label frequency of the edge updated. We sample edges with high, middle, and low frequency on the label from \emph{ls} as the inserted edges and put all the remaining edges in \emph{ls} to the initial graphs. Then, we get three dynamic datasets, of which each inserted edge have a label of high, middle, and low frequency in the data graph, denoted as \emph{High}, \emph{Middle}, and \emph{Low}. Figure \ref{fig:exp-varying-label-freq} shows the query time of the algorithms. The performance trends keep consistent in different datasets.

\end{document}